\documentclass[referee]{aa} 
\usepackage{graphicx}
\usepackage{natbib}
\bibpunct{(}{)}{;}{a}{}{,}    
\newcommand{\beq}{\begin{equation}}
\newcommand{\eeq}{\end{equation}}
\newcommand{\bea}{\begin{eqnarray}}
\newcommand{\eea}{\end{eqnarray}}

\newcommand{\subscr}[1]{_\mathrm{#1}}

\newcommand{\url}[1]{{\tt #1}}

\newcommand{\doverd}[2]{\frac{\partial #1}{\partial #2}}
\newcommand{\doverdt}[1]{\frac{\partial #1}{\partial t}}
\newcommand{\lsim}{\mathrel{\rlap{\raise -.3ex\hbox{${\scriptstyle\sim}$}}%
                   \raise .6ex\hbox{${\scriptstyle <}$}}}%
\newcommand{\gsim}{\mathrel{\rlap{\raise -.3ex\hbox{${\scriptstyle\sim}$}}%
                   \raise .6ex\hbox{${\scriptstyle >}$}}}%
\def\gapp{\lower 3pt\hbox{${\buildrel > \over \sim}$}\ }
\def\lapp{\lower 3pt\hbox{${\buildrel < \over \sim}$}\ }

\begin{document}
\title{Simulations of eccentric disks in close binary systems}
\author{
Wilhelm Kley \inst{1},
John~C.B. Papaloizou \inst{2},
\and
Gordon I. Ogilvie \inst{2}
}
\offprints{W. Kley,\\ \email{wilhelm.kley@uni-tuebingen.de}}
\institute{
     Institut f\"ur Astronomie \& Astrophysik, 
     Universit\"at T\"ubingen,
     Auf der Morgenstelle 10, D-72076 T\"ubingen, Germany
\and
     Department of Applied Mathematics and Theoretical Physics,
     University of Cambridge,
     Centre for Mathematical Sciences,
     Wilberforce Road,
     Cambridge, CB3 0WA, United Kingdom
}
\abstract
{
Eccentric accretion disks in superoutbursting cataclysmic 
and other binary systems. 
}
{
We study the development of finite eccentricity
in  accretion disks in close binary systems using a grid-based
numerical scheme. We perform detailed parameter studies to explore the dependence on viscosity,
disk aspect ratio, the inclusion of a mass-transfer stream and the
role of the  boundary conditions.
}
{
Using a two-dimensional grid-based scheme we study the instability
of accretion disks in close binary systems that causes them to attain
a quasi-steady state with finite eccentricity. Mass ratios $0.05 \le q \le 0.3$
appropriate to superoutbursting cataclysmic binary systems are considered.}
{
Our grid-based scheme enables us to study the development of eccentric disks
for disk aspect ratio $h$ in the range $0.01-0.06$ and
dimensionless kinematic viscosity $\nu$ in the range  
$3.3 \times 10^{-6} - 10^{-4}.$ Previous studies using particle-based methods
were limited to the largest values for these parameters on account of their diffusive
nature. Instability to the formation of a precessing eccentric disk
that attains a quasi-steady state with mean eccentricity 
in the range $0.3 - 0.5$ occurs readily. The shortest growth times
are $\sim 15$ binary orbits for the largest viscosities and the instability
mechanism is for the most part 
consistent with the mode-coupling mechanism associated with the 3:1 resonance
 proposed by Lubow.
  However, the results
are sensitive to the treatment of the inner boundary and to the incorporation
of the mass-transfer stream. In the presence of a stream we found a critical viscosity
below which the disk remains circular. 
}
{Eccentric disks readily develop in close binary systems with $ 0.05 \le q \le 0.3.$
Incorporation of a mass-transfer stream tends to impart stability for small
enough viscosity (or, equivalently, mass-transfer rate through the disk) and to
assist in obtaining a  prograde precession rate that is 
in agreement with observations.
 For the larger $q$ the location of the 3:1 resonance is pushed outwards towards
 the Roche lobe where higher-order mode couplings and nonlinearity occur.
 It is likely that three-dimensional simulations that properly resolve the disk's vertical structure are required  to make significant progress in this case.
 }
\maketitle
\markboth
{Kley at al.: Eccentric disks in binary stars}
{Kley at al.: Eccentric disks in binary stars}
\section{Introduction}
\label{sec:introduction}

SU Ursae Majoris systems are a class of cataclysmic variable that
in addition to undergoing ordinary outbursts undergo less frequent 
larger superoutbursts. During these, their  light curves show superhumps which 
have a period generally slightly longer than the orbital period
\citep [for a discussion of relevant observations see for example][]
{{2002A&A...389..478A},{2005PASP..117.1204P}}.

A natural explanation of this phenomenon is that it is associated
with a rigidly precessing eccentric accretion disk.
Superoutbursts and superhumps are found to be limited to cataclysmic binaries 
with mass ratio $q < 0.35.$
Assuming that the difference between the superhump period  and the orbital period
represents the solid body precession of an eccentric disk, \citet{{2005PASP..117.1204P}}
discuss the  establishment of
the expected correlation of this  period with $q.$
   
Studies of the light curve and effects on it associated with the  mass-transfer stream
impacting the disk 
have enabled an eccentricity of $0.38\pm 0.1$ to be  found for the disk in 
 OY Car by
\citet{1992A&A...263..147H}, who also argue that the
normal superhumps are produced by tidal stresses after the maximally extended disk edge
approaches the companion. 
Similar results  for 
 IY UMa, which has a similar mass
ratio, were obtained by
\citet{2001MNRAS.324..529R}.
These authors also noted the potential importance of the impacting accretion 
or mass-transfer stream
on the eccentric disk with its varying outer radius  for understanding  the late superhump light.

These observations motivate an interest in developing
good numerical schemes for simulating accretion disks undergoing tidal
interaction with a secondary binary component.

Although, starting  with \citet{1988MNRAS.232...35W}, there have been many simulations of eccentric disks
using particle-based methods
\citep[e.g.][ and references therein]{1997MNRAS.289..889K, 1998MNRAS.297..323M, 2007MNRAS.378..785S} 
there has been almost no grid-based work
to date that is applicable to close binary systems,
 even though such an approach, apart from constituting a check,
is promising in the context of being able to gain access to a wider region
of parameter space that may be difficult to access with particle-based methods.
\citet{1994A&A...288..807H} performed grid-based simulations, but without
including disk viscosity and was only able to
obtain an eccentric disk when  only the $m=3$ component 
of the companion's tidal potential was retained.
Subsequent grid-based simulations by \citet{1999MNRAS.304..687S} and \citet{2000AcA....50..163K}
also did not show clear evidence for the onset of an eccentric instability.
\citet{2005A&AS..432..757P} solved the linearized equations for the
$m=1$ mode of a free disk
without a binary companion  and considered the possible development of 
a parametric instability when the vertical structure was incorporated.

A study of eccentric disks also has application to other areas of astrophysics
where disks and orbiting companions interact.
Thus \citet{2006A&A...447..369K} have  studied  eccentric disks
in the context of the very small mass ratios appropriate to disk--planet interactions
where their existence may have important consequences for theories
of planet formation.

Motivated by these considerations, the aim of this paper is
to  make an extensive study of the dynamics of two-dimensional disks that lie
in the orbital plane of a binary system, which consists of a primary star and a
lower-mass secondary star in a circular orbit.
The disk is centred on the primary star in a binary system with mass ratio in the range
$0.05 \le q \le 0.3$ thought to be 
relevant for superoutbursting cataclysmic binaries.
We perform grid-based simulations of disks
 that are initialised with circular motion, in order to study
 the onset of an eccentric instability
 and the properties of the resulting quasi-stationary eccentric disk.
Use of  a  grid-based rather than a particle-based
method enables consideration of disks with a wider range of viscosities
and more realistic aspect ratios in the range $0.01-0.05$ that are
presently difficult to access for particle-based methods such as SPH
due to their inherently diffusive nature \citep[but see][]{2004A&A...418..325S}.
We establish numerical convergence and
explore the dependence of the results on the physical parameters of the disk.
We also consider variations of the inner
 boundary conditions and the presence of a mass-transfer stream.
Due to the global nature of the eccentric disk mode, results are found to be sensitive
to these.
When a stream is present we are able to find a transition to stability when the
viscosity is reduced below a critical value. We also find that incorporation
of a mass-transfer stream is important for obtaining values for the
disk precession frequency that are compatible with observations. 

The plan of the paper is as follows.  In Sect.~\ref{sec:hydro-model} we describe the hydrodynamic model.
We then go on to describe the numerical methods we use in Sect.~\ref{subsect:numerics}.
Numerical simulations showing the growth of eccentricity
for a typical disk model are then presented in Sect.~\ref{sec:standard}.
It is found that a quasi-stationary eccentric disk with mean eccentricity in the range
$0.3-0.5$ is often produced.
We explore the dependence of this phenomenon on 
disk viscosity, 
 aspect ratio and the inner boundary condition in Sect.~\ref{sec:physical},
showing that the development of an eccentric disk
is favoured at high viscosity but, because it is a global phenomenon,
it is sensitive to the treatment of the disk inner boundary.
The quasi-stationary eccentric disk is found to precess
as a rigid body. In this section we also study 
the dependence of the disk precession rate
on the disk aspect ratio and boundary conditions showing that
the precession rate is prograde for cool enough disks. 
We then  investigate the dependence of the tendency of the
disk to become  eccentric on the binary mass ratio in
Sect.~\ref{subsec:mass} and discuss numerical resolution demonstrating
the numerical convergence of our results  in Sect.~\ref{subsec:resolution}.

In Sect.~\ref{Instability mechanism} we explore the instability 
mechanism and establish the importance of the $m=3$ component
of the tidal potential for causing the instability
through the mode-coupling mechanism proposed by \citet{1991ApJ...381..259L}.  
Comparison with a linear analysis of eccentric modes is made in Sect.~\ref{sec:linear}.

We go on to investigate the effect of a 
 mass-transfer stream in Sect.~\ref{sec:stream}. 
This is shown to make an important contribution to the disk precession rate
and also to cause the disk to become stable to developing
eccentricity for sufficiently small viscosity.
Finally in Sect.~\ref{Concl} we summarize and discuss our results.

\section{Hydrodynamical model}
\label{sec:hydro-model}
To model a flat two-dimensional disk we generally use a cylindrical ($r, \varphi$) coordinate system centred
on the primary star that corotates with the binary.
The latter is assumed to have a circular orbit and thus to rotate uniformly.
This coordinate system  is 
convenient because the positions 
of the two stars remain fixed during the simulations. 
We have checked that the same results
are obtained  when simulations are run in the inertial frame.
The disk material is treated as a viscous fluid with a constant kinematic shear viscosity and 
(except where noted below) vanishing bulk viscosity.

For a flat disk located
in the orbital plane $z=0,$ the
velocity components are ${\bf u} = (u_r, u_{\varphi}, 0).$
Thus  $u=u_r$ is the radial velocity
and the local disk angular velocity is $\Omega = u_{\varphi}/r$ as measured in the corotating frame which rotates with the constant angular
velocity, $\omega$,  of the binary.
 
The vertically integrated  continuity equation  is
\begin {equation} 
 \doverdt{\Sigma} + \nabla (\Sigma {\bf u} )
                =  0.
\label{Sigma}
\end{equation}
The vertically integrated radial component of the equation of motion is
\begin {equation} 
 \doverdt{(\Sigma u_r)} + \nabla (\Sigma u_r {\bf u} )
  = \Sigma \, r ( \omega + \Omega)^2
        - \doverd{p}{r} - \Sigma \doverd{\Phi}{r} - \Sigma b_{r} + f_{r}\, ,
      \label{u_r}
\end{equation}
and the azimuthal component can be written in the form
\begin {equation}
 \doverd{[\Sigma r^2 (\omega + \Omega)]}{t} 
   + \nabla [\Sigma r^2 (\omega + \Omega) {\bf u} ]
         =
        - \doverd{p}{\varphi} - \Sigma \doverd{\Phi}{\varphi}
   - r \Sigma b\subscr{\varphi} + r f\subscr{\varphi}.
      \label{u_phi}
\end{equation}
Here $\Sigma$ denotes the surface density
\[  \Sigma = \int^\infty_{-\infty} \rho dz,\]
where $\rho$ is the density, $p$ the
vertically integrated (two-dimensional) pressure. The
gravitational potential $\Phi$ is due  the primary
with mass $M_1$ and the secondary having mass $M_2$ and  is given by
\[
    \Phi = - \frac{G M_1}{| {\bf r} - {\bf r}_1 |}
       -  \frac{G M_2}{| {\bf r} - {\bf r}_2 |},
\]
where $G$ is the gravitational constant and ${\bf r}_1$ and
${\bf r}_2$ are the position vectors of $M_1$ and $M_2,$ respectively.
Since the mass of the disk in cataclysmic variables is only about
$10^{-10} M_{\odot}$ we can safely neglect the self-gravity of the disk
as well as its influence on the binary orbit.
The additional terms $\vec{b} = (b_r, b_\varphi)$ take into account the
 acceleration of the origin of the coordinate system. 
Rotation of the coordinate axes 
is  taken account  of through the terms involving $\omega.$
The  viscous force per unit area is ${\bf f} = (f_r,f_{\varphi}).$
The explicit form of the components of  ${\bf f}$ is given in \citet{1999MNRAS.303..696K}.

For simplicity we use a locally isothermal equation of state
where the vertically integrated
pressure $p$ is related to the surface density and
sound speed through
\begin{equation}
     p = \Sigma \, c\subscr{s}^2.
\end{equation}
The local isothermal sound speed $c\subscr{s}$ is taken to have 
a fixed  dependence on radius
and is given 
by
\beq   \label{eq:cs}
    c\subscr{s} = \frac{H}{r} \, u\subscr{Kep} \, \equiv h \, u\subscr{Kep},
\eeq
where $u\subscr{Kep} = \sqrt{G M_1/r}$ denotes the Keplerian orbital velocity
of the unperturbed disk that would exist in the absence of the binary companion.
Equation (\ref{eq:cs}) follows from vertical hydrostatic equilibrium.
The constant ratio $h$ of the vertical height $H$ to the
radial distance to the primary, $r,$ is taken as a fixed input parameter.
Here we adopt a
standard value of
\beq 
     h =  H/r =  0.05  \label{hr}
\eeq
This is somewhat large when compared  to values expected for
disks in cataclysmic binary systems \citep{2007MNRAS.378..785S},
 but smaller values are also considered below.

\subsection{Dimensionless units}
  We  adopt a system of units for which 
the unit of length is taken to be the orbital separation of the 
binary,
$a\subscr{bin}$.
The unit of time, $t_0,$  is taken to be 
 the reciprocal of the  orbital angular frequency
$\omega$ of the binary. Thus
\beq  \label{t_0}
  t_0 = \omega^{-1} =
  \left( \frac{a\subscr{bin}^3}{G (M\subscr{1} + M\subscr{2})} \right)^{1/2},
\eeq
 The orbital period of the binary is then given by
\beq
   P\subscr{orb} = 2 \, \pi t_0 \label{eq:Pbin}.
\eeq
The results of  calculations 
will usually be expressed as a function of the evolution time
 in units of $P\subscr{orb}$. 
The unit of velocity is 
given by $u_0 = a\subscr{bin}/t_0$.
Setting the gravitational constant $G$ to be unity,
Eq.~(\ref{t_0}) can be also viewed as  specifying the unit of mass
to be the total mass of the system $M = M_1 + M_2.$
The surface density can be scaled by any constant value    
$\Sigma_0,$ because this drops out of the governing equations.  Accordingly
the surface density may be  scaled to   make the total disk mass
be any specified value. 
The kinematic viscosity coefficient, $\nu,$ is  expressed in units of 
$\nu_0 = a\subscr{bin} u_0 = \omega a\subscr{bin}^2.$

\subsection{Numerical methods}
\label{subsect:numerics}
The   governing Eqs.~(\ref{Sigma})--(\ref{u_phi})   determining the
evolution of  $\Sigma$, $u$ and $\Omega$ are
solved using an Eulerian finite-difference scheme.
The computational domain $[r\subscr{min}, r\subscr{max}] \times
[\varphi\subscr{min}, \varphi\subscr{max}]$
is subdivided into $N_r \times N_\varphi$ 
grid cells that are equally spaced in each coordinate  direction. 
For  typical runs we use $N_r = 200, N_\varphi = 200,$
though many have been checked at the higher resolution  $N_r = 300, N_\varphi = 300.$  

The numerical method utilises a staggered mesh, 
where scalar quantities are located at  cell centres and vector quantities
at cell interfaces. We adopt an
operator-split scheme for which  advection is based on the 
spatially 2nd order monotonic transport algorithm of \citet{1977JCoPh..23..276V}, and
the viscosity can be treated either explicitly or implicitly.
The basic features of   the two-dimensional code {\tt RH2D}
that we use  have been described
in more detail in \citet{1989A&A...208...98K}.  The 
use of the code  in cylindrical  $(r,\varphi)$ coordinates has been described 
in \citet{1999MNRAS.303..696K}.

  Implementation in a rotating coordinate system
requires special treatment of the Coriolis terms 
to ensure angular momentum conservation
\citep{1998A&A...338L..37K}.
To prevent possible instabilities due to shocks,  that could
for example be generated by the
tidal perturbation of the 
secondary, we use 
an artificial {\it bulk} viscosity as described in \citet{1999MNRAS.303..696K}.
This is in addition to the physical shear viscosity $\nu.$ 
We note, that an excellent test of the correct implementation of viscosity in
numerical simulations of accretion disks is the study of non-axisymmetric instabilities
in the viscously spreading ring problem \citep{2003A&A...399..395S}.

In addition, to prevent numerical instabilities due to very low densities (or very steep density
gradients) near the outer  boundary of the disk, where it is tidally 
truncated by the secondary,
we introduce a `density floor', $\Sigma\subscr{floor}.$ This is applied such that
whenever evolution causes  the density at any location
to fall below $\Sigma\subscr{floor},$ it is reset to this value.
 Formally, this  implementation
does not conserve mass exactly, but for sufficiently low values it does not  produce
any influence on the dynamics. 
In all simulations we use a value of $\Sigma\subscr{floor}=10^{-8}$
expressed  in  units  such that the initial value of  
$\Sigma$ at $r=a\subscr{bin}$ would be unity in the absence of tidal truncation (i.e.~in units of $\Sigma_0$ defined in Eq.~\ref{sigma0} below).
We  test the dynamical influence of the floor
in section~\ref{subsec:inner} below.

Since this is an explicit code, the timestep limitation is given by the Courant condition
\beq
    \Delta t = f_c \, \min_{grid} \frac{(\Delta r, r \Delta \varphi)} { | \vec{u} | + c_s }
\eeq
where the Courant number $f_c$ has to be smaller than one, and we typically use $f_c = 0.70$. 
After a few test simulations it soon became clear that, due to the large
range in radius with $r\subscr{max}/r\subscr{min}$ typically larger than 10,
the required computational times become excessive because of the stringent timestep limitation
arising from  the inner boundary.   This occurs because  the  Keplerian
 angular velocity  increases  rapidly as the primary star is approached according to
$\Omega_K \propto r^{-3/2}$. This results in approximately
25,000 timesteps being required  for  one binary orbit with our standard resolution.
As typically a few hundred orbital times need to be calculated per run, this necessitates
about 10 million timesteps for each run.
  Exploration of a large parameter space  would be prohibitive.
For that purpose we implemented the {\tt FARGO} algorithm
\citep{2000A&AS..141..165M} into our {\tt RH2D} code.
In  this algorithm, for each annular grid ring,
the mean angular velocity $\bar{\Omega}$ is calculated
and all variables are simply shifted by an integer number of grid cells, corresponding
to an  advection velocity as close as possible to $r\bar{\Omega}.$
Advection is then only performed  using the residual velocity, leading to a substantially
increased timestep.
This method  leads,
for our standard parameters using an equidistant grid, to a speed-up by a  factor of about 5.

Additional test simulations have shown that  for cases  that do not use the
{\tt FARGO} algorithm, numerical instabilities due to  very small
densities are likely to occur  in the outer region.   These could only be avoided
by   further reducing the timestep or adopting  an implicit treatment of  viscosity, both of which   
result in  additional  computational cost.
A higher density floor also helps to reduce these instabilities but to avoid them
entirely (and still use a large timestep) one would need unreasonably large values that
possibly influence the dynamics; see the discussion on boundary conditions in Sect.~\ref{subsec:inner}
below. 

 Here, the use of the faster advection algorithm enhances the
stability properties,  allowing an explicit treatment of viscosity.
A few comparison simulations that do not use {\tt FARGO}, but allowed a reduced
timestep,  were performed
 to check the accuracy of our method 
 using the {\tt NIRVANA} code \citep{1998CoPhC.109..111Z}.
Finally, we note that the order of the applied numerical scheme matters in obtaining the eccentric state
of the disk. Switching to a simple first order upwind scheme does not result in eccentric
disks no matter how fine the grid resolution is chosen. For the 2nd order upwind scheme the
different choices of the slope-limiter all yield very similar results.

\begin{figure}[ht]
\begin{center}
\rotatebox{0}{
\resizebox{0.98\linewidth}{!}{%
\includegraphics{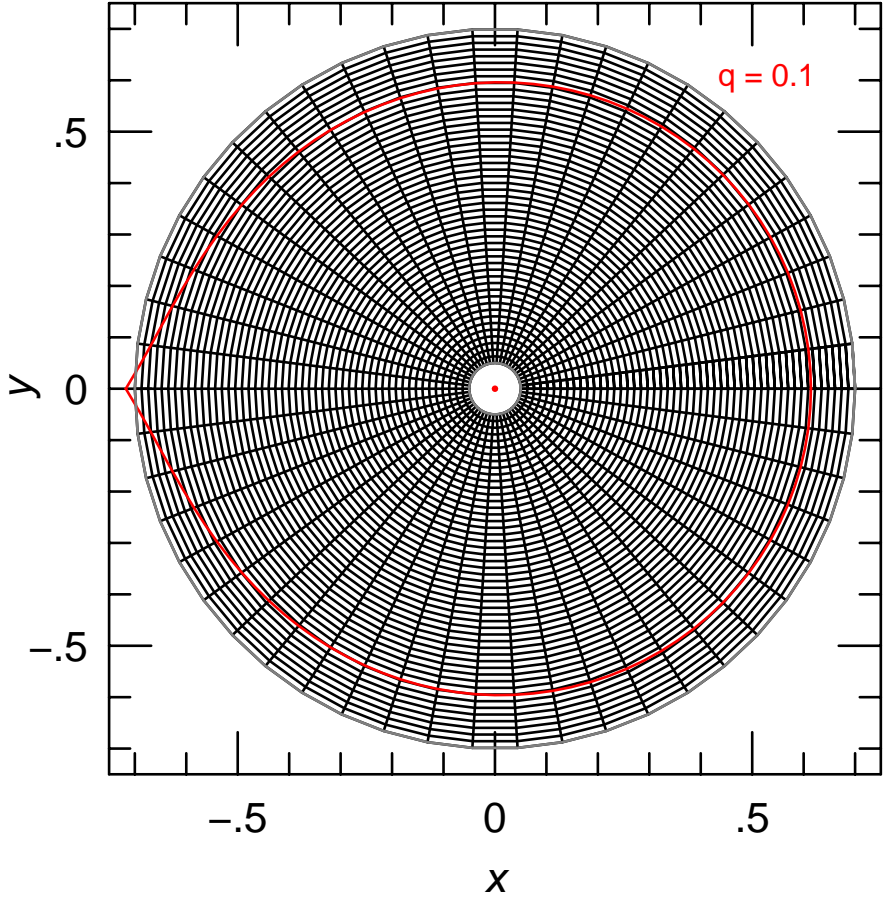}}}
\end{center}
 \caption{The grid structure  used in the computations. The radial domain is  
$[r\subscr{min}, r\subscr{max}] = [0.05, 0.70]$ and   the $\varphi$ domain 
is the whole annulus $[0, 2\pi].$
For our standard case this is covered with $N\subscr{r} \times N_\varphi$ = $200\times200$ 
grid cells that are equally spaced in each direction. The Roche lobe of the primary, for 
our standard mass ratio $q = M_2/M_1 = 0.1,$ is
defined by the additional solid (red) line. The primary and secondary star are located at 
($x=x_p=0, y=y_p=0$) and ($x=x_s=-1,y =y_s=0$) respectively. 
Note that in this figure, for illustrative purposes,
we display a low-resolution $50 \times 50$ grid.
}
   \label{fig:grid}
\end{figure}

\subsection{Setup and boundary conditions}
\label{subsec:setupbc}
The coordinate system is centred on the primary star and, because it
corotates with the binary system, the position of the secondary in Cartesian coordinates is
fixed at $x=-1, y=0$ (see Fig. \ref{fig:grid}).   
For the radial domain we adopt $[r\subscr{min}, r\subscr{max}]$ where
$r\subscr{min}=0.05$ and $r\subscr{max}=0.70$ for all models 
in units where the binary separation is unity.
The azimuthal domain is taken to be  
 $[ 0, 2\pi].$ 

For the initial surface density profile, for $r <0.4$
in dimensionless units, we adopt a power law of the form
\beq
      \Sigma(r, \varphi) = \Sigma_0 \, r^{-1/2}. 
\label{sigma0}
\eeq
For $r>0.4,$ $\Sigma$ is reduced smoothly towards
a minimum value of $10^{-8} \Sigma_0$. 
The initial velocity is taken to be  pure Keplerian rotation around the primary star
as would be measured in an inertial frame
($u_r=0, r\Omega = (G M_1/r)^{1/2} - r\omega$).   No correction is  made
for pressure effects or the non-axisymmetric binary potential. The
 presence of the secondary star is allowed to cause the surface density
 distribution to evolve away from the initial
profile  on a dynamical timescale.
This normally causes a rapid adjustment of the
outer profile through tidal truncation of the disk.
The fixed temperature stratification is always given by $T(r) \propto r^{-1}$
which follows from the assumed constant aspect ratio  $h = H/r = {\rm constant}.$  

In the azimuthal ($\varphi$) direction, periodic boundary conditions
for all variables are imposed. At the outer radial boundary of the grid
which extends somewhat beyond the Roche lobe of the primary 
(see Fig.~\ref{fig:grid}) we use standard outflow
boundary conditions. This imposes
zero gradient conditions on all variables apart from the radial velocity, i.e. the values in the ambient ghost cells
have the same values as the last active grid point.
The radial velocity at the ambient ghost cell
is set equal to the value at $r = r\subscr{max}$
if the flow direction is outward (zero gradient condition) and zero otherwise.
This outflow condition allows material to leave the system but inflow is
not permitted.

As we found that  the form of the boundary condition  applied at the inner boundary
may play an important role in determining the characteristics of the
flow, we implemented four different boundary conditions at $r\subscr{min}$: 
\begin{itemize}
\item {\it Rigid reflecting}: Here the inner boundary acts as a solid wall
where the radial velocity is set to zero, while for the other variables mirror symmetry
conditions are applied to determine values in ghost cells.
\item {\it Viscous outflow}: Here an axisymmetric form for the radial velocity at
the inner boundary is specified. This is directed inward with a magnitude
given by
\beq
     u(r\subscr{min}) = - \frac{3}{2} \, \frac{\nu}{r\subscr{min}}.
\eeq
 This boundary condition imposes  a
steady-state accretion
disk flow.
\item {\it Damping condition}: All variables within a certain region near the
inner boundary are relaxed on the orbital timescale at $r\subscr{min}$
towards their initial values. This approach reduces wave
reflection from the inner boundary and is described in
\citet{2006MNRAS.370..529D}.
In addition $u_r(r\subscr{min})$ is fixed at  zero for all time.
\item {\it Free outflow}: This boundary condition corresponds to the outflow 
condition applied at $r\subscr{max}.$  The radial flow velocity at $r\subscr{min}$ can only
be directed inwards and no inflow from the inner regions into the 
computational domain is allowed. 
\end{itemize}
We found from our numerical simulations that
the form of the inner boundary condition can make quite a difference
in the results concerning the eccentricity and precession rate of the disk.

\section{Standard model}
\label{sec:standard}
To study the effect of varying the physical and numerical
parameters defining the simulations, we first consider in detail a standard model 
for use 
as a reference for the extensive parameter studies  we have performed.
The parameters defining this model are summarized in Table~\ref{tab:standard}.
\begin{table}[ht]
\centerline{
\begin{tabular}{ll}
\hline
Mass ratio ($q =M_1/M_2$)  &   0.1  \\ 
Disk thickness  ($h=H/r$)  &   0.05  \\ 
Viscosity   ($\nu$)  &   $10^{-5}$   \\ 
Radial range ($[r\subscr{min}, r\subscr{max}]$)  &  $[0.05, 0.70]$  \\ 
Grid  ($N_r \times N_\varphi$)  &  $200 \times 200$   \\ 
Inner boundary   &  reflecting  \\ 
\hline
\end{tabular}
}
\caption{Parameters of the standard model. Values are quoted in dimensionless
units.
} 
\label{tab:standard}
\end{table}

\begin{figure}[ht]
\begin{center}
\rotatebox{0}{
\resizebox{0.98\linewidth}{!}{%
\includegraphics{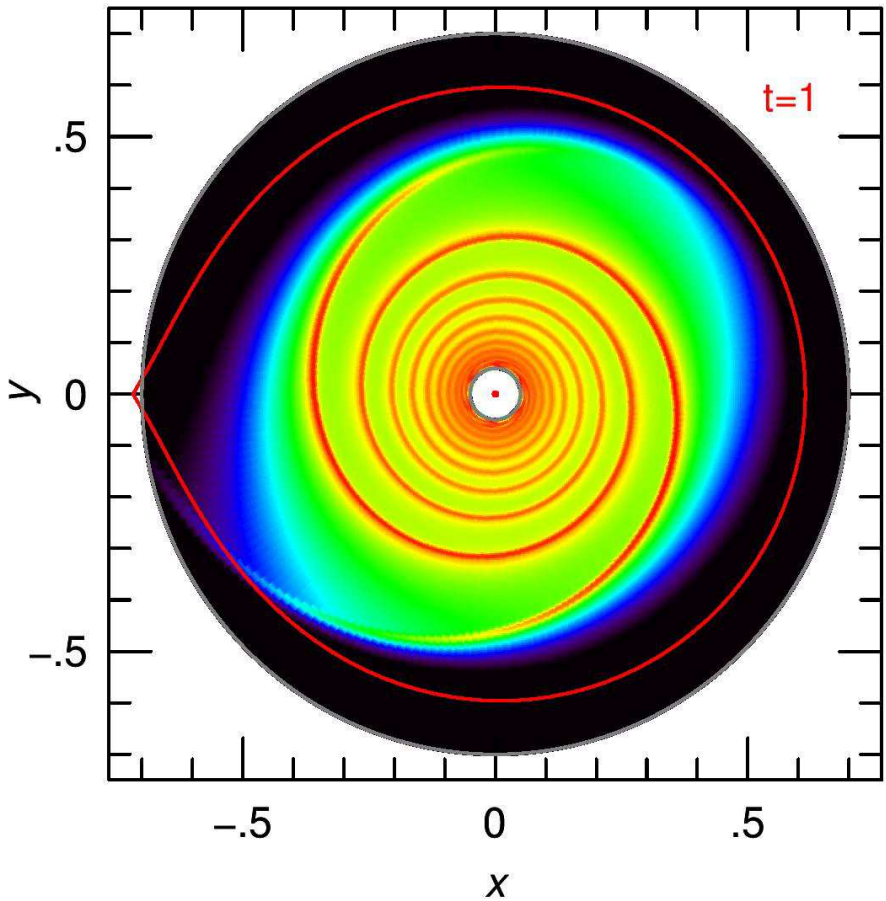}}}
\end{center}
\caption{Greyscale plot of the  surface density of the disk for the
standard model after 1 binary orbit.
The solid (red) curve indicates the Roche lobe of the primary star (central red dot).
}
   \label{fig:m01f-sig1xy}
\end{figure}

\begin{figure}[ht]
\begin{center}
\rotatebox{0}{
\resizebox{0.98\linewidth}{!}{%
\includegraphics{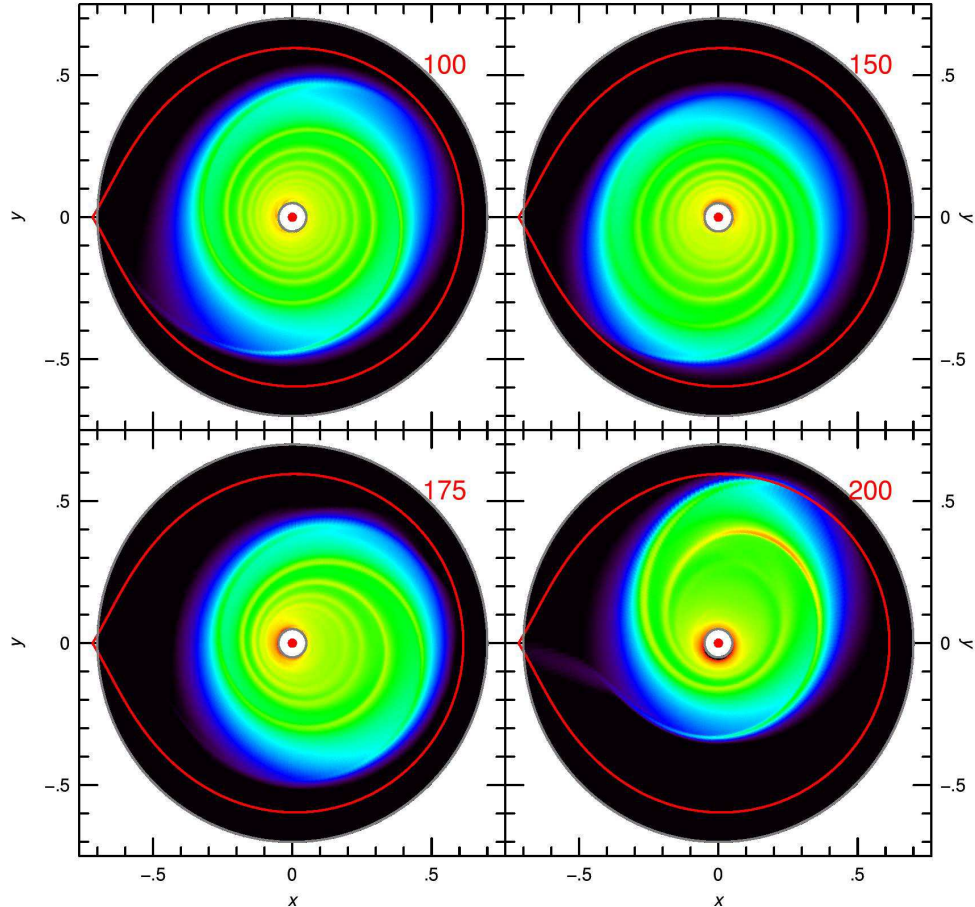}}}
\end{center}
  \caption{Greyscale plot of the  surface density of the disk for the
standard model at $100, 150, 175$ and $200$ orbital periods. The solid (red) curve
indicates the Roche lobe of the primary star (central red dot).
}
   \label{fig:m01f-sig4xy}
\end{figure}

We  now describe the characteristic behaviour  of the standard model.
Due to the strong gravitational perturbation induced by the  secondary,
 the initial axisymmetry of the disk is rapidly destroyed and a strong two-armed spiral density
  response is formed. This is   shown in  Fig.~\ref{fig:m01f-sig1xy} where the surface  density
is displayed after the system has evolved for  one orbital period of the binary. 
The two-armed spiral which is strongest in the
outer disk region but extends all the way through the disk down to the primary star
is clearly visible.
The surface density shows some evidence for a non-axisymmetric pattern
near the inner boundary of the disk, confined to the innermost
two radial gridcells. Such a feature (here a $m=6$ mode) is generated
typically in grid-based simulations with reflecting inner boundary conditions, and is
not a consequence of using the {\tt FARGO}-algorithm.
It can be reduced or minimised by using special,
non-reflecting or damping, boundary conditions.
After the onset of the eccentric instability this feature dissapears in the simulation.

In the subsequent evolution the disk slowly  becomes  eccentric.  This is indicated
in Fig.~\ref{fig:m01f-sig4xy} where we display  
the two-dimensional surface density structure at four different times.
The distortion of the disk is evident at later times. While in the first snapshot
taken after 100 orbits there is only a slight indication of disk eccentricity, the subsequent panels
demonstrate a growing eccentric disk mode and one can infer the presence 
of  an apsidal line joining
the locations of minimum and maximum displacements from the primary. Additionally there is
indication for precession of this apsidal line. 

\begin{figure}[ht]
\begin{center}
\rotatebox{0}{
\resizebox{0.98\linewidth}{!}{%
\includegraphics{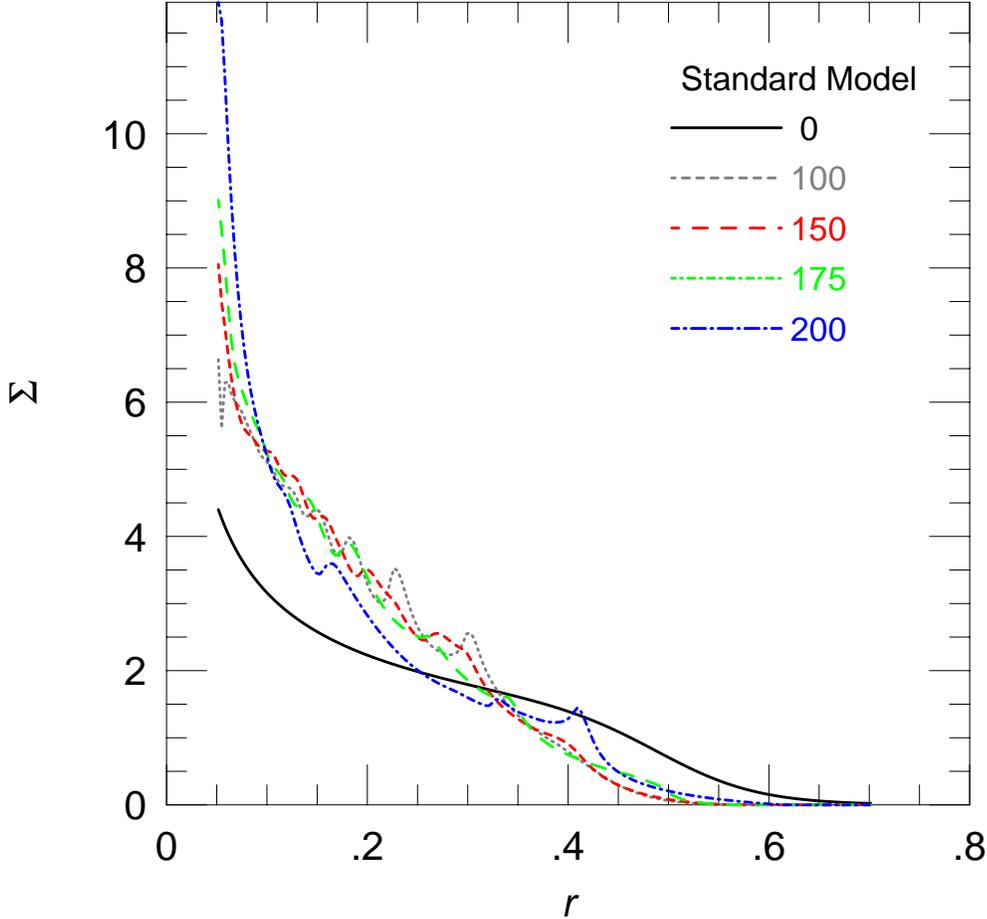}}}
\end{center}
  \caption{Radial distribution of the azimuthally averaged surface density $\Sigma(r)$
at different times. The solid black curve indicates the tapered initial
profile.
}
   \label{fig:m01f-sig4r}
\end{figure}

The tidal torques   induced by the secondary also lead to a truncation of the disk
and a rearrangement of the (mean) radial surface  density profile. This
is demonstrated in Fig.~\ref{fig:m01f-sig4r} where the azimuthally averaged
density is shown at 5 different times. The solid black curve indicates
the initial (axisymmetric)   profile, $\Sigma = r^{-1/2},$ with
a taper applied for  $r>0.4$. Clearly, the disk becomes truncated at a radius around
$r \approx 0.5$   where the surface  density starts to  drop much more  steeply. After about 250 orbits
the average surface density distribution  attains  a new  quasi-stationary profile (see below).

\begin{figure}[ht]
\begin{center}
\rotatebox{0}{
\resizebox{0.47\linewidth}{!}{%
\includegraphics{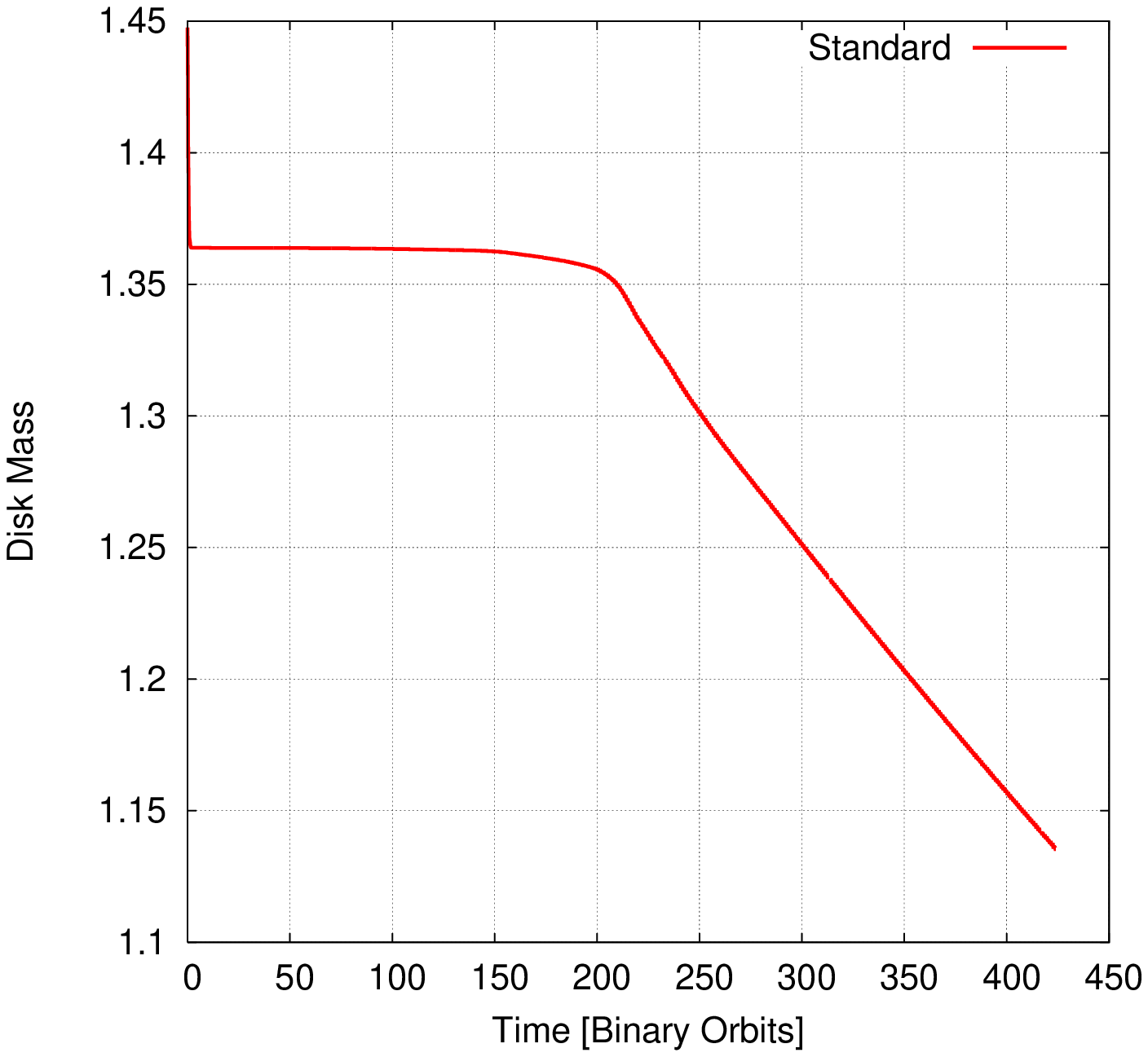}}}
\rotatebox{0}{
\resizebox{0.47\linewidth}{!}{%
\includegraphics{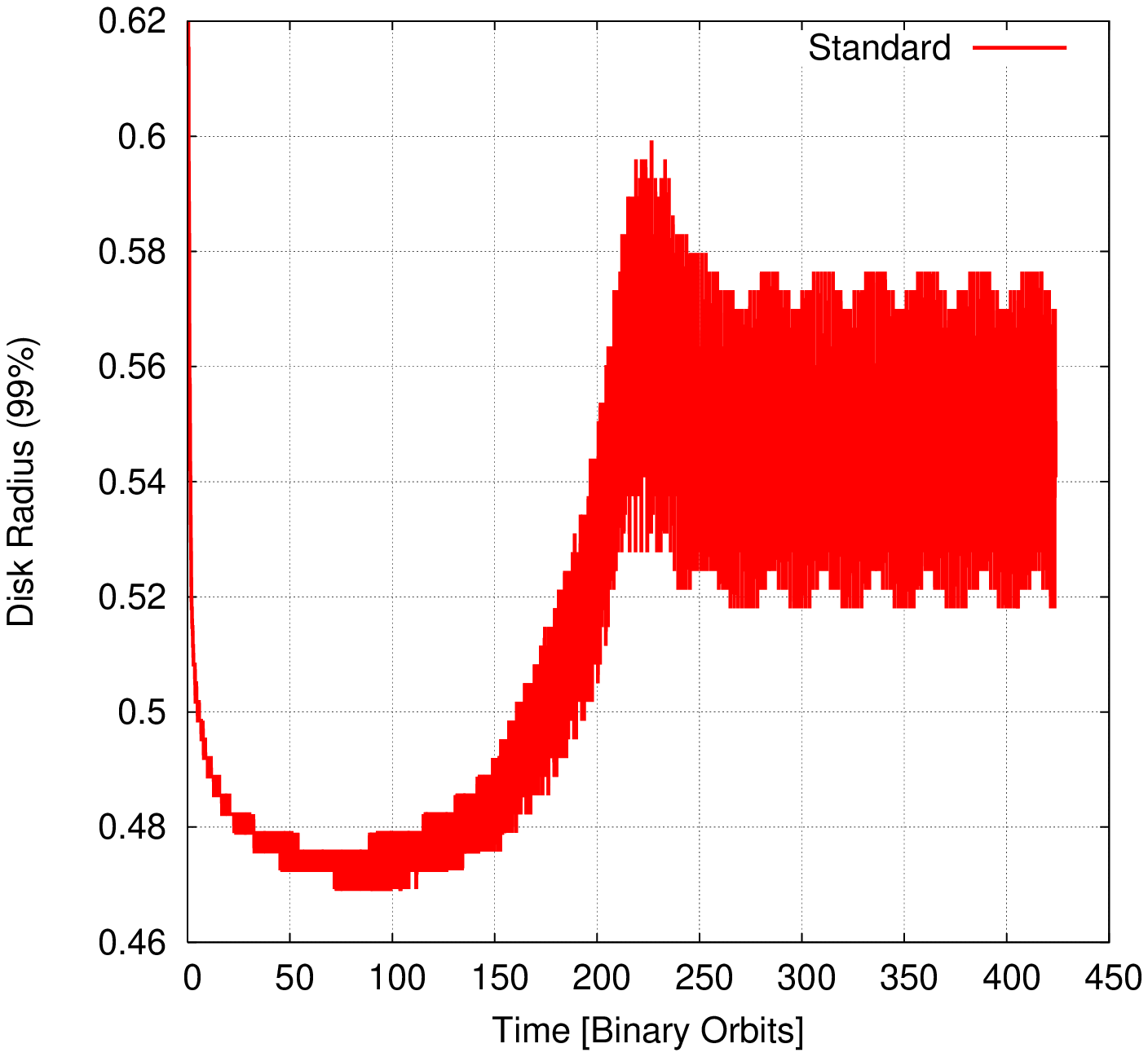}}} \\
\rotatebox{0}{
\resizebox{0.47\linewidth}{!}{%
\includegraphics{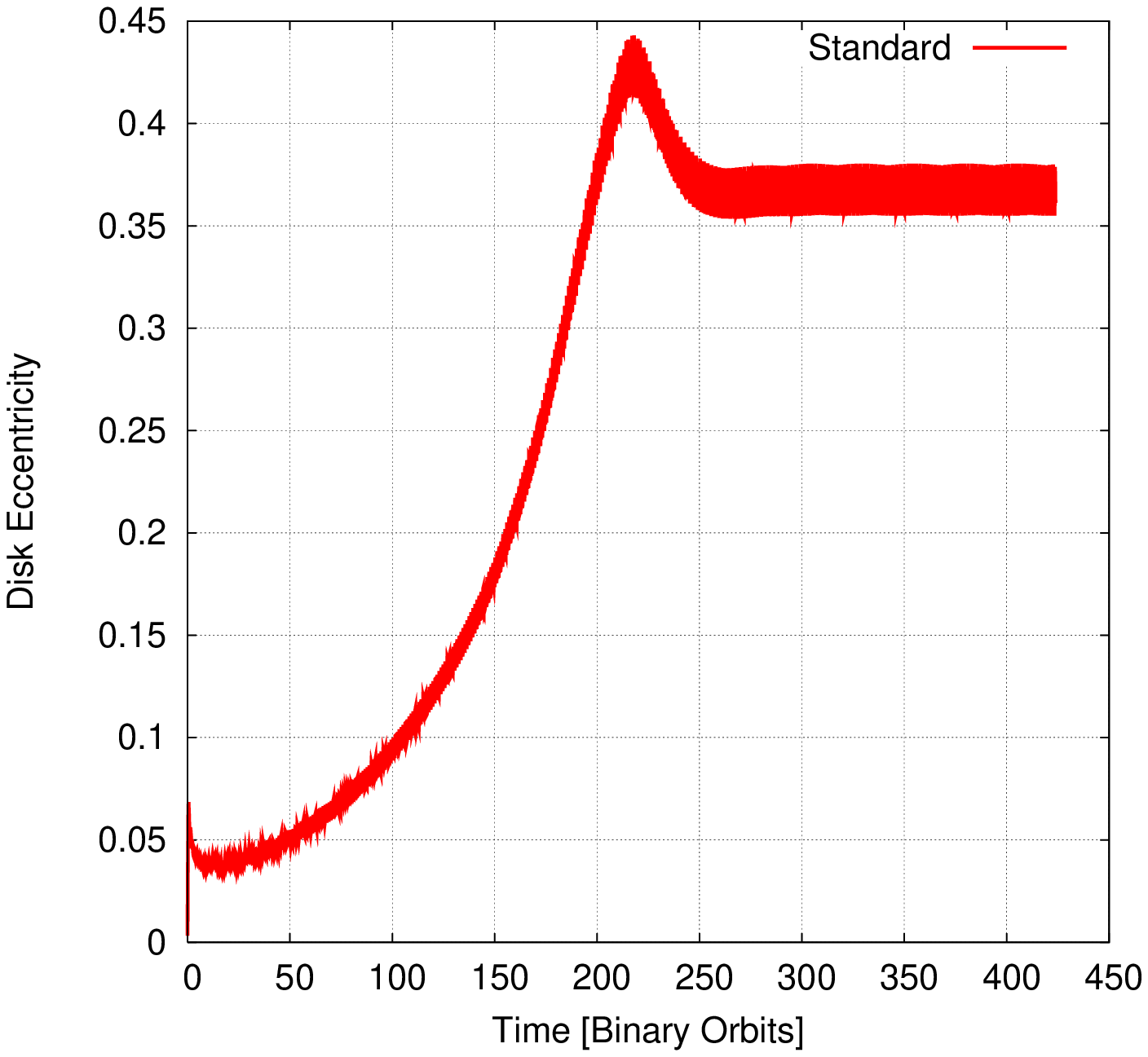}}}
\rotatebox{0}{
\resizebox{0.47\linewidth}{!}{%
\includegraphics{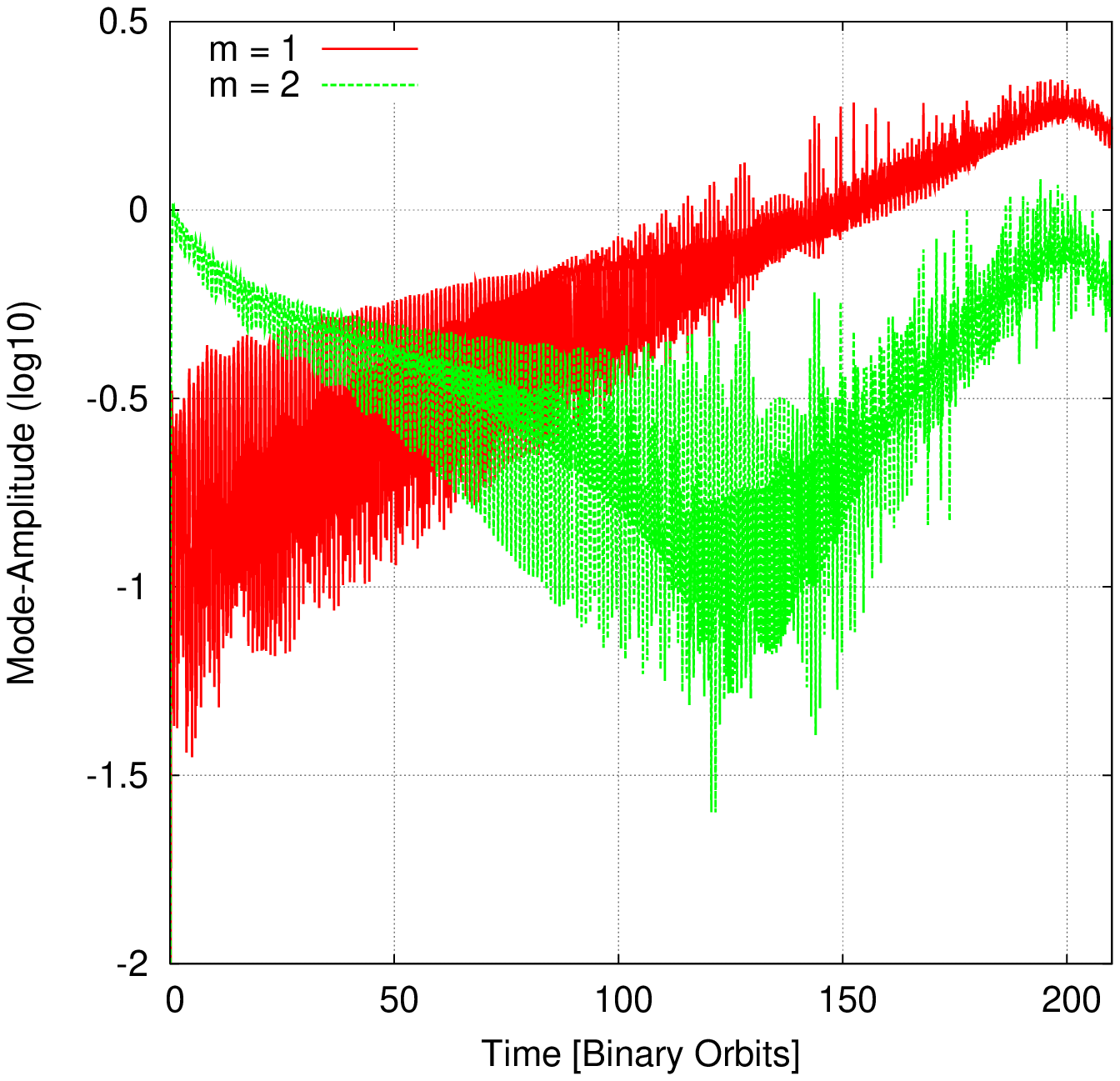}}} 
\end{center}
  \caption{Evolution of various global disk properties of the standard model as functions of time. The total mass in the computational domain, or equivalently the
disk mass, is shown in the upper left panel. The disk radius, defined as the 
radius containing $99\%$ of the mass, is shown in the upper right panel.
The disk eccentricity is shown in the lower left panel and the
$m=1$ and $m=2$ Fourier amplitudes are shown in the lower right panel.}
   \label{fig:m01f-diskpara}
\end{figure}

In Fig.~\ref{fig:m01f-diskpara} we display the time-evolution of several global disk
quantities (in dimensionless units). 
The total mass in the computational domain is seen to drop off rapidly within the first 2 orbits due to the initial mass loss
through the outer boundary induced by the tidal perturbation of the secondary
(see also Fig.~\ref{fig:m01f-sig1xy}).
After this initial
burst of mass loss, the disk is tidally truncated, such that
as seen in Fig.~\ref{fig:m01f-diskpara}, the disk radius, defined to be the radius
containing 99\% of the mass, is reduced from $0.61$ to $0.48.$ This is then smaller
than the Roche lobe.  The mass subsequently remains constant until $t\approx 200.$
The disk radius reaches a minimum at around $t\approx 80$, and increases slowly thereafter.
This increase of the radius of the disk is coupled with an increase in the
mean eccentricity $e\subscr{d}$ of the disk which is displayed in the bottom left panel of Fig.~\ref{fig:m01f-diskpara}.
The quantity $e\subscr{d}$ is determined by first calculating the eccentricity
for each individual gridcell assuming that it is in an unperturbed orbit around
the central star, with the position and velocity given by the actual grid values
averaged to the cell centres. To obtain the global value $e\subscr{d}$ we then
perform a mass-weighted average over the individual cell values.
The longitude of the disk's periastron $\varpi\subscr{d}$ is calculated in
the same way. The explicit formulae are given for example in \citet{1999ssd..book.....M}.  

As seen in Fig.~\ref{fig:m01f-diskpara}, $e\subscr{d}$  begins to exponentiate after $t\approx20$ with
an $e$-folding time of about 80 orbits. After about 250 orbits the eccentricity
has settled to a final value of  $e\subscr{d}=0.37$ with an average
outer disk radius $r\subscr{d}\approx 0.55$. The increase in $e\subscr{d}$ is
also associated with a further decrease in  the disk mass. 
Beyond $t\approx250$
mass spills over the Roche lobe and is lost from the system.
This mass loss does not affect the dynamics of the system (i.e. it
does not change the
equilibrium values of $e\subscr{d}$ and $r\subscr{d}$) 
because the absolute value of the
mass density scales out of the evolution equations. 
This feature has been confirmed numerically
in additional test calculations where the total mass in the system is kept constant by rescaling
the density of all grid cells at each timestep in order
to account for the lost mass. 
In the bottom right panel of Fig.~\ref{fig:m01f-diskpara} the time-evolution
of the $m=1$ and $m=2$ surface density Fourier amplitudes
calculated at a single
disk radius ($r=0.37$) are displayed. 
The $m=1$ mode associated with the eccentricity
increases exponentially from the beginning of the run  up to $t\approx 200$ after
which induced additional  mass loss leads to a reduction
in the amplitude. To show the evolution more clearly we have plotted here
only the first 200 orbits.
The only other Fourier amplitude that plays a significant role is
that associated with $m=2,$ which at early times (when
the spiral arm feature dominates the disk) is much larger than the  $m=1$
amplitude, but it  becomes weaker
at later times. In the final stages
it increases again, saturating eventually at a value still
smaller than the $m=1$ amplitude.
\begin{figure}[ht]
\begin{center}
\rotatebox{0}{
\resizebox{0.98\linewidth}{!}{%
\includegraphics{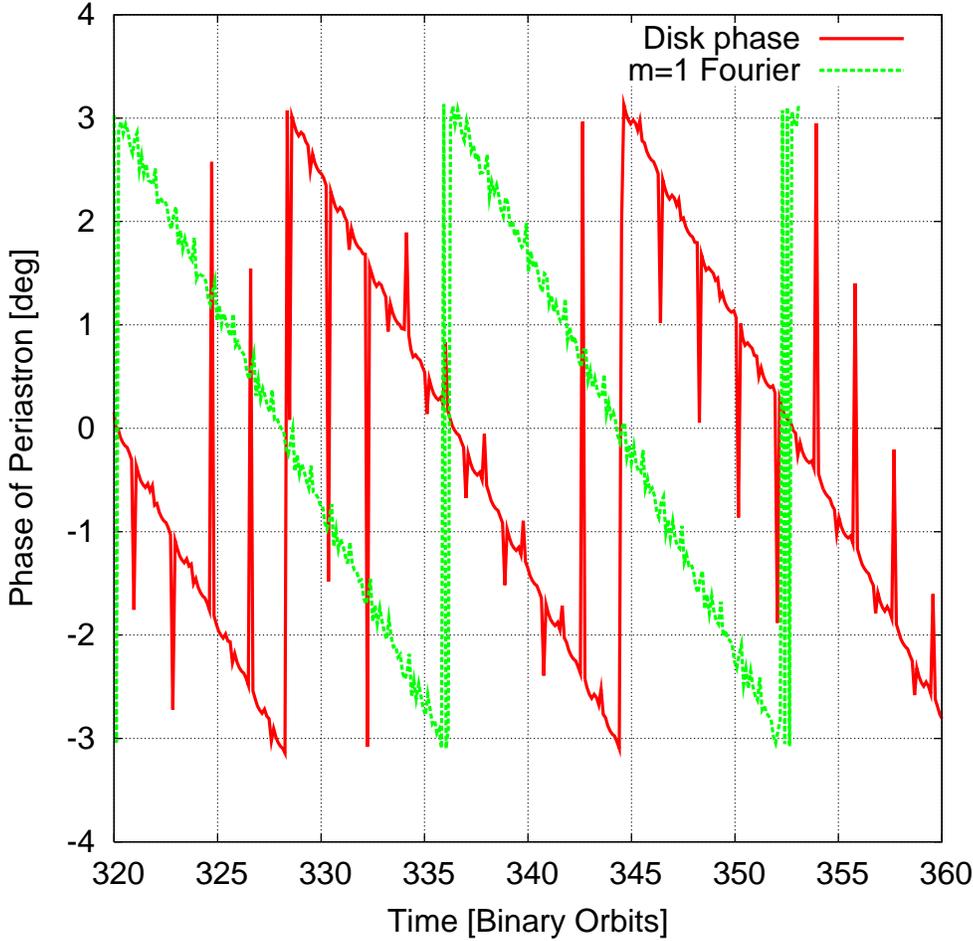}}}
\end{center}
  \caption{Time-evolution of the longitude of pericentre (periastron) of the disk.
The solid darker (red) curve is the mass-weighted mean pericentre over the whole disk, calculated
as described in the text. The lighter dashed (green) curve corresponds to a phase of the
$m=1$ Fourier mode at the single radius $r=0.37$. Note that the phases are calculated in the
{\it inertial} frame.
}
   \label{fig:m01f-phase1}
\end{figure}

As mentioned above, Fig.~\ref{fig:m01f-sig4xy} indicates 
global precession of the eccentric disk.  
This is quantified in Fig.~\ref{fig:m01f-phase1} where the solid
(red) curve corresponds to the time-evolution of the longitude of pericentre
(periastron) of the disk, $\varpi\subscr{d}$, calculated in the same manner as $e\subscr{d}$.
Obviously the disk precesses as a whole at a constant rate, a feature that is supported by the additional
light (green) curve which corresponds to the phase of the $m=1$ Fourier component at
one given radius $r=0.37$. Both angles precess at the same rate,
confirming our interpretation. The periodic disturbances
in $\varpi\subscr{d}$ coincide with the times when the disk's apocentre 
is closest to the
secondary star. In this case the outer parts of the eccentric disk are strongly disturbed
which renders the calculation of $\varpi\subscr{d}$ more inaccurate.
From the plot, which shows the phases in the inertial frame, we conclude that the disk
experiences a slow {\it retrograde} precession with a period $P\subscr{d} \approx 16.4$ orbits.  

\begin{figure}[ht]
\begin{center}
\rotatebox{0}{
\resizebox{0.47\linewidth}{!}{%
\includegraphics{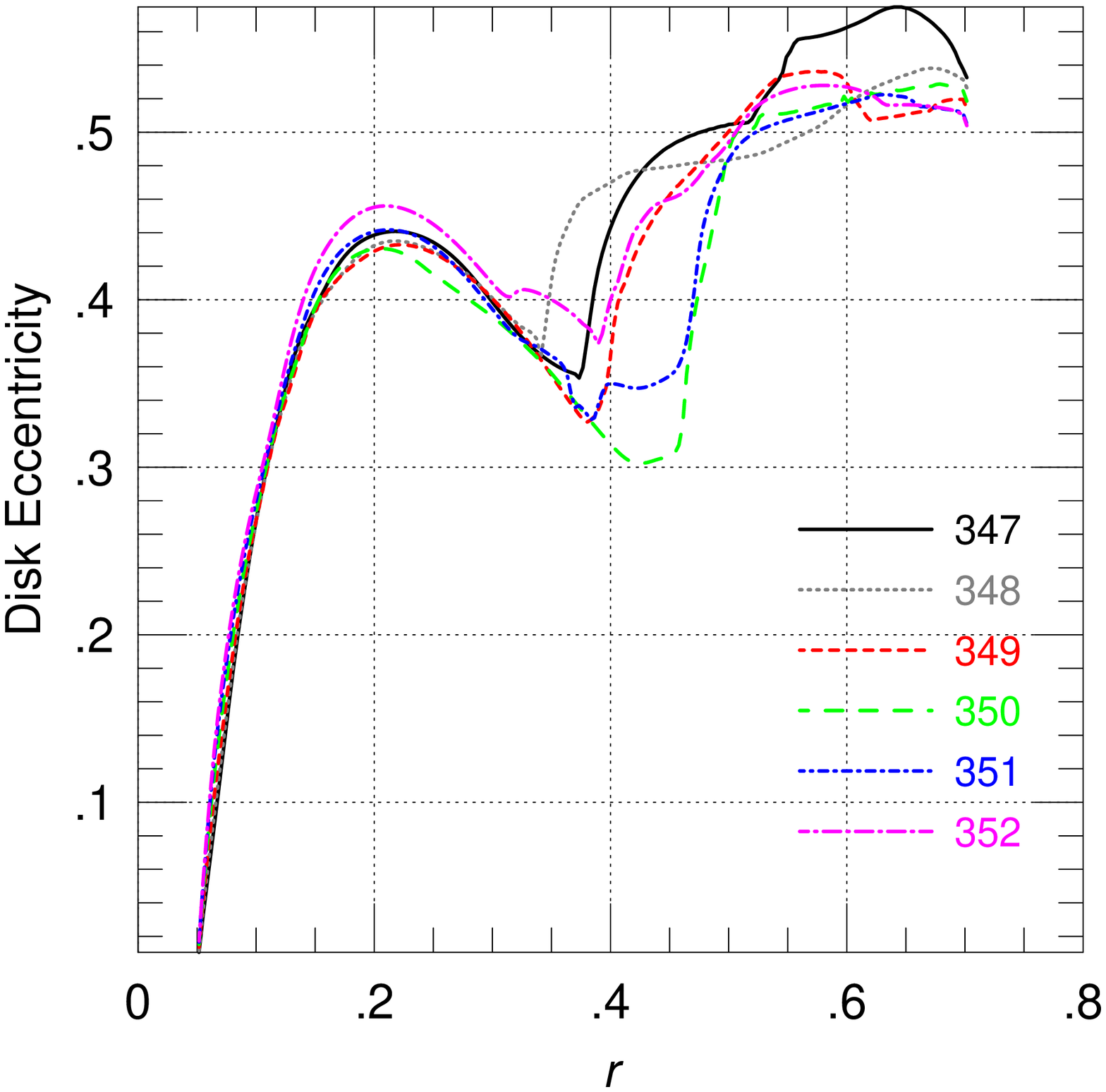}}}
\rotatebox{0}{
\resizebox{0.47\linewidth}{!}{%
\includegraphics{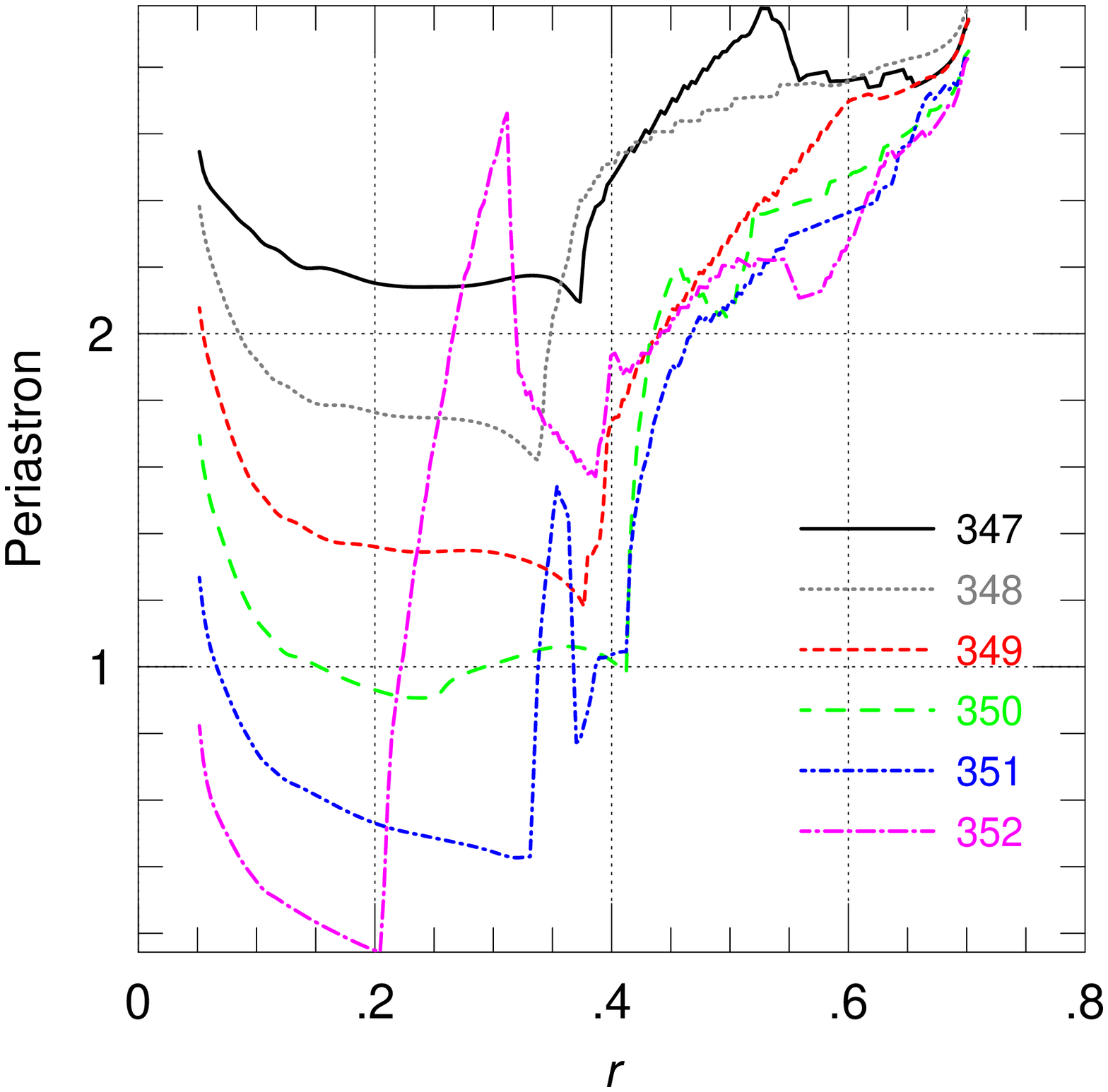}}} 
\end{center}
  \caption{Radial variation of the eccentricity and longitude of pericentre of the
disk at various times (quoted in binary orbits).
}
   \label{fig:m01f-ecc-peri}
\end{figure}

To study the internal structure of the disk we show in Fig.~\ref{fig:m01f-ecc-peri} the disk's
radial distribution of  eccentricity and  the longitude of pericentre 
at different times (indicated by the
legend) which lie within the time interval plotted in Fig.~\ref{fig:m01f-phase1}.
The distribution of eccentricity within the disk does not vary much with 
the binary orbital phase and drops
smoothly towards zero in the centre of the disk, due to the rigid boundary at the inner radius.
In the main part of the
disk the eccentricity lies around $e=0.4$ which is consistent with two-dimensional
surface plots of the density (see the snapshot at $t=200$ in Fig.~\ref{fig:m01f-sig4xy}).
The radial variation of the pericentre indicates  a small twist within the 
innermost parts of the disk,
which is larger  when the disk's apocentre is closest to the secondary star
(at times 351 and 352) when it  can reach about 45 degrees.  
In addition, the fact that the  curves 
at different times appear vertically shifted indicates again the
precession of the disk as a single entity.
\section{Dependence on viscosity, temperature and the inner boundary condition}
\label{sec:physical}

\subsection{Viscosity}
\label{subsec:viscosity}
The first parameter we investigate is the kinematic  viscosity $\nu.$ 
To study dependence of the results on the magnitude of the kinematic
viscosity we started from our standard model ($\nu = 10^{-5}$) and varied
only the value of $\nu$ keeping all other parameters  unchanged. 
\begin{figure}[ht]
\begin{center}
\rotatebox{0}{
\resizebox{0.98\linewidth}{!}{%
\includegraphics{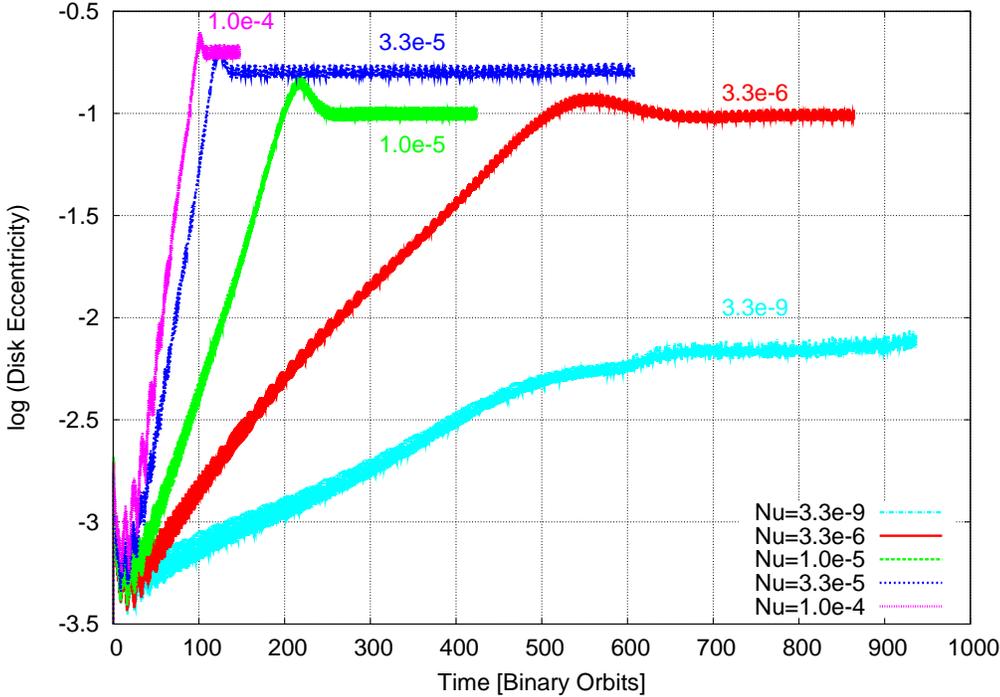}}}
\end{center}
  \caption{Time-evolution of the mean eccentricity of the disk 
 (natural logarithm) for
different kinematic viscosities (in dimensionless units).
}
   \label{fig:m05fm-eccdcl1}
\end{figure}
\begin{figure}[ht]
\begin{center}
\rotatebox{0}{
\resizebox{0.48\linewidth}{!}{%
\includegraphics{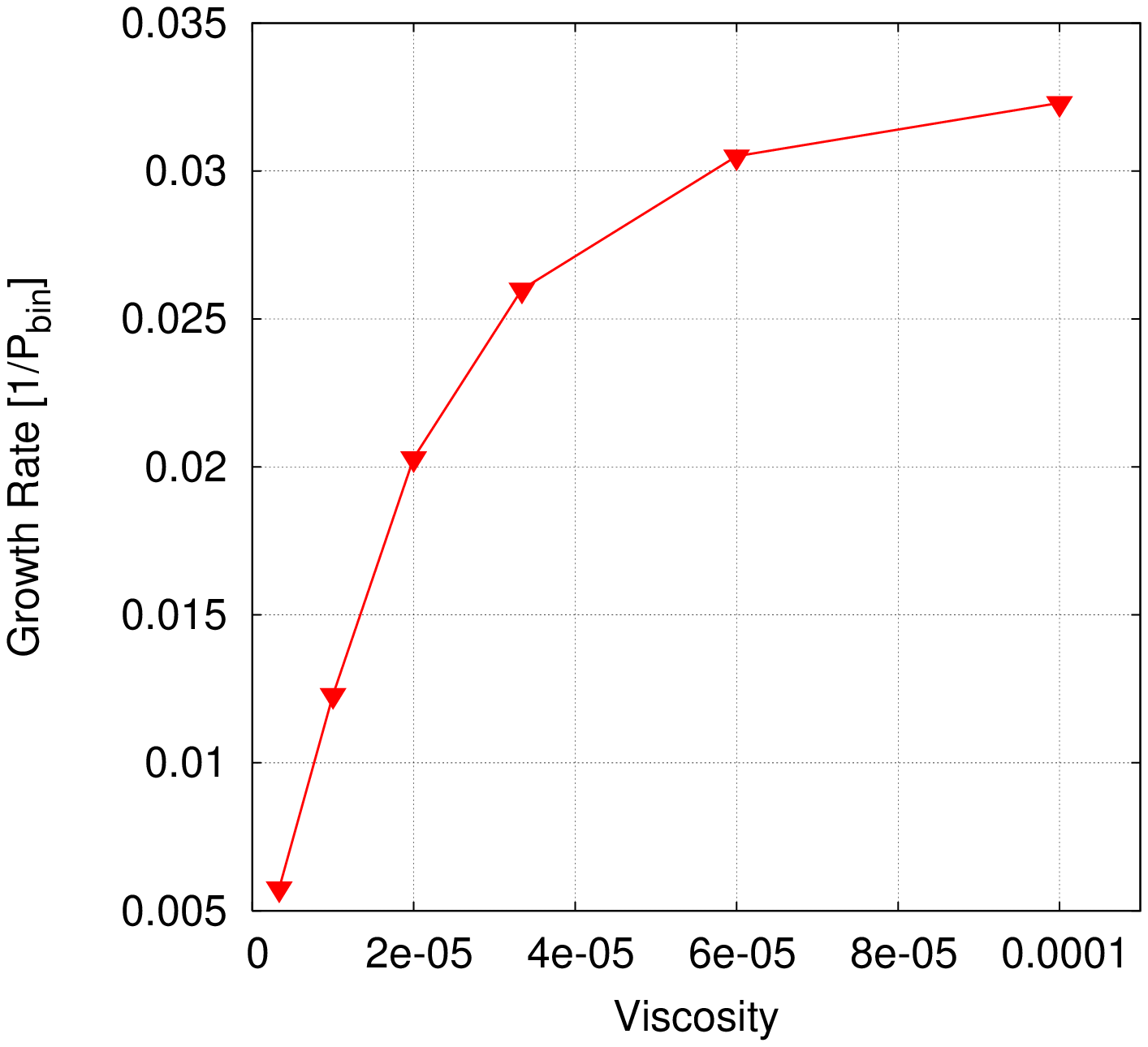}}}
\rotatebox{0}{
\resizebox{0.48\linewidth}{!}{%
\includegraphics{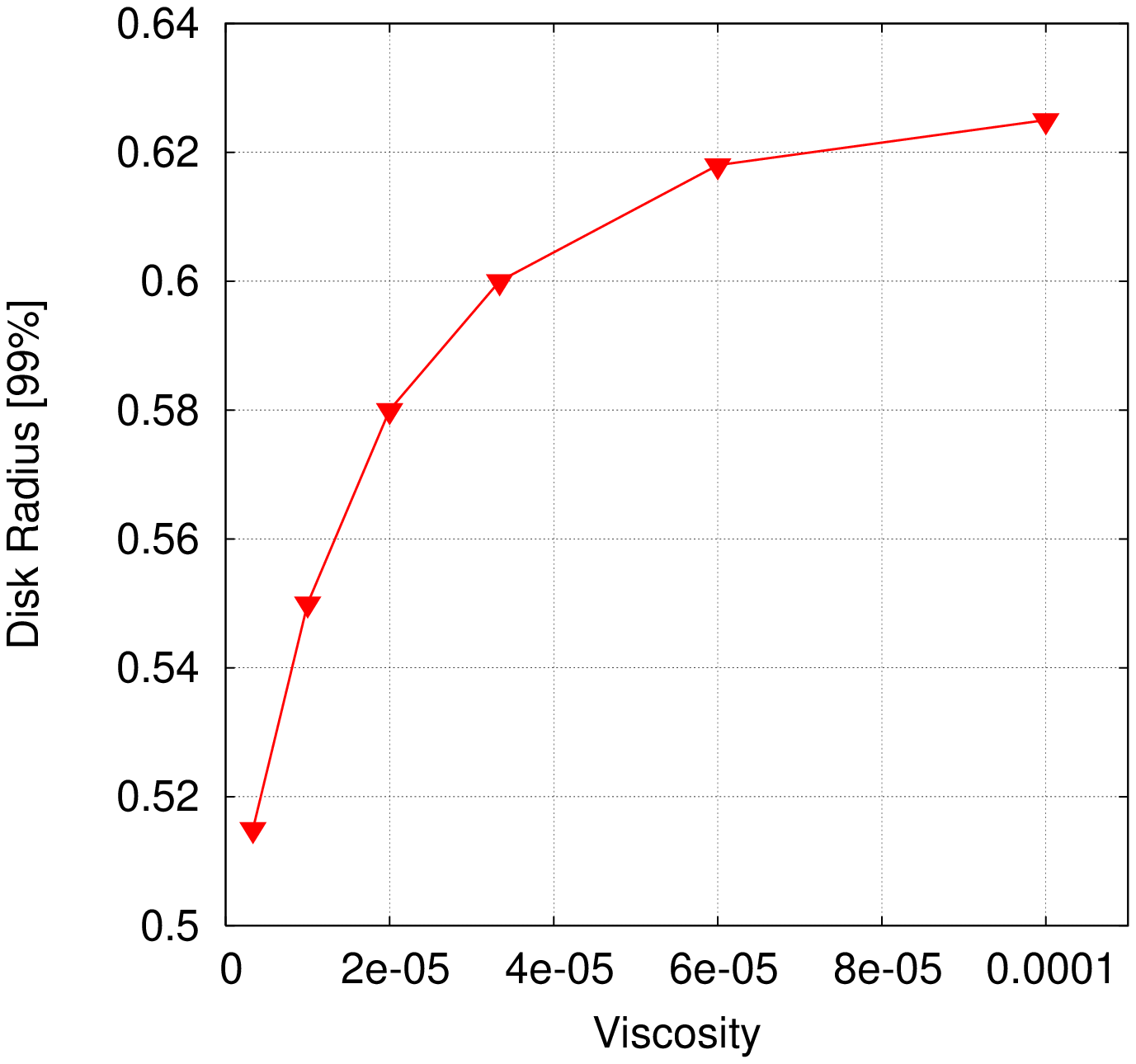}}} \\
\rotatebox{0}{
\resizebox{0.48\linewidth}{!}{%
\includegraphics{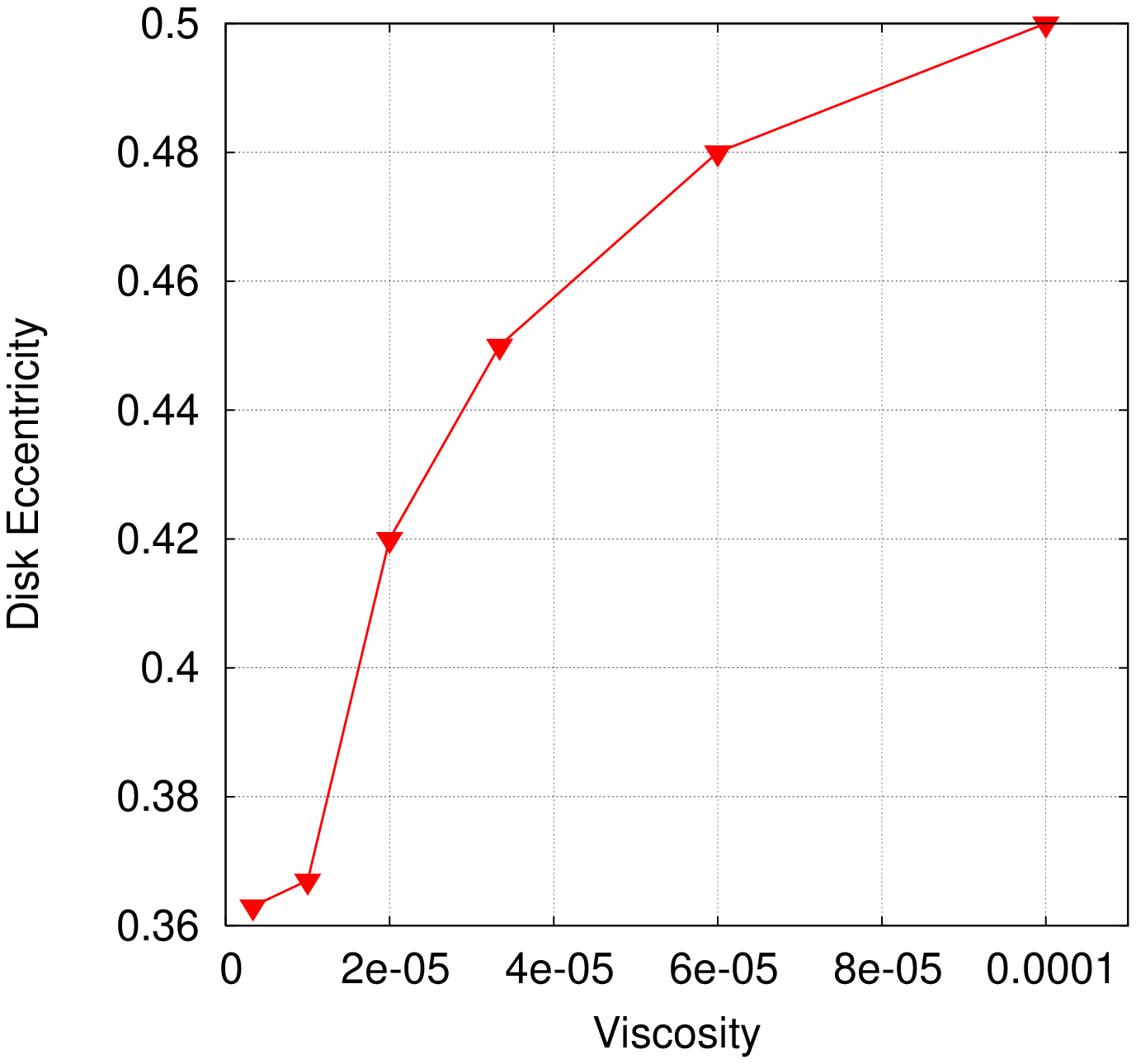}}}
\rotatebox{0}{
\resizebox{0.48\linewidth}{!}{%
\includegraphics{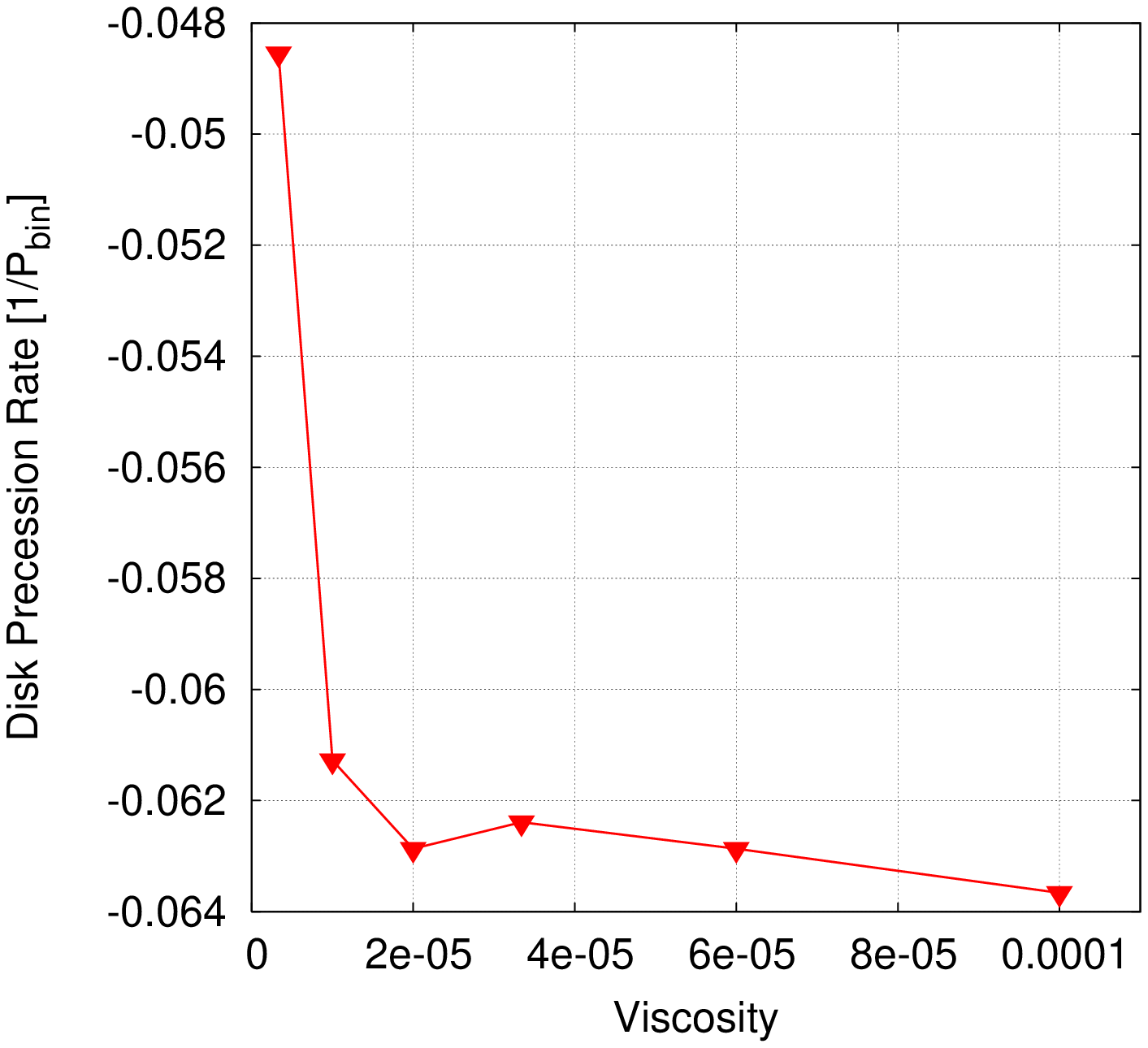}}}
\end{center}
  \caption{Dependence of various global disk parameters on the magnitude of the viscosity.
Upper left panel: Growth rate of the eccentricity of the disk; Upper right
panel: The radius, $r\subscr{d},$  within
which 99\% of the mass of the disk is contained; Lower left panel: The mass-weighted
mean eccentricity of the disk $e\subscr{d}$; Lower right panel: Precession rate of the disk $\dot\varpi\subscr{d},$
negative values indicating retrograde precession. The values for $r\subscr{d}$, $e\subscr{d}$ and $\dot\varpi\subscr{d}$ are
determined at a time at which the  system reaches its final quasi-steady state.
}
   \label{fig:visc}
\end{figure}

The variation of the mean disk eccentricity 
with time is displayed for different kinematic
viscosities in Fig.~\ref{fig:m05fm-eccdcl1}, where the middle (green) curve refers
to the standard model which (for reference) has been displayed already in 
Fig.~\ref{fig:m01f-diskpara} at early times. 
The growth rate of the eccentricity depends strongly on the magnitude
of the viscosity: increasing the viscosity  shortens the growth time substantially,
while for smaller viscosities it takes much longer for the disk to become eccentric.
For the lowest-viscosity model ($\nu = 3 \times 10^{-9}$), which corresponds essentially
to an inviscid calculation, the disk does not  attain significant eccentricity 
in a run time of $1000$ orbits. However, there is an indication
that there may be a very long-term eccentricity growth produced by numerical effects.  From the upper left
panel of Fig.~\ref{fig:visc} a simple extrapolation indicates a limiting 
value of $\nu$ for which the disk  becomes eccentric  of  $\nu\sim2\times 10^{-6}.$
Already the growth time for  $\nu=3.3 \times 10^{-6}$ is $\sim 162$ orbits.
In general the models settle to a quasi-stationary state with a constant eccentricity, which
decreases as the viscosity decreases. However, quasi-steady states with mean
eccentricities below $e =0.36$ do not seem to be reached for this
model  by decreasing the viscosity alone.  But then  a longer time  is
required for this value to be attained.

\begin{figure}[ht]
\begin{center}
\rotatebox{0}{
\resizebox{0.98\linewidth}{!}{%
\includegraphics{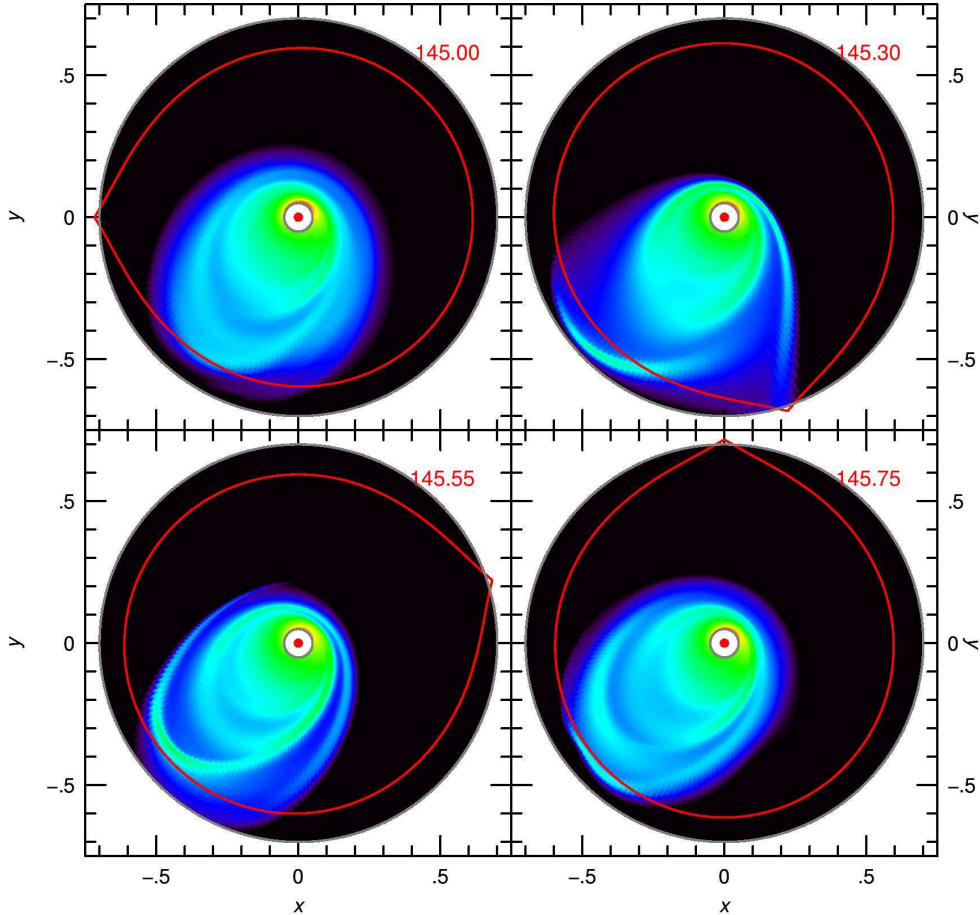}}}
\end{center}
  \caption{
Four snapshots of the disk  surface  density contours during
one orbital period as viewed in the non-rotating frame.
Time increases from top to bottom and left to right
The solid (red) curve
indicates the Roche lobe of the primary star.
The viscosity is $10^{-4},$ the mass ratio is $0.1$ and
$H/r =0.05.$ As the companion has a close approach to the disk outer
edge a tidal tail is pulled out (top right). This subsequently rejoins the
disk (bottom left) which remains relatively unperturbed until the
next close approach.
}
   \label{fig:m02f-inert}
\end{figure}

Clearly the initial  growth of $e\subscr{d}$ in these models is exponential
and  growth rates,  $\sigma\subscr{d},$  defined through
\beq
     e\subscr{d}(t) \propto \exp( \sigma\subscr{d} t),
\eeq
can be determined by fitting simulation data to the above relation.
The results are plotted in the upper left panel of Fig.~\ref{fig:visc}
where the variation of $\sigma\subscr{d}$ with viscosity is plotted.
 In the remaining 3 panels we plot the
equilibrium values of the radius $r\subscr{d}$, the eccentricity $e\subscr{d}$ and the
precession rate $\dot\varpi\subscr{d}$ of the disk. Each of the  quantities ($\sigma\subscr{d}$, $r\subscr{d}$, $e\subscr{d}$)
shows a strikingly similar dependence on $\nu.$
This behaviour can be understood in terms of the diffusive influence of viscosity.
For larger viscosity the disk material will tend to spread outwards at a faster rate 
with more material being
pushed to the outer regions of the disk where it feels a stronger
(eccentricity enhancing) tidal perturbation due to the secondary. Hence
the increase of the growth rate with viscosity. For the very same
reason, the equilibrium value of the radius $r\subscr{d}$ 
as well as the final eccentricity will increase with $\nu.$
 The fact that all
these quantities respond similarly to variations of $\nu$ indicates that
they respond in a similar manner to the outward expansion of the disk.
On the contrary, the disk precession rate $\dot\varpi\subscr{d}$ depends only weakly on the
viscosity (for $\nu \gsim 5 \times 10^{-6}$)~. 
From the lower two panels ($e\subscr{d}$, $\dot\varpi\subscr{d}$) there is an indication that for
 $\nu  \le 10^{-5}$ the  disk shrinks rapidly
 causing  corresponding changes to the precession rate while the eccentricity levels off.
However, we did not consider simulations with $\nu < 3.3 \times 10^{-6}$ as it is likely
that at this stage they  become resolution-limited
such that the numerical viscosity cannot be neglected.

Finally we illustrate the behaviour of the eccentric  disk 
once it has attained its final quasi-steady state for the
case with $\nu=10^{-4}$ in Fig. \ref{fig:m02f-inert}. Cases with other values
of $\nu$ show similar behaviour. In Fig. \ref{fig:m02f-inert} we show the
 surface  density contours during
one orbital period as viewed in the non-rotating frame.
  In this frame the secondary (moving counter clock wise) has a close approach to the
 slowly precessing disk's apocentre approximately once every orbit.
 As the companion  approaches the disk's outer
edge a tidal tail is pulled out. This subsequently rejoins the
disk which remains relatively unperturbed until the
next close approach. As even at apocentre, the disk matter rotates
faster than the binary, so the tidal tail leads the secondary and
is clearly responsible for angular momentum transfer to the orbit.
This must also be associated with enhanced tidal dissipation.
As the disk appears relatively unperturbed away from these closest approaches,
most of the tidal angular momentum transfer may occur as a result of these
close approaches.

\begin{figure}[ht]
\begin{center}
\rotatebox{0}{
\resizebox{0.48\linewidth}{!}{%
\includegraphics{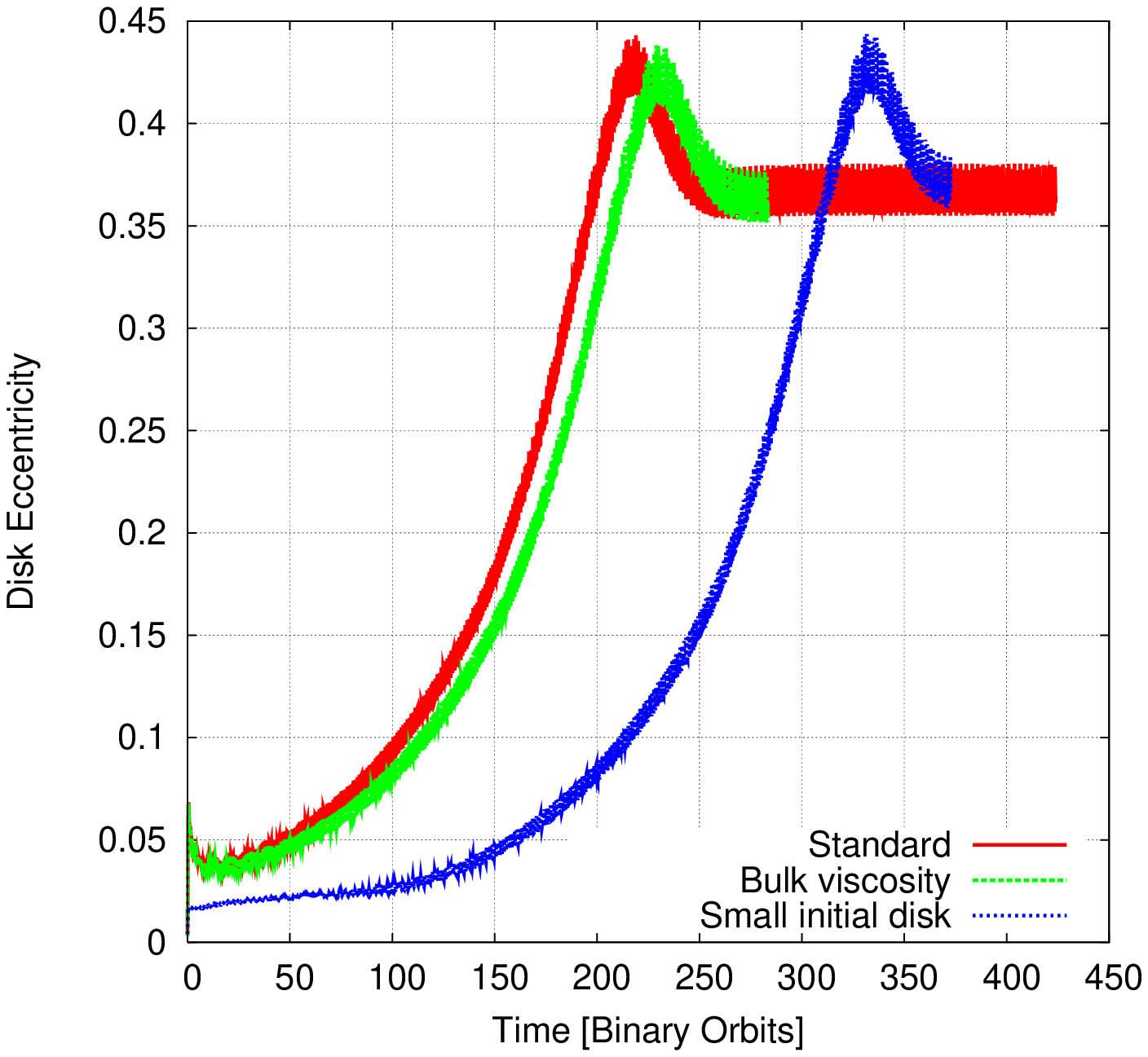}}}
\rotatebox{0}{
\resizebox{0.48\linewidth}{!}{%
\includegraphics{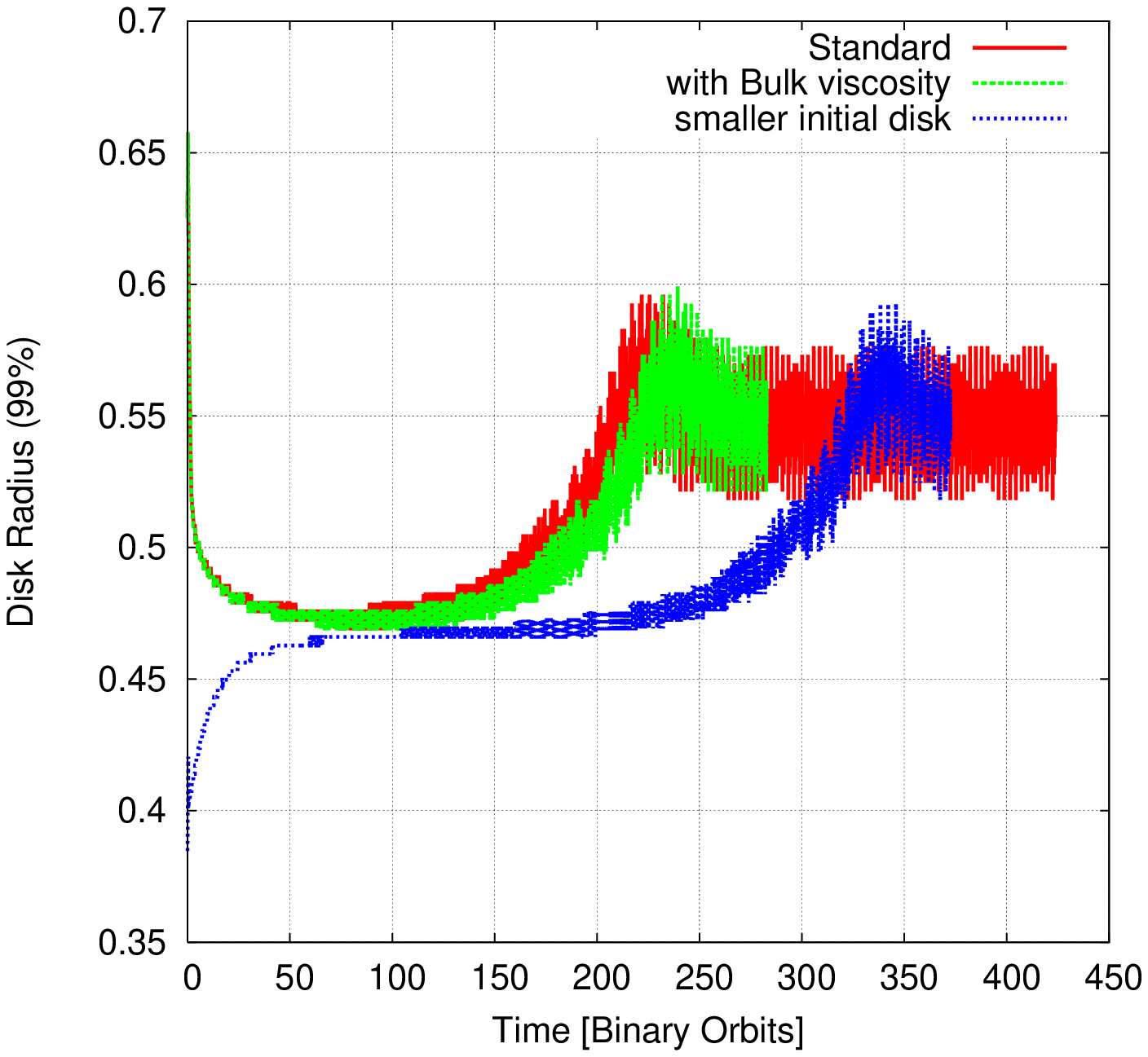}}}
\end{center}
 \caption{Mean disk eccentricity $e\subscr{d}$ and disk radius $r\subscr{d}$ as functions of time
  for  the standard model (red curve),
  the same case but with an added  explicit bulk viscosity $\nu\subscr{bulk}=2\nu$
  (green curve) and a case
with standard parameters, but
for which the initial disk radius is taken to be $r\subscr{d}(0) = 0.3$ (blue curve).
}
\label{fig:m01j-mult}
\end{figure}

\subsubsection{Effect of bulk viscosity}

The imposed kinematic viscosity, $\nu,$ is intended to represent  the effects of
some form of disk  turbulence, probably arising from the
magnetorotational instability \citep{1998RvMP...70....1B}. It is possible that an effective bulk viscosity
may also result which could also act as a stabilizing mechanism. 
In addition, in typical SPH simulations on this problem an explicit bulk viscosity
is included; see for example the recent study by \citet{2007MNRAS.378..785S}.
 In order to investigate its effects, we performed runs for which  
 an explicit bulk viscosity 
$\zeta = \Sigma \nu\subscr{bulk}$ was added.
In this study we used a constant $\nu\subscr{bulk}$ and present in Fig.~\ref{fig:m01j-mult}
results for the case when the bulk viscosity coefficient
 is twice as large as the shear viscosity coefficient:
$\nu\subscr{bulk} = 2 \nu$.
The influence of the added bulk viscosity on the simulations  is very small:
the only effect is a slightly increased damping, as indicated for example
by the longer growth time. Otherwise an added bulk viscosity has no dynamical
influence. This indicates that 
the direct effect of viscosity on the eccentric mode is unimportant
and implies that mechanisms such as viscous overstability cannot be 
responsible for the eccentric instability seen here.
We  also plot the evolution for a case where the initial disk
radius  $r\subscr{d}(0)=0.3$, rather than the usual  $r\subscr{d}(0)=0.5$,  in   Fig.~\ref{fig:m01j-mult}.
 As expected the initial growth
is slowed down due to the small outer disk radius which remains around $r\subscr{d}=0.47$ until
$t=150$. Only when the eccentric instability, which appears to grow (though  more slowly)
even for this small disk, has developed sufficiently does the radius of the disk
increase again. (It should be noted that, owing to the way in which it is defined, an increase of $r\subscr{d}$ may be associated purely with a growth of the eccentricity of the disk rather than an increase of the semimajor axis at which the disk is tidally truncated.)  At later times the behaviour is in fact identical to the standard
model. Hence,  different initial conditions lead to the same outcome,  as one might expect
for viscous flows.

\subsection{Temperature}
\label{subsec:temp}
\begin{figure}[ht]
\begin{center}
\rotatebox{0}{
\resizebox{0.98\linewidth}{!}{%
\includegraphics{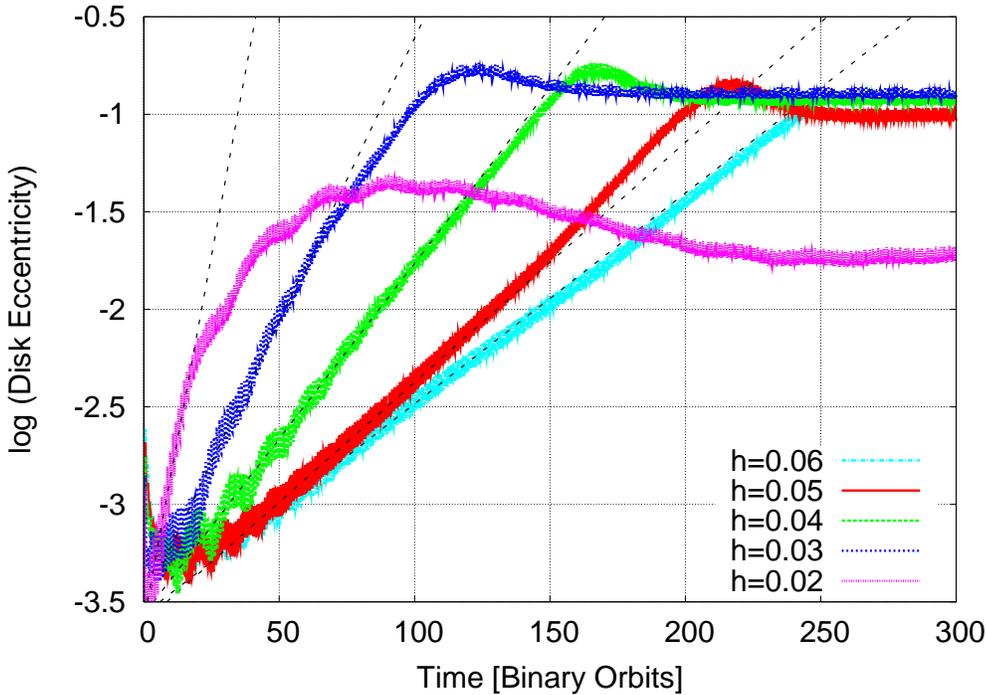}}}
\end{center}
  \caption{Time-evolution of the mean eccentricity of the disk  for
different disk aspect ratios  $h \equiv H/r.$
}
   \label{fig:eccdcl1-hr}
\end{figure}

In this section  we study the effect of variations of  the disk temperature or, equivalently, the relative scaleheight $h=H/r$ of the
disk. 
Starting again from the standard model we vary $h$ from $0.01$ to $0.06.$
The variation of $e\subscr{d}(t)$ is displayed in Fig.~\ref{fig:eccdcl1-hr}, where the
thick (coloured) lines refer to different values of $h$, and the dashed lines are
eyeball fits made during the initial growth period.  The growth timescale
depends on the disk temperature and it is found to be shorter for cooler disks.
The final disk eccentricities do not depend on the value of $h$ as long as 
$h\geq 0.03$. For the smallest value displayed here, $h=0.02$, the standard grid resolution
($200\times200$) is not sufficient. In additional simulations we  found that
runs (for $h=0.02$) with higher resolution yield larger values of 
$e\subscr{d}$ more consistent with the other runs.
The final radius of the disk, $r\subscr{d}$,  also does not depend on the disk aspect ratio 
for  $h > 0.02.$ All runs yield an outer disk radius $r\subscr{d} \approx 0.55$, 
with a slight tendency for larger disks to be associated with larger $h.$
\begin{figure}[ht]
\begin{center} 
\rotatebox{0}{
\resizebox{0.48\linewidth}{!}{%
\includegraphics{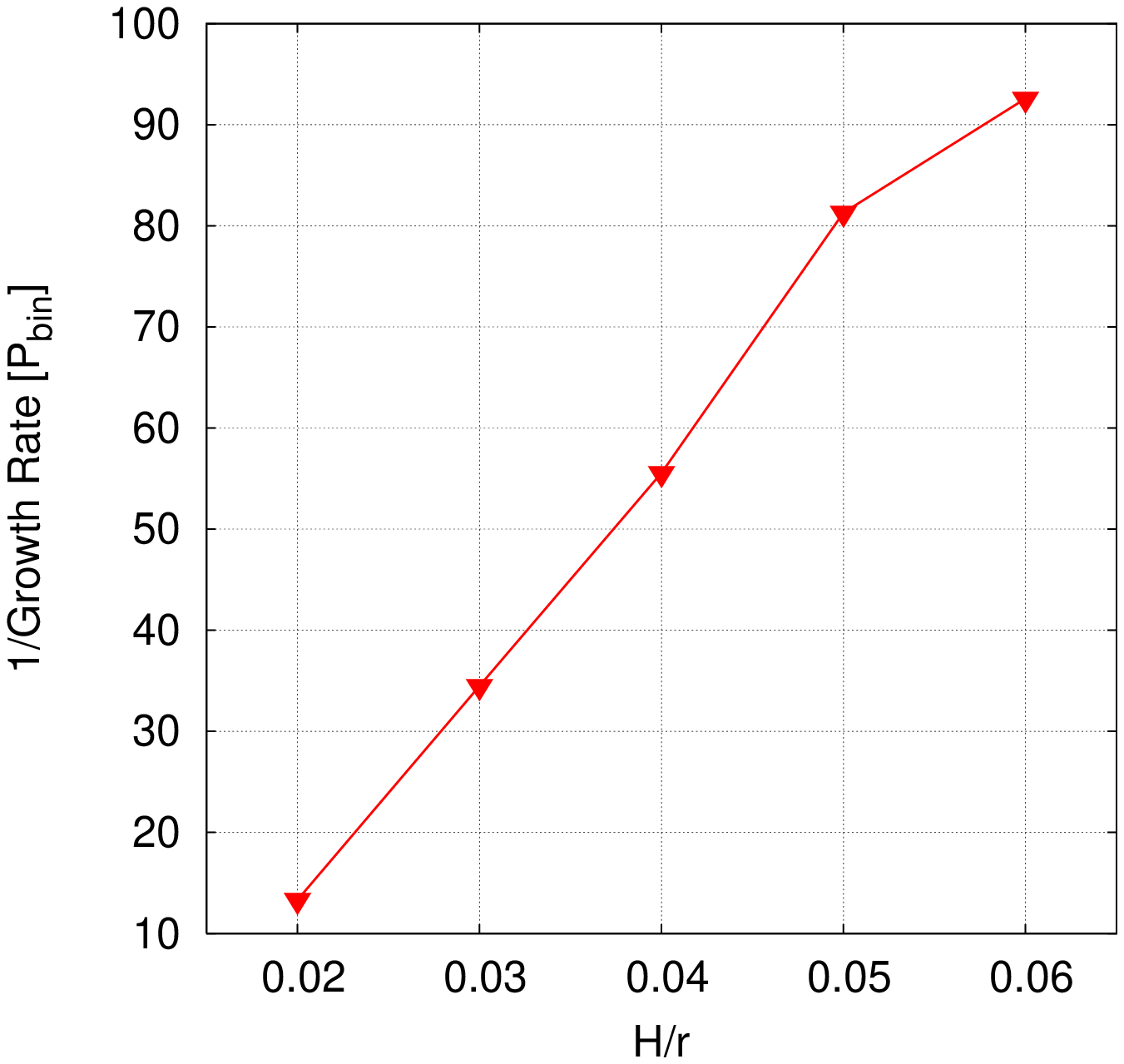}}}
\rotatebox{0}{ 
\resizebox{0.48\linewidth}{!}{%
\includegraphics{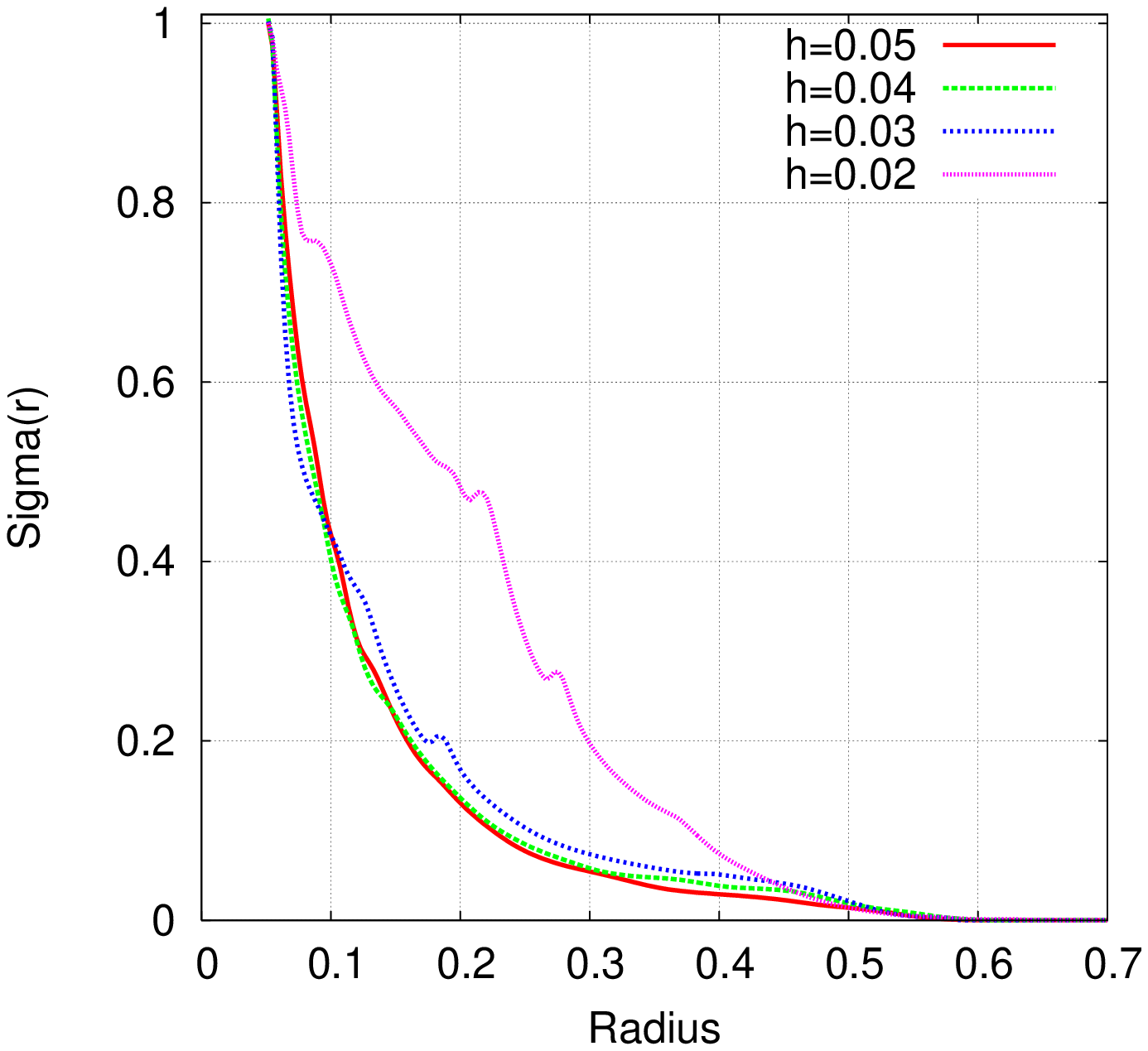}}} 
\end{center}
  \caption{{\bf Left}: Dependence of the growth time ($\sigma\subscr{d}^{-1}$) of disk
eccentricity on $h=H/r$. Quoted is the time (in orbital periods) to increase the
eccentricity by a factor of $e;$
{\bf Right}: The equilibrium  surface density profile of the disk normalized to the
central value for different $h.$
}
   \label{fig:hr-mult1}
\end{figure}

The inferred growth times as a function of $h=H/r$ are plotted in the left panel
of Fig.~\ref{fig:hr-mult1}. The growth is fastest for cooler disks and the growth time
$1/\sigma\subscr{d}$ increases {\it linearly} with increasing $h$. For the thinnest disks we study with
$h = 0.02, $ the $e$-folding time is around 10 orbits (for the standard
viscosity).
In the right panel of Fig.~\ref{fig:hr-mult1} the equilibrium surface densities are
displayed for various $h$. Obviously the value of $h$ does not play a role
in determining the equilibrium structure. At higher resolution, the surface
density  profile of the
cool $h=0.02$ disk is in better agreement with the others as well.
 
\begin{figure}[ht]
\begin{center}
\rotatebox{0}{
\resizebox{0.98\linewidth}{!}{%
\includegraphics{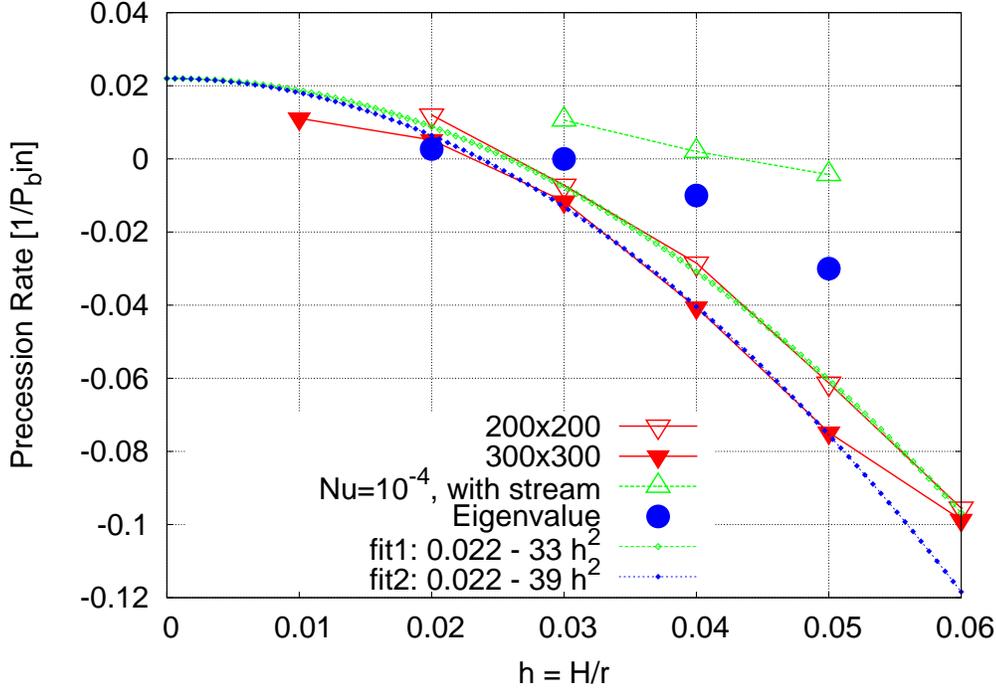}}}
\end{center}
  \caption{Dependence of the equilibrium disk precession rate $\dot\varpi\subscr{d}$ on
disk temperature. The two separate lines with the triangles refer to two different
grid resolutions, the upper to the standard case and the lower to $300\times300$.
The additional curves with the small open and solid diamonds are approximate parabolic
fits to the data points. Values obtained by solving the linear eigenvalue
problem using the azimuthally averaged  surface density profiles
are also indicated by the filled circles. The open triangles are obtained from models
where the disk is impacted by a mass-transfer stream (see Sect.~\ref{sec:stream} below).
}
   \label{fig:hr-om}
\end{figure}

\begin{figure}[ht]
\begin{center}
\rotatebox{0}{
\resizebox{0.98\linewidth}{!}{%
\includegraphics{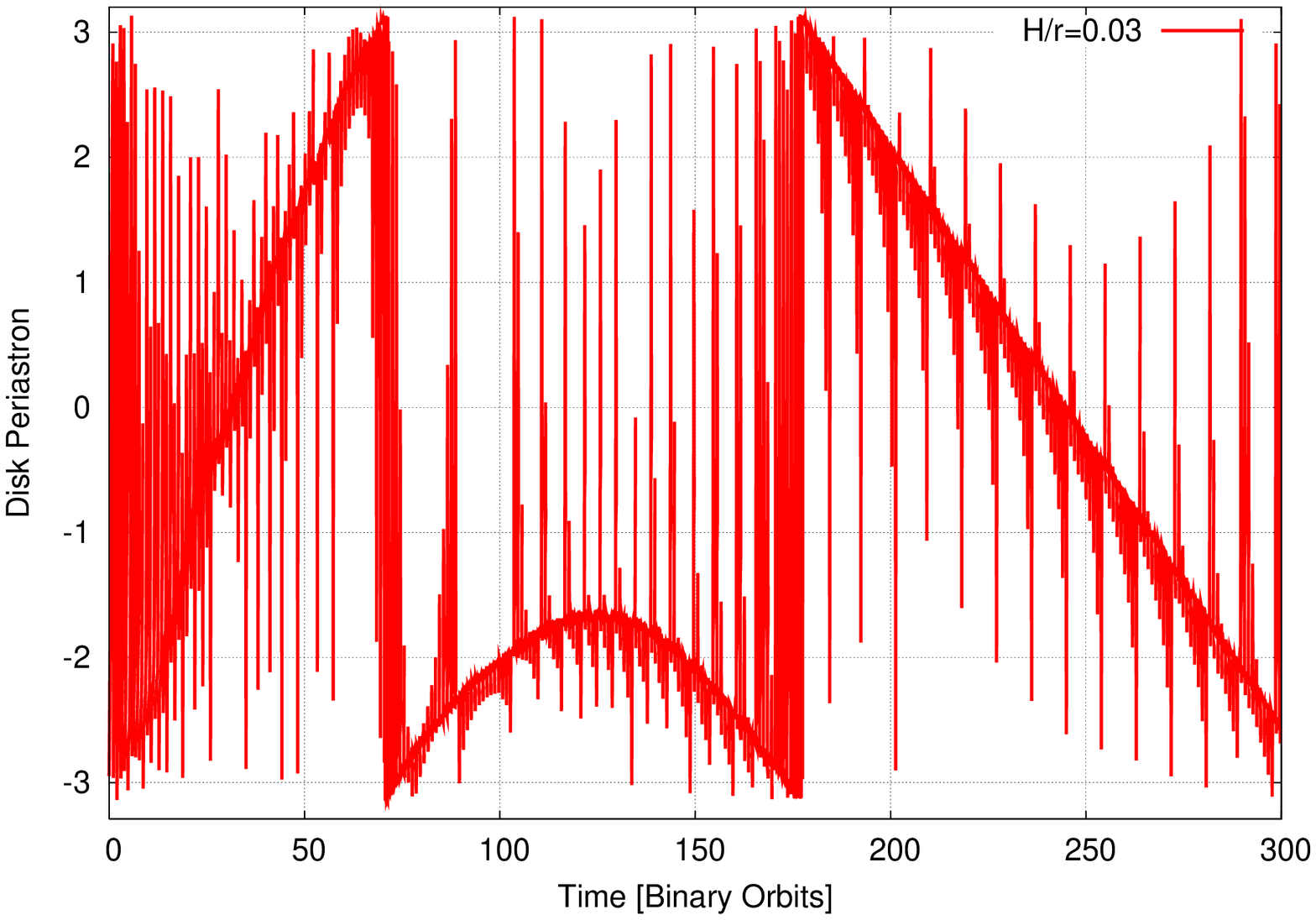}}}
\end{center}
  \caption{Time-evolution of the pericentre (in the inertial frame)
of the disk, $\varpi\subscr{d}(t)$, for
a model with an intermediate $H/r=0.03$.
}
   \label{fig:m01g-omd1}
\end{figure}
We plot the equilibrium precession rate of the disk
as a function of $h$ in  Fig.~\ref{fig:hr-om}.
The inverse of this quantity gives the period of one
complete turn of the disk (in the inertial frame).
Values obtained by solving the linear
free $m=1$ mode eigenvalue problem using the
mean surface density distributions shown in Fig. \ref{fig:hr-mult1}
and assuming a free outer boundary
are also plotted 
(for a discussion of this approach see \citet{2005A&AS..432..757P} or  \citet{2006MNRAS.368.1123G}).
These  are in good agreement with the results derived from the nonlinear
simulations.
For very cool disks, $h \lsim 0.025,$
the precession rate is positive while for hotter disks the disk precesses
in  a retrograde sense.
The precession period at the  transition is  infinite,
corresponding to the disk
being stationary in the inertial frame.
In general the  precession rate is determined by a combination of different effects.
There is a contribution, $\dot\varpi\subscr{dyn},$ due to the axisymmetric component of the binary potential, and there is also a pressure contribution, $\dot\varpi\subscr{press}.$ The
first part $\dot\varpi\subscr{dyn}$ is positive, giving a prograde contribution, while the 
latter part $\dot\varpi\subscr{press}$ is negative, tending to produce retrograde precession.
Our results confirm the existence of these 
 contributions and indicate a  parabolic
dependence  of $\dot\varpi\subscr{d}$ on $h$ for locally isothermal disks.
The influence of the viscosity on the precession rate is negligible as we have indicated above.

A change of the direction of precession sometimes occurs during the
earlier growth phases of the simulations and is illustrated 
in Fig.~\ref{fig:m01g-omd1} where we plot the time-evolution of the apsidal line
  of the disk
for  $h=0.03.$ Initially, during the period of eccentricity growth, the disk precesses
in a prograde sense and at time $t \approx 130$, as it settles into a quasi-stationary state,
it turns around and undergoes
retrograde precession. Thus this value of $h$ lies close to the borderline separating  these two possibilities for the final state.

We also remark that the precession rate is sensitive to the mass distribution
in the outer regions of the disk and it can be changed from retrograde to prograde
by supplying matter from the secondary through the $L_1$ point
(see Fig.~\ref{fig:hr-om} and Sect.~\ref{sec:stream} below).

\subsection{Effect of the inner boundary condition}
\label{subsec:inner}

\begin{figure}[ht]  
\begin{center}
\rotatebox{0}{
\resizebox{0.98\linewidth}{!}{%
\includegraphics{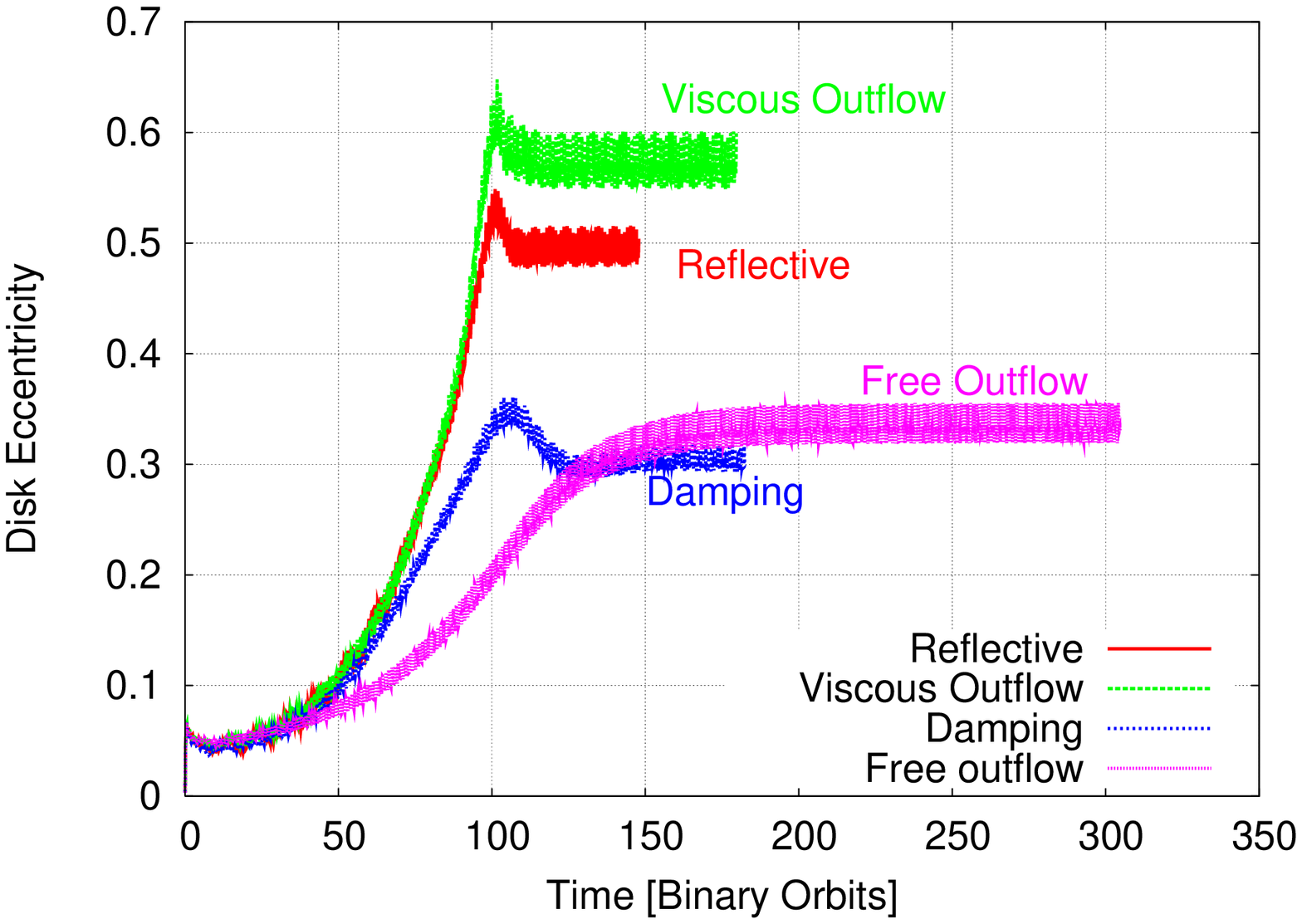}}}
\end{center}
  \caption{Time-evolution of the mean eccentricity of the disk for
different inner boundary conditions for a kinematic viscosity of $\nu = 10^{-4}$
(ten times the standard value).
}
   \label{fig:eccdc1-bc}
\end{figure}

We here discuss the effects of the different
kinds of boundary conditions 
that we have implemented
as described in Sect.~\ref{subsec:setupbc},
and discuss briefly the influence of a higher density floor.
The evolution of the mean eccentricity of the disk for the
different inner boundary conditions for a kinematic viscosity of $\nu = 10^{-4}$
is shown in Fig. \ref{fig:eccdc1-bc}.

%

\begin{figure}[ht]
\begin{center}
\rotatebox{0}{
\resizebox{0.98\linewidth}{!}{%
\includegraphics{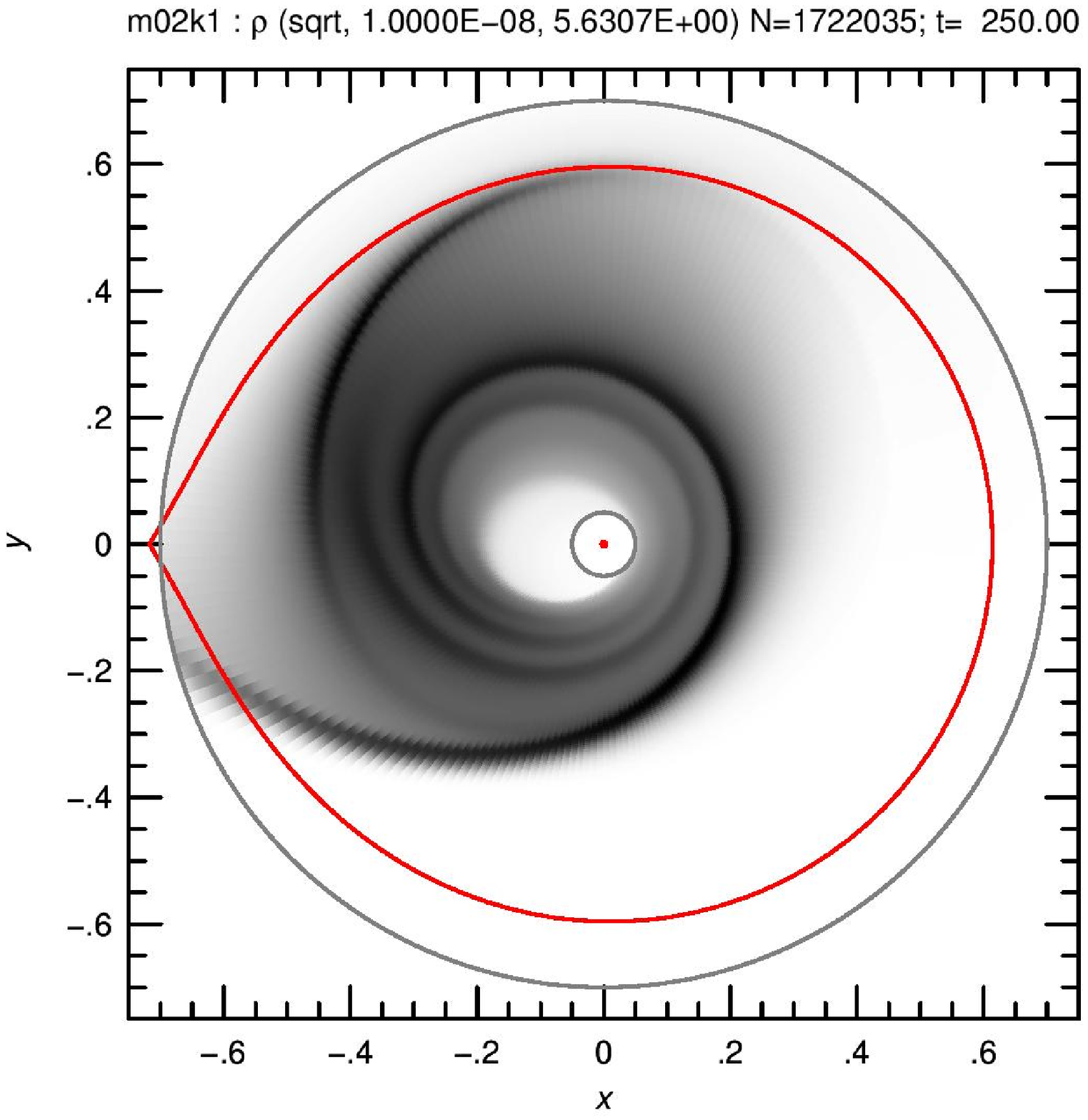}}}
\end{center}
  \caption{Two-dimensional surface density structure for the model with the open outflow
inner boundary condition after 250 orbital periods. 
}
   \label{fig:m02k1-xy-250}
\end{figure}

The results shown  in Fig.~\ref{fig:eccdc1-bc} indicate that the adoption of an inner boundary condition
with damping properties (such as the damping boundary conditions and the outflow condition)
lead to configurations that differ from the standard case. Both the growth rate and 
the mean equilibrium eccentricity of the disk are strongly affected.
Increased damping yields  smaller growth rates and smaller final $e\subscr{d}.$
The outer disk radius shows similar tendency but it varies within a much smaller range
from $r\subscr{d} \approx 0.60$ for the damping boundary condition to  $r\subscr{d} \approx 0.64$ for the viscous
outflow condition (see Table~\ref{tab:res-bc}).
The viscous outflow condition and the standard reflecting condition behave very
similarly because both have  strong reflective
properties. 
The most pronounced influence of the boundary is on the precession rate of the disk,
as is indicated in Table~\ref{tab:res-bc}.  
For the free outflow condition we find
a very slow positive, i.e. {\it prograde}, precession of the disk.
However, as is indicated in Fig.~\ref{fig:m02k1-xy-250},  a very pronounced
eccentric inner hole forms in the centre of the disk which does not happen in  cases
where the inner boundary is forced to be axisymmetric.
When an open boundary condition is used,  material on an eccentric orbit can leave the system freely and is not `reflected back' into the computational domain.
A side-effect of the use of the open boundary condition
(and also the damping condition)
is a reduction in the overall eccentricity of the disk.
Clearly, the treatment of the inner boundary condition is an important
issue that needs to be considered in more detail.
In the case of a central compact stellar object,
 such as a white dwarf or  neutron star, which has a
very small pressure scaleheight in its upper layers, it might behave 
approximately as a rigid wall with typical reflecting wall conditions
for the normal velocity
and a no-slip boundary condition for the tangential velocity.
This is the situation of the  boundary layer problem in the theory of accretion disks
which has an associated radial lengthscale that is too small to make inclusion
tractable in a simulation of the type considered here.
However, it appears that the overall disk properties may be significantly affected by the 
behaviour of this small innermost region of the disk. 
The final line in Tab.~\ref{tab:res-bc} refers to a model with a reflecting inner boundary
and a high density floor of $10^{-3}$ (instead of $10^{-8}$). While the final values of the
disk's eccentricity and radius are not changed much, there is a significant difference in
the precession rate. The higher density in the outer parts of the primary's Roche lobe
slows down the precession by a factor of about 2.8. Hence, to accurately determine the 
important rate of disk precession a small floor value is clearly required. 

\begin{table}[ht]
\centerline{
\begin{tabular}{llll}
\hline
Boundary condition   &  $e\subscr{d}$   &  $r\subscr{d}$   &  $\dot\varpi\subscr{d} [1/P\subscr{orb}]$    \\
\hline
Reflecting           &   0.50   &  0.640   &  -0.064  \\
Viscous outflow      &   0.56   &  0.651   &  -0.060  \\    
Damping              &   0.31   &  0.598   &  -0.046  \\
Free Outflow         &   0.33   &  0.597   &  +0.0037  \\
Reflecting, high floor  &   0.52   &  0.620   &  -0.022  \\  
\hline
\end{tabular}
}
\caption{Results obtained using the different inner
 boundary conditions and density floor.
}
\label{tab:res-bc}
\end{table}

\section{Mass ratio}
\label{subsec:mass}

We  now investigate the influence of the mass ratio $q = M_2/M_1$ on the
disk dynamics.
To speed up the simulations all of the runs in this section have been performed
with $\nu=10^{-4}$ which is ten times larger than adopted for the standard simulation
(see Table~1).

\begin{figure}[ht]
\begin{center}
\rotatebox{0}{
\resizebox{0.98\linewidth}{!}{%
\includegraphics{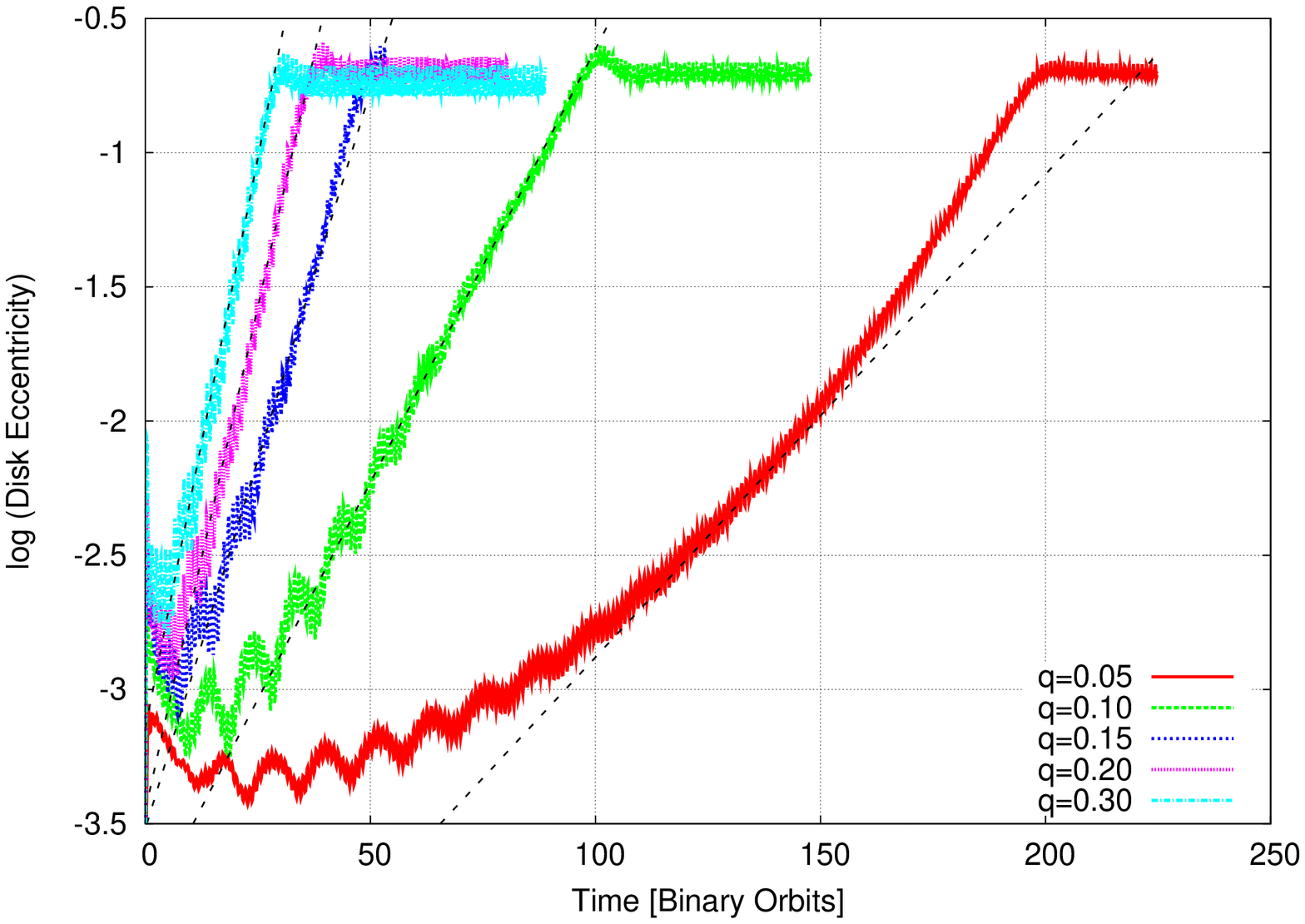}}}
\end{center}
  \caption{Time-evolution of the mean eccentricity of the disk (logarithmic) for
different mass ratios $q=M_2/M_1$ of the binary for a kinematic viscosity of $\nu = 10^{-4}$
(ten times the standard value). The thinner black dashed lines indicate approximate linear
fits.
}
\label{fig:eccdcl1-q}
\end{figure}

\begin{figure}[ht]
\begin{center} 
\rotatebox{0}{
\resizebox{0.48\linewidth}{!}{%
\includegraphics{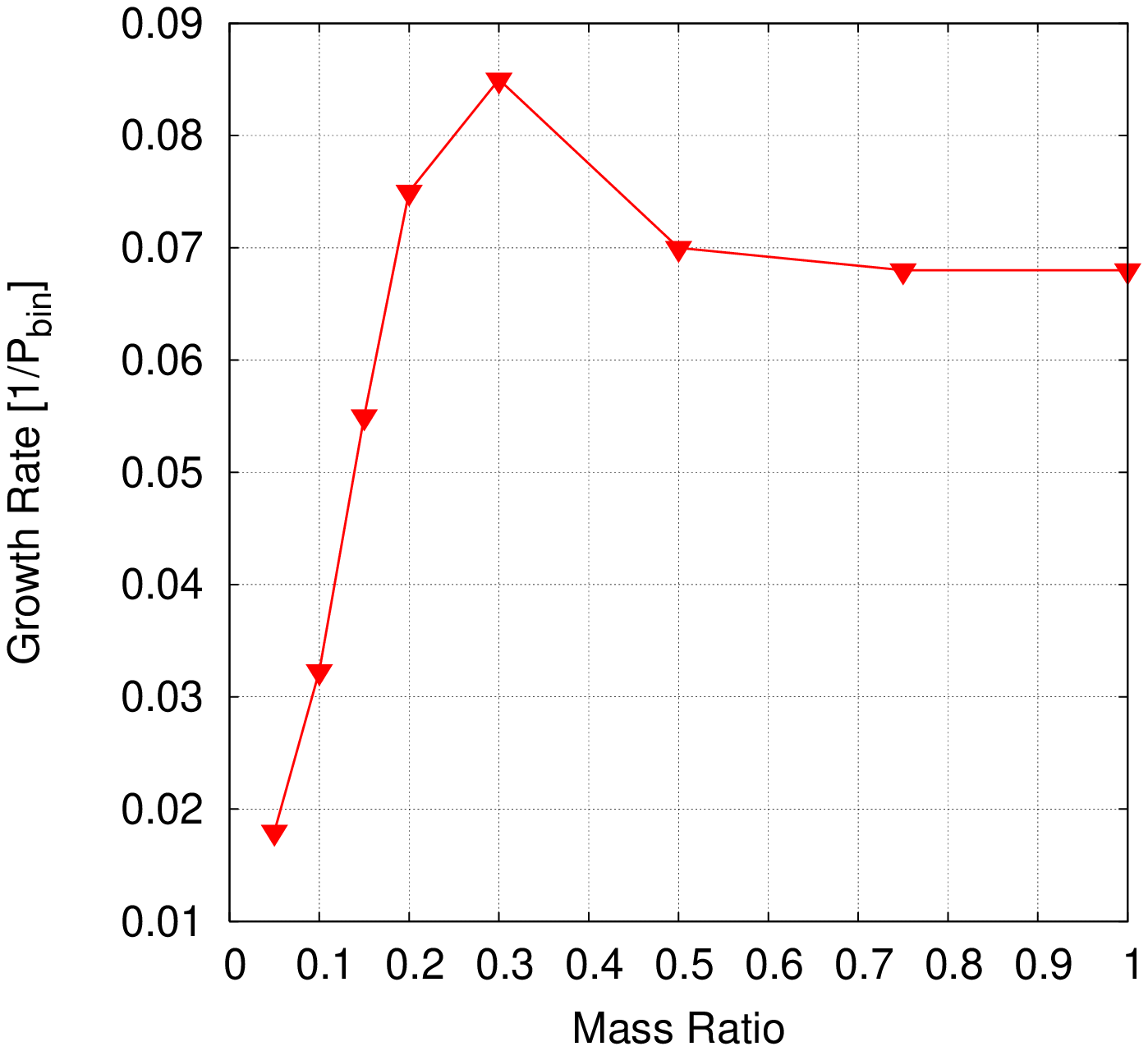}}}
\rotatebox{0}{
\resizebox{0.48\linewidth}{!}{%
\includegraphics{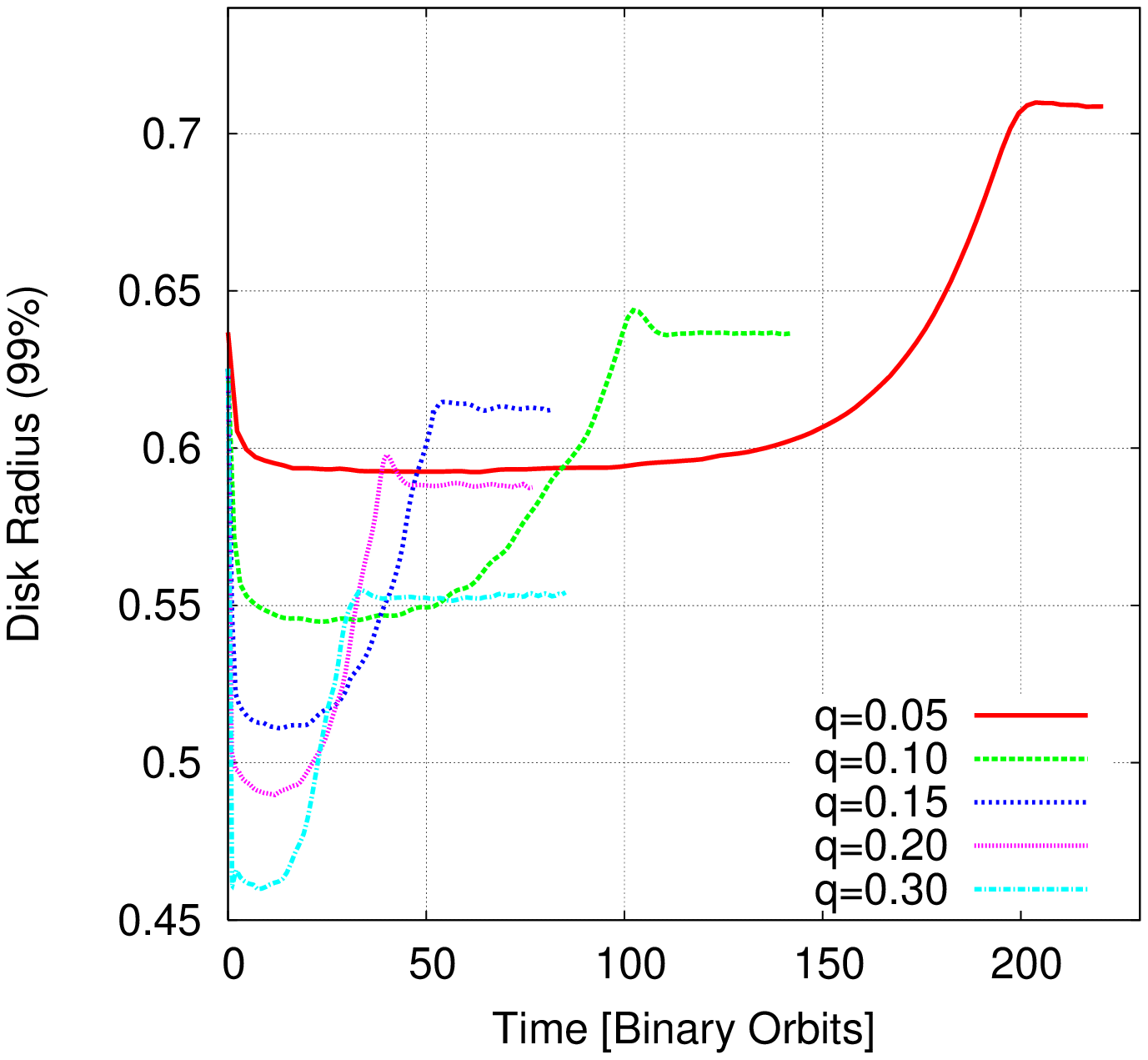}}}
\end{center}
  \caption{{\bf Left}: Dependence of the growth rate ($\sigma\subscr{d}$) of the disk
eccentricity on the mass ratio $q$.
{\bf Right}: The disk radius $r\subscr{d}$ as a function of time for different $q$. The curves
are smoothed. 
} 
   \label{fig:q-mult1}
\end{figure}

The   eccentricity as a function of time is shown in Fig.~\ref{fig:eccdcl1-q}
for different $q$ with over-plotted straight-line fits indicating  exponential growth.
For larger mass ratios $q$ the growth is strongly enhanced, but all models settle to
approximately the same final eccentricity $e\subscr{d}$. For the smallest mass ratio ($q=0.05$) there
is no direct indication of an exponential growth phase and we have drawn an
approximate  straight-line fit applicable to the middle of the growth phase. 
The extracted  growth rates are plotted as a function of mass ratio in
the left panel of Fig.~\ref{fig:q-mult1}. For small $q$ the growth rate
increases linearly with $q$ but it slows  down at the largest value $q=0.3$
plotted. 
We have performed explorative runs for larger $q$ and find eccentricity growth all the
way to $q=1$ with fastest grow occuring at $q\approx 0.3$ (see Fig.~\ref{fig:q-mult1}).
Below it is argued that this behaviour may be related to higher order non-linear mode coupling
which becomes more important for larger $q$, but in this work we do not elaborate on this issue
any further.

The variation of the disk radius with time (right panel of Fig.~\ref{fig:q-mult1})
shows a sharp drop in $r\subscr{d}$  just after the start of the simulations due to the tidal
effect of the secondary which is stronger for larger $q.$
After attaining a  minimum value, as low as $0.46$ for the largest $q,$ the disk expands
again on the eccentricity 
growth timescale. The quasi-stationary disk radius is
smaller for larger $q$ due to the enhanced tidal effects.
Note that for the $q=0.05$ case we have used a larger outer boundary radius $r\subscr{max} = 0.77$. 

\begin{figure}[ht]
\begin{center} 
\rotatebox{0}{
\resizebox{0.48\linewidth}{!}{%
\includegraphics{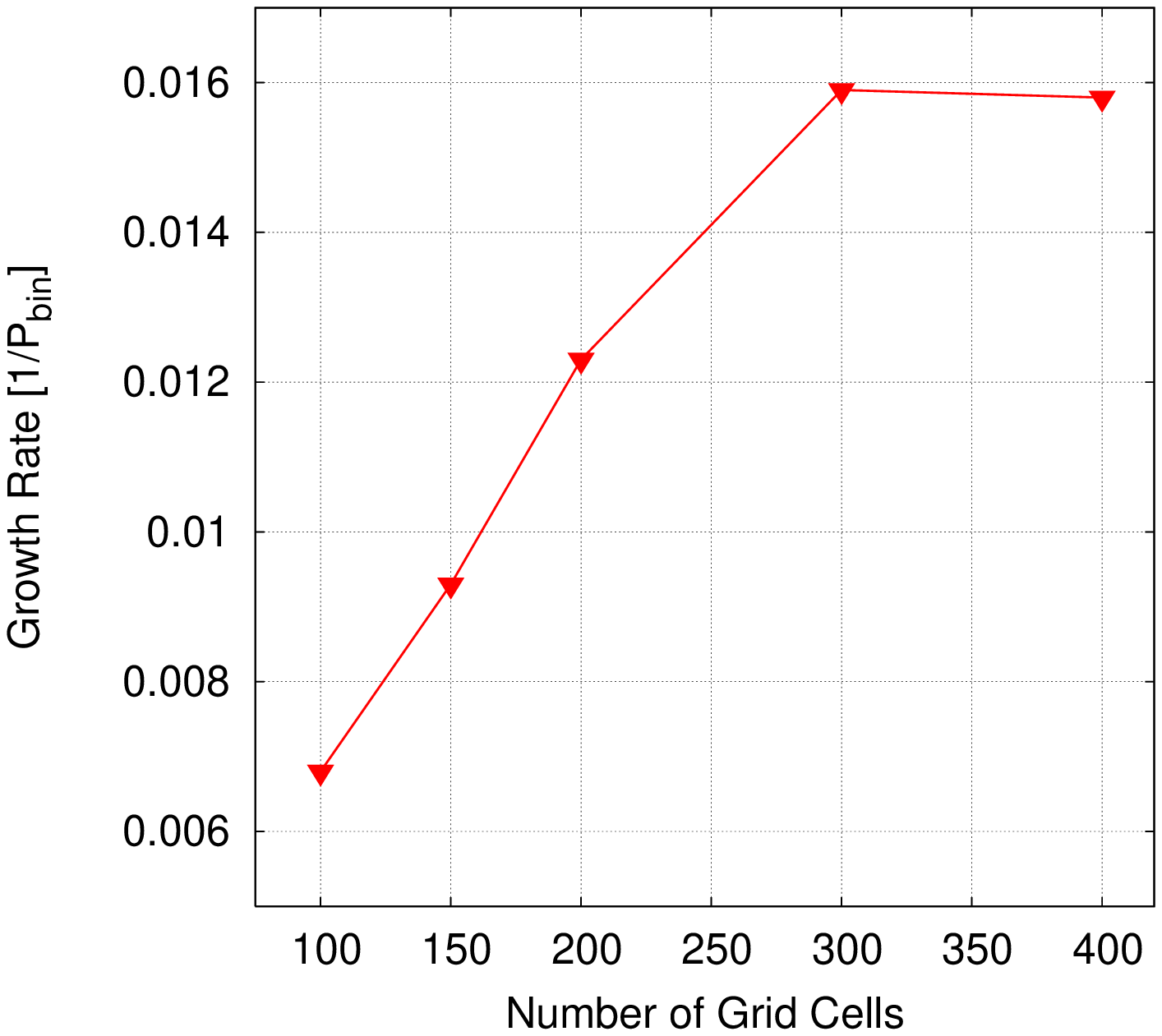}}}
\rotatebox{0}{
\resizebox{0.48\linewidth}{!}{%
\includegraphics{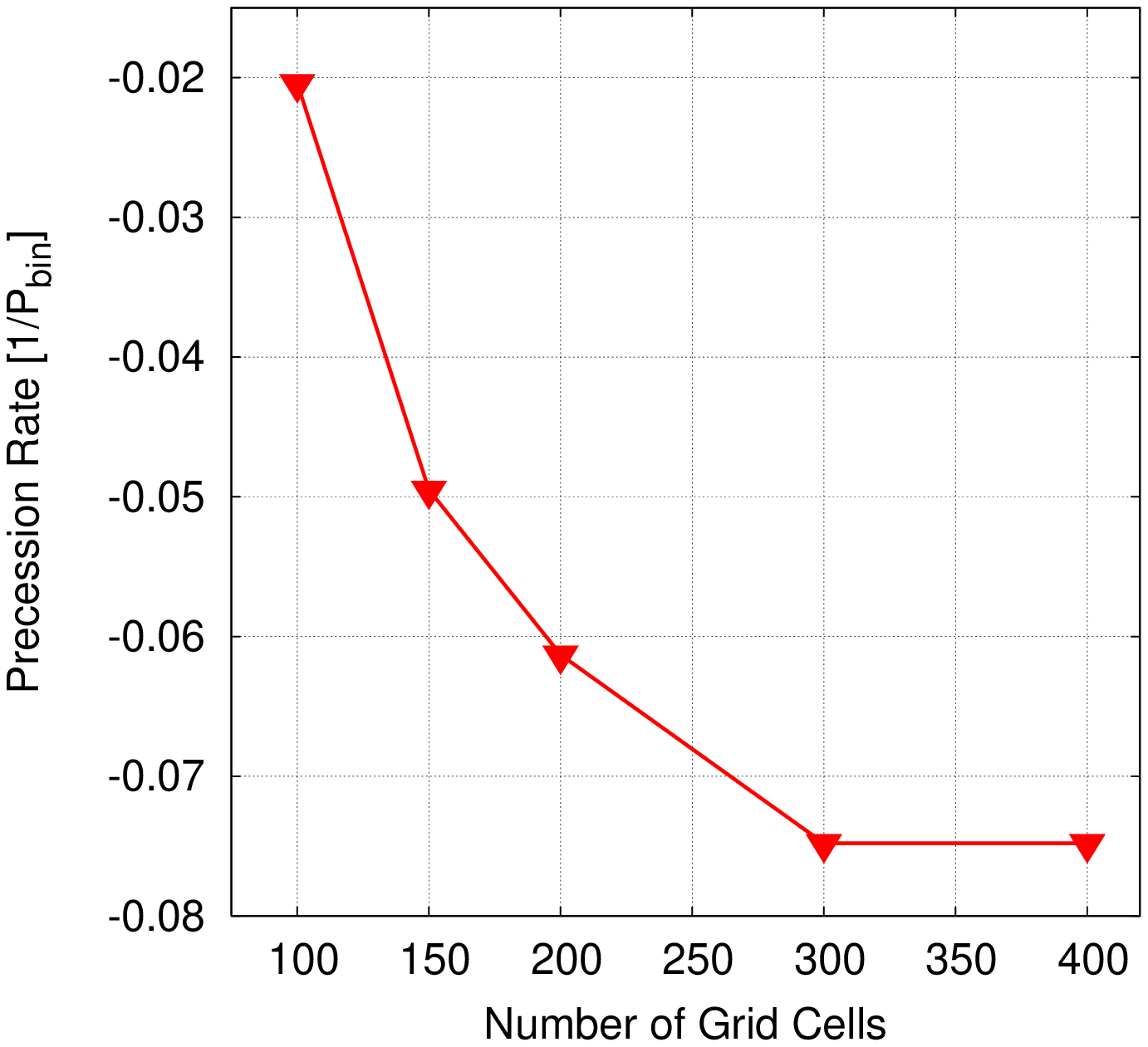}}}
\end{center}
  \caption{{\bf Left}: Dependence of the growth rate ($\sigma\subscr{d}$) of the disk
eccentricity on the numerical resolution. The two-dimensional grid always has an equal
number of
grid points in each direction. Thus in the standard case $200$ 
corresponds to $200\times200.$ 
{\bf Right}: The precession rate of the disk as a function of resolution.} 
   \label{fig:res-mult1}
\end{figure}

%

%
\section{Numerical resolution}
\label{subsec:resolution}
An important issue in (multi-dimensional) simulations is the question of numerical
resolution. To study its effect we have varied the grid resolution
in our standard model from $100\times100$ up to $400\times400.$ 
The grids we use are always uniform.
  The models are typically started at $t=0$ to study the whole
growth phase of the evolution. In some cases 
higher-resolution simulations have  been started using
results obtained from coarser grids to provide initial data
with the object of investigating the quasi-stationary state.
Our main results are contained in Fig.~\ref{fig:res-mult1} where the dependence of the
growth and precession rates of the disk are indicated for different grid resolutions.
The physical parameters used are those of the standard model.
At low resolution, the growth rate increases  approximately as a linear
function of the number
of grid cells. Beyond $300\times 300$ it levels off which indicates
that we have reached numerical convergence. This finding is corroborated
by the fact that the precession rate apparently levels off at the same resolution
(right panel). 
The converged growth rate  
$\sigma\subscr{d} \approx 0.016$
differs from the value for our standard resolution ($\sigma\subscr{d} \approx 0.012$) by
approximately $30\%$; similarly the value of $\dot\varpi\subscr{d}$ differs by about $25\%$.
While the absolute values for the growth and precession rate change, the qualitative behaviour 
of the disk, in particular the prograde slow precession in this case, does not change.

\section{Instability mechanism}\label{Instability mechanism}
\subsection{Mode coupling}

\begin{figure}[ht]
\begin{center}
\rotatebox{0}{
\resizebox{0.98\linewidth}{!}{%
\includegraphics{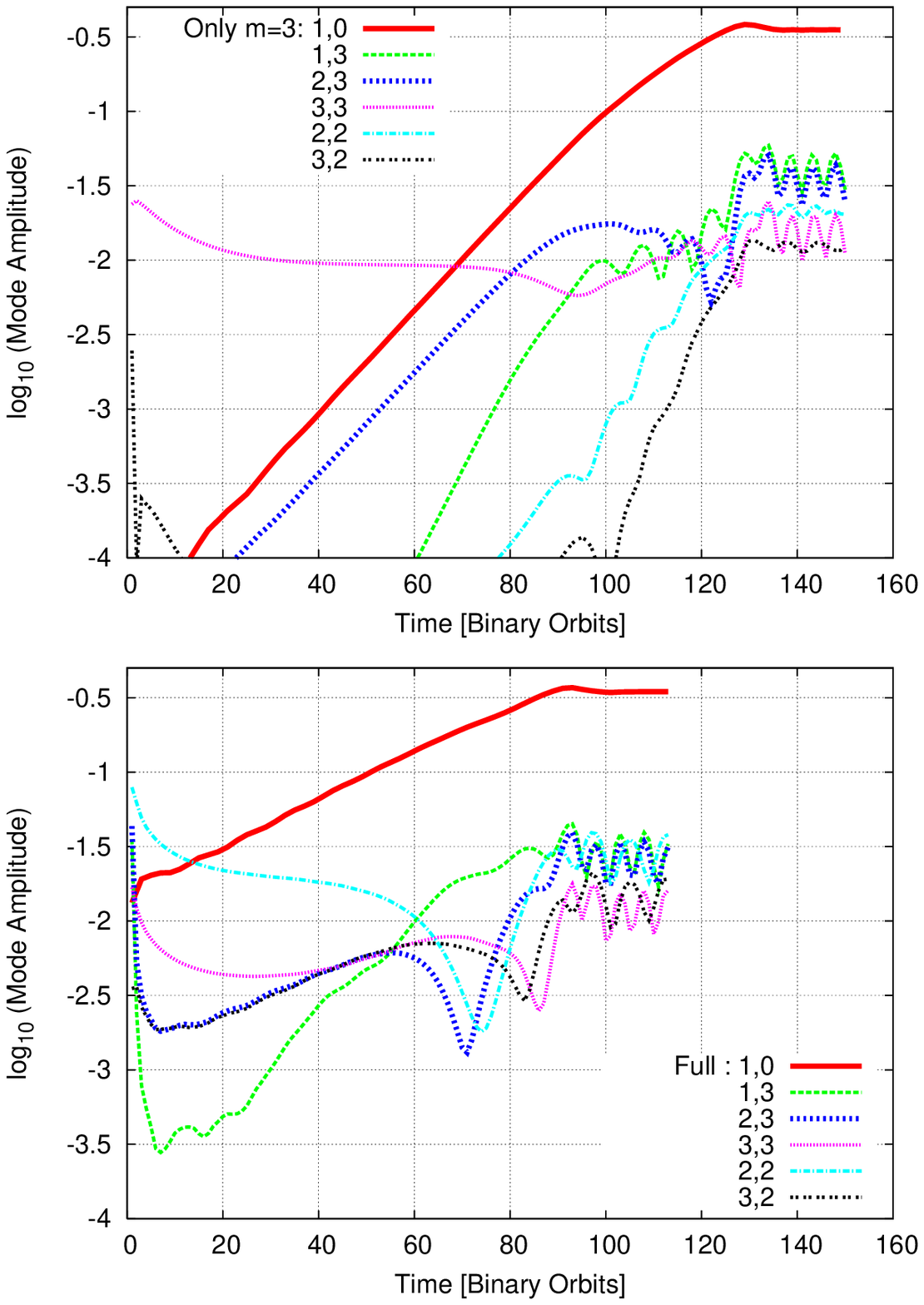}}}
\end{center}
\caption{Time-evolution of the  mode amplitudes $S_{k,l}$ 
(see Eq.~\ref{eq:mode-ampl}) for
$q=0.1$ with a kinematic viscosity  $\nu = 10^{-4}$ plotted on a logarithmic scale.
The upper panel gives results
for the case when only the component of the tidal potential with $m=3$ is
retained. The lower panel gives results corresponding to the case when the full tidal potential
is used. In both cases the mode with $k=2$ and $l=3$ grows at the same rate as the mode with
$k=1$ and $l=0,$ the latter mode being responsible for the disk eccentricity.
The growth is slower when the full potential is used.
}
   \label{fig:q01-modes-log}
\end{figure}

\begin{figure}[ht]
\begin{center}
\rotatebox{0}{
\resizebox{0.98\linewidth}{!}{%
\includegraphics{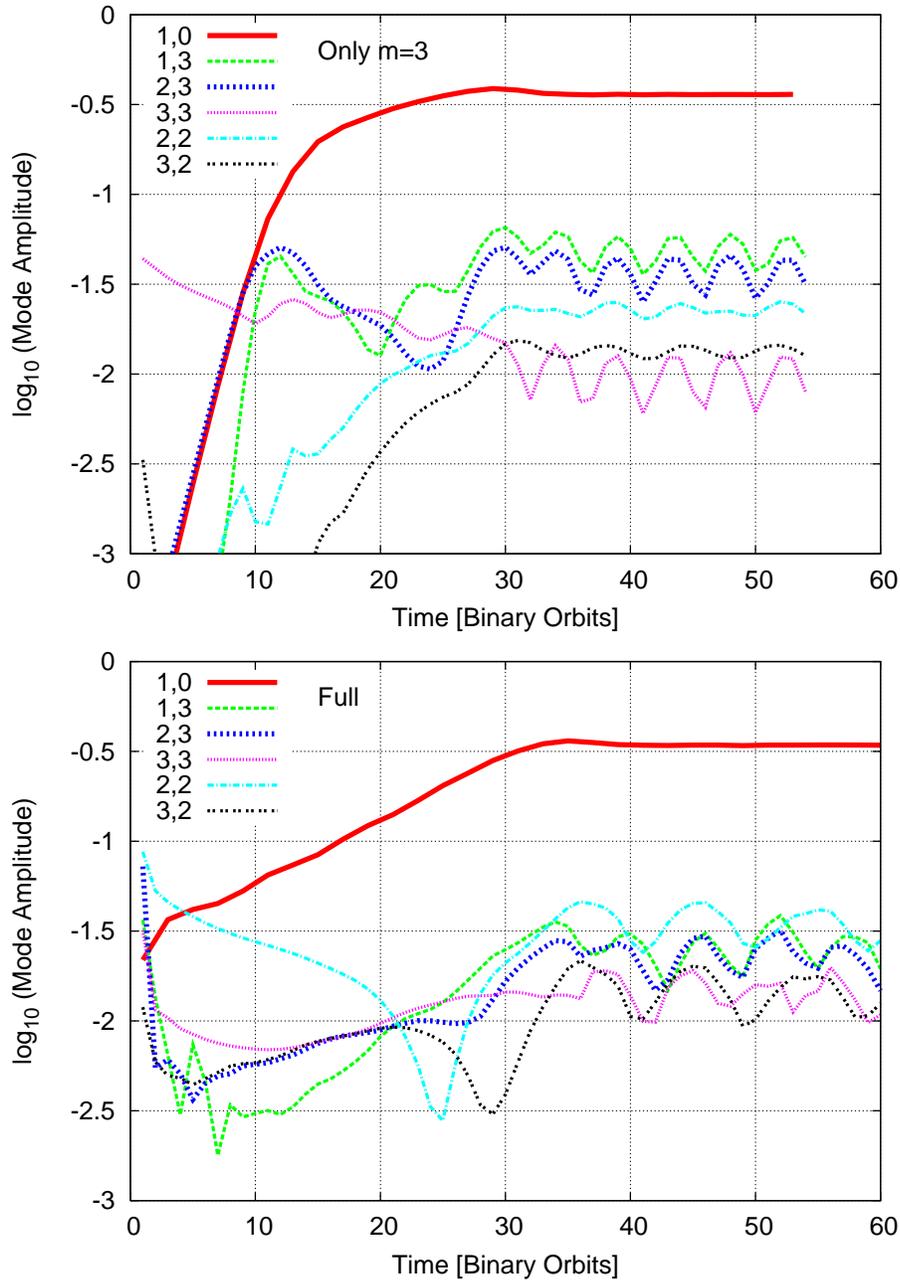}}}
\end{center}
  \caption{ As in Figure \ref{fig:q01-modes-log} but for $q=0.2.$
Note that there is a very close correspondence between  $S_{2,3}$ and $S_{1,0}$
when only the $m=3$ component of the tidal potential is used.
}
\label{fig:q02-modes-log}
\end{figure}

\begin{figure}[ht]
\begin{center}
\rotatebox{0}{
\resizebox{0.98\linewidth}{!}{%
\includegraphics{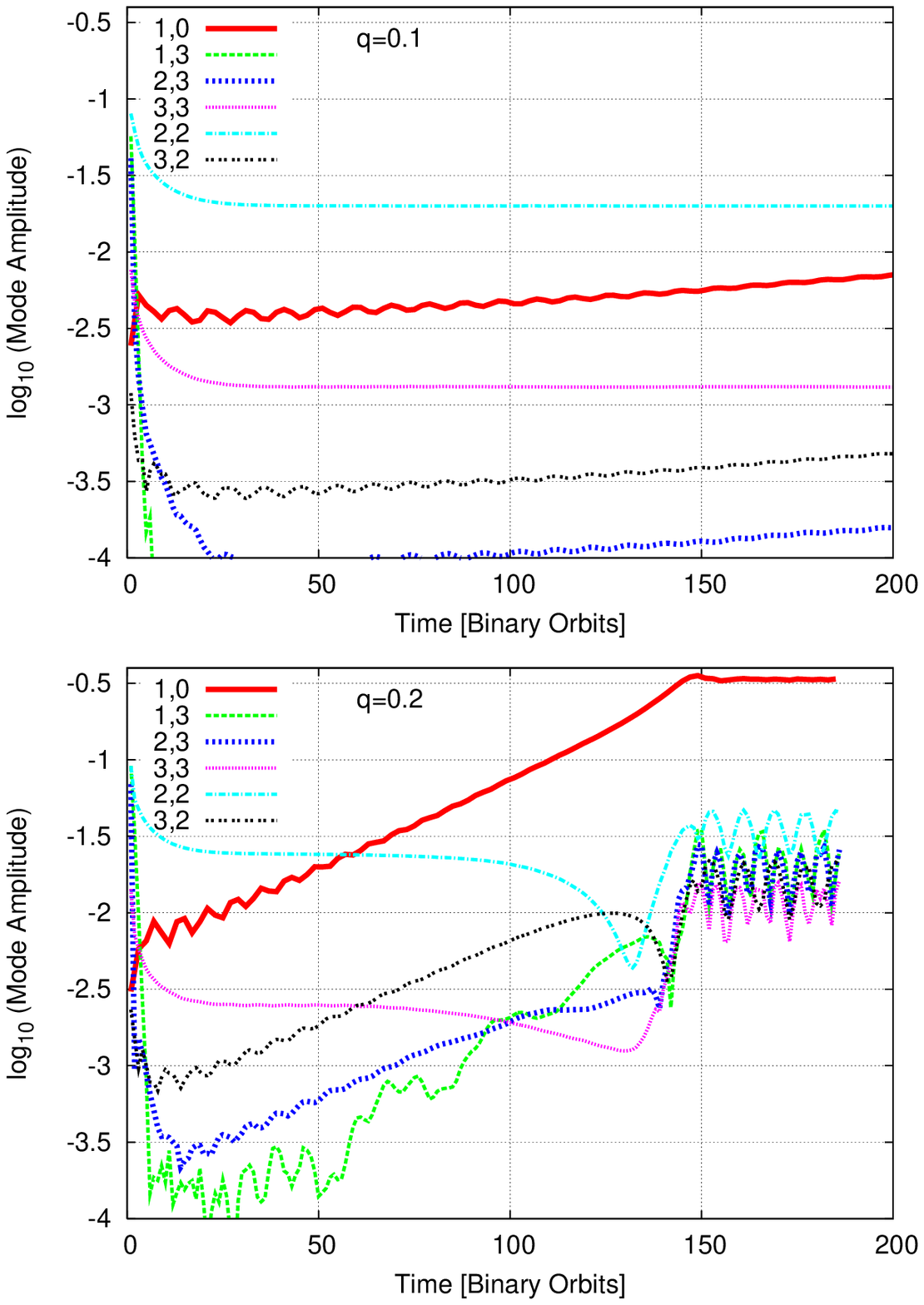}}}
\end{center}
 \caption{ Time-evolution of the mode amplitudes $S_{k,l},$
plotted on a logarithmic scale,  for
$q=0.1$ (upper panel) and $q=0.2$ (lower panel). For these cases, the tidal potential
with the $m=3$ component removed is used  and
the kinematic viscosity  $\nu = 10^{-4}.$  When $q=0.1$ little growth of the disk eccentricity
or $S_{1,0}$ is seen. When $q=0.2$ some growth  occurs but at a significantly slower rate than when
the full potential is used (see Fig. \ref{fig:q02-modes-log}).}
\label{fig:modes-nom3-log}
\end{figure}
In this section we  investigate the mechanism of 
excitation of the disk eccentricity.
A  mechanism responsible for this
 that involves the coupling
of velocity (or surface density) perturbations  with different azimuthal mode numbers through
the companion's tidal potential  has been described
by \citet{1991ApJ...381..259L}.
This was later confirmed by  \citet{1991ApJ...381..268L}    by carrying out and analysing
appropriate SPH simulations. 

Although incorporating the slow precession of the disk
as viewed in the inertial frame
in the discussion below would introduce minor adjustments, for simplicity
we neglect it.
 Then when viewed from this frame,
 the  eccentricity of the disk may be considered to be associated
with a stationary surface density perturbation with azimuthal mode number
$m=1.$ A disk in which the gas orbits in circles about the primary
is subject to a tidal potential  with components which vary as trigonometric functions
of $m(\varphi -\omega t),$ $m= 0,1,2,3,\dots$
These produce corresponding responses in the state variables that
scale with the mass ratio, $q.$

When the disk becomes
eccentric through the excitation of an $m=1$ mode that is approximately stationary in the inertial frame, nonlinear
coupling between the response to the tidal potential and this mode
produces an effective forcing which will be a sum of components
that are trigonometric functions of $(m\pm 1)\varphi - m\omega t,$ $m = 0, 1,2,3\dots$
These will have amplitudes that scale as $eq,$
$e$ being the characteristic disk eccentricity.
More generally a coupling of order $j$ produces effective forcing
$(m\pm j)\varphi - m\omega t,$ $m = 0, 1,2,3\dots$ with
pattern speed $m\omega/(m\pm j)$ and amplitude scaling as
$e^j.$

\subsection{Energy and angular momentum exchange with the companion}
Each of these effective forcings produces a surface density
response that results in energy and angular momentum exchange with the
companion. This exchange may in some cases
produce a feedback causing the eccentricity
to grow. This works because a companion in fixed circular orbit
{\it must} undergo angular momentum  changes, $\Delta J,$ and energy changes,
$\Delta E$, such that $\Delta J=\Delta E/\omega.$
However, the angular momentum and energy changes associated  with
a disturbance with pattern speed
$m\omega/(m\pm j)$ are such that
$\Delta J= \Delta E(m\pm j)/(m\omega).$
Thus for surface density disturbances with the $-$ alternative, that rotate
faster than the companion and are thus expected to transfer orbital angular
momentum to it, too little
is transferred per unit of energy transferred.
The difference has to come from the disk orbits
which accordingly become eccentric.
 Energy and angular momentum transfer to the companion
 is expected to be especially effective if there is an associated
 Lindblad resonance in the disk \citep{1980ApJ...241..425G}.

 Similarly disturbances
 with the $+$ alternative, which rotate more slowly
 than the companion and are expected to gain angular momentum,
 may cause eccentricity growth. However, such disturbances
 are less closely matched to the rotation of the disk
 and in particular have no Lindblad resonances within it
  \citep{1979ApJ...233..857G}
 and are thus not expected to be significant.

 Accordingly, disturbances rotating with pattern speeds
 $m\omega/(m - j),$  with $m$ and $j \ne m$ being integers,
 may be associated with eccentricity growth
 if there is an associated significant energy and angular momentum
 transfer. The presence of a Lindblad resonance is expected to
 enhance this. Although in principle any value of $j$
 might contribute to eccentricity growth, linear instability
  could result  only when   $j=1$ \citep{1991ApJ...381..259L}
and the forcing as a result of coupling is linear in $e.$
In that case the only value of $m$  leading to a disturbance
with a Lindblad resonance within the disk is $m=3.$
From the above discussion this disturbance has
azimuthal mode number $m-j = k = 2$ and pattern speed $3\omega/2.$
The associated Lindblad resonance corresponds to the $3:1$ resonance
at which the disk orbital period is one third of that of the binary.
When this is present the mode coupling may lead to a linear
instability to eccentricity growth \citep{2006MNRAS.368.1123G}.

According to the above view of the mode coupling leading to eccentricity growth,
the surface density perturbation corresponding to the disk eccentricity
should grow together with a surface density perturbation
 with $k=2$ and angular frequency $l\omega,$ with $l=3.$
This disturbance is derived from a coupling between the disk eccentricity
and the $m=3$ component of the tidal potential, which accordingly plays an important part
in the process.

\subsection{Mode amplitudes}
Following  \citet{1991ApJ...381..268L}, in order
to demonstrate the operation of this mechanism and the significance
of the $m=3$ component of the tidal potential,
 we define the mode amplitude
associated with the pair $(k,l)$ as
\beq 
\label{eq:mode-ampl}
S_{kl}  =  \sqrt{(a_{kl}^2+b_{kl}^2+c_{kl}^2+d_{kl}^2)}.
\eeq
In this expression
\bea
& a_{kl} =  \int^{t+2\pi}_t\int_A \Sigma(r,\varphi,t')\cos(k\varphi) \cos(l\omega t')\,dA\,dt'& \\
& b_{kl}=  \int^{t+2\pi}_t\int_A \Sigma(r,\varphi,t')  \sin(k\varphi)\sin(l\omega t')\,dA\,dt'& \\
& c_{kl} =  \int^{t+2\pi}_t \int_A \Sigma(r,\varphi,t')\cos(k\varphi) \sin(l\omega t')\,dA\,dt'& \\
& d_{kl}= \int^{t+2\pi}_t \int_A \Sigma(r,\varphi,t')  \sin(k\varphi)\cos(l\omega t')\,dA\,dt'&,\\ \nonumber
\eea
where the integration is over  the surface area of the disk, $A.$
Note that $S_{kl}$ also involves integration over the
domain $(t, t + 2\pi)$  and is thus  a function of the local time $t.$

In order to test the predictions of the above mode-coupling
analysis, we evaluate the time-evolution of  the mode amplitudes for the $(k,l)$ pairs
$(1,0),(1,3),(2,3),(3,3),(2,2),(3,2)$ for runs with $q=0.1$ and $q=0.2.$
We describe below results obtained with  the kinematic viscosity $\nu=10^{-4}.$
However, similar results are obtained for $\nu=10^{-5}.$
In order to establish the importance of the component
of the tidal potential with $m=3,$
we consider cases for which the full tidal potential is used,
only the $m=3$ component of the tidal potential is used and finally  for which the
full tidal potential with the $m=3$ component removed is used.
By studying the time-evolution of the mode amplitudes in these cases,
the importance of the $m=3$ component of the tidal potential
could be confirmed. Note that because the defined mode amplitudes
involve integrals over the disk radius, the significance
of their absolute values is unclear because of cancellation effects.
However, they may still be used to relate to the growth
rates of linear modes.

The time-evolution of the  mode amplitudes $S_{k,l}$ for
$q=0.1$  is plotted in
Fig. \ref{fig:q01-modes-log}.
The upper panel gives results
for the case when only the component of the tidal potential with $m=3$ is
retained. The lower panel gives results corresponding to the  case when the full tidal potential
is used. In both cases the mode with $k=2$ and $l=3$ grows at the same rate as the mode with
$k=1$ and $l=0,$ the latter mode being responsible for the disk eccentricity.
Although it is difficult to demonstrate the causal nature of this connection, this behaviour would certainly be expected from the discussion of mode coupling given above.
We also find that the final state and magnitude of the eccentricity of the disk
is similar in all cases for which the disk is unstable to becoming eccentric.
However, the growth rate  is measured to be  smaller when the full potential is used.
Fig. \ref{fig:q02-modes-log} gives similar plots to those
shown in in Figure \ref{fig:q01-modes-log} but for $q=0.2.$
Note that, as expected from the mode-coupling theory,
there is a very close correspondence between the mode with $k=2$ and $l=3$
and the mode with $k=1$ and $l=0$ when only the $m=3$ component of the tidal potential is used
for this mass ratio.

\subsection{Effectiveness of the component of the tidal potential with $m=3$}
In order to determine whether the $m=3$ component of the tidal potential
is essential for producing eccentricity growth, we performed simulations  using
the full tidal potential with the  $m=3$ component  removed.
The time-evolution of the mode amplitudes $S_{k,l}$ is shown
in Fig. \ref{fig:modes-nom3-log}
for $q=0.1$ and $q=0.2.$
When $q=0.1$ little growth of the disk eccentricity
is seen confirming the importance
of the $m=3$ component.
However, when $q=0.2$ some growth is seen but at a significantly slower rate than when
the full potential is used.

Although the results discussed above indicate the importance of the $m=3$
component of the tidal potential, they also indicate that it is
not always needed. From Fig. \ref{fig:modes-nom3-log} it
is apparent that although eccentricity growth is
very small when $q=0.1,$ it is still present
but reduced when $q=0.2.$ The time required to attain saturation
is increased from about $30$ to $150$ orbits in that case.

As in the other cases the mode with $k=2$ and $l=3$ grows at the same rate as the mode with
$k=1$ and $l=0$ indicating that this is still responsible for causing eccentricity growth.
Also a response with $k=3$ and $l=3$ is still present even though
it cannot be driven by the $m=3$ component of the potential.
The most natural explanation is that this response is produced
by a forcing originating from a coupling between the response to the dominant $m=2$ tide
and the $m=1$ tide. As before that can then couple to the $S_{1,0}$ mode producing eccentricity growth.
Thus in this case one has replaced the   $(3,3),(1,0)$ pairwise coupling by a
$(2,2), (1,1),(1,0)$ triple coupling. In so doing the strength
is reduced by a factor $q$ making the process less significant in the
smaller mass ratio case as is seen in the simulations.
This process thus contributes more effectively for larger $q.$
On the other hand the $3:1$ resonance which enables effective angular momentum
transport between disk and companion is pushed towards the Roche lobe
requiring further disk expansion and resulting in
 the process becoming increasingly nonlinear.

\section{Comparison with linear theory}
\label{sec:linear}

In an attempt to understand the linear phase of the simulations, during which the eccentricity is small and grows exponentially in time, we analysed the full set of linear equations governing small perturbations, with azimuthal mode number $m=1$, of a steady, axisymmetric disk.  (This is in addition to the approach mentioned in section~\ref{subsec:temp}, in which approximations are made that are appropriate for slowly evolving global eccentric modes.)  For the purposes of comparison with the standard model in which a reflecting inner boundary condition is used, the basic state of the linear calculation is a non-accreting solution with zero radial velocity.

This approach yields appropriate solutions corresponding to slowly precessing eccentric modes of the disk, of which the `lowest-order' mode with the simplest radial structure might be expected to be the most relevant.  The precession rate can be positive or negative depending on the mass ratio of the binary, the value of $H/r$, the surface density profile and the boundary conditions.  While it is possible, using this approach, to account for a number of aspects of the simulation results, the sensitivity of the precession rate to the exact surface density profile and the boundary conditions is a matter of concern.

Such modes also grow or decay slowly as a result of viscosity.  It is known that shear viscosity can render modes unstable through the mechanism of viscous overstability \citep{1978MNRAS.185..629K}, in which some of the energy being drawn by viscous stresses from the differential rotation of the disk is diverted into growing oscillations.  Viscous overstability has been identified as an important issue in the theory of eccentric disks \citep{2001MNRAS.325..231O}.
A local analysis in the case of an isothermal two-dimensional disk with constant kinematic shear viscosity and no bulk viscosity indicates overstability for radial wavelengths $\ga9H$ and a maximum local growth rate of approximately $0.034\nu/H^2$ for a radial wavelength of approximately $13H$ \citep{2006MNRAS.372.1829L}.  This behaviour is confirmed in our global linear calculations.  For sufficiently thin disks, growing modes are obtained.  However, for the standard model with $H/r=0.05$, the preferred local wavelength of the viscous overstability is as long as $0.66r$ and in fact we do not find overstable global modes.  In addition, we confirm by this method that the inclusion of a bulk viscosity with $\nu\subscr{bulk}=2 \times 10^{-5}$ makes a negative contribution to the growth rate that is quite negligible in comparison with those found in the simulations.  Therefore viscous overstability is not relevant to the standard model but might be an important issue when thinner disks are considered.

It is possible within this linear approach to model the nonlinear mode-coupling process by adding terms that correspond to a localised growth of the eccentric perturbations in the vicinity of the 3:1 resonance.  Rather than localising the growth in the form of a delta function as done by \citet{2006MNRAS.368.1123G}, this effect can be represented using a Gaussian or Lorentzian function of radius centred on the resonance and with a width appropriate to a Lindblad resonance broadened by pressure.  Including these terms allows us to obtain growing eccentric modes with growth rates comparable to those obtained in the simulations.  However, it is again found that a detailed comparison is difficult because of the sensitivity of the results to the exact surface density profile and the treatment of the resonance.

According to the angular momentum equation in the case of a uniform shear viscosity, the surface density in the non-accreting basic state should deviate from the initial $r^{-1/2}$ profile only as a result of the tidal torque applied to the disc.  \citet{1977MNRAS.181..441P} provide a method to calculate the tidal torque per unit radius, which gives a result proportional to the viscosity of the disk.  When this is balanced with the viscous torque it predicts a surface density profile that is independent of the viscosity and depends only on the mass ratio $q$.  Unfortunately this approach does not give accurate agreement with the surface density profile shown in Fig.~\ref{fig:m01f-sig4r}, indicating that the angular momentum balance is nonlocal and involves propagating density waves not considered by \citet{1977MNRAS.181..441P}.  The standard model disk is therefore smaller than predicted by \citet{1977MNRAS.181..441P}.  We note that a two-dimensional isothermal disk may allow radial wave propagation to a greater extent than more realistic models.  It may also be relevant that SPH simulations are less likely to allow these waves to propagate and may therefore produce larger disks that are more susceptible to eccentric instability.  Clearly the tidal truncation of disks needs to be better understood in order to make accurate predictions of the growth and precession rates of eccentric modes.

\section{Effect of a mass-transfer stream }
\label{sec:stream}
To make better contact  with possibly more realistic cases with mass overflow from the
secondary we have performed some studies where we modify the
outer boundary condition to allow for mass inflow  through the $L_1$ point.
The model parameters are dealt with in the same way as before. We adopt the
mass ratio $q =0.1$ which allows a scaling to be made such that  
binary parameters match very closely those of  OY~Car which has
a primary of mass $M_1 = 0.685M_\odot$ and $P\subscr{orb} = 0.063$~d~. The
physical separation of the two stars is thus $a\subscr{bin}=0.627 R_\odot.$
Mass is injected into the system at a constant rate. 
Note that in the case of locally isothermal disk models with  
constant kinematic viscosity as adopted here, the independence of the dynamics of the flow,
and in particular the final values of eccentricity and disk radius in a quasi-steady state,
to the surface density scale
means that the results are 
independent of the magnitude of the inflow rate!
This may thus be scaled to a value appropriate for
cataclysmic variables, typically
$\dot{M} = 10^{-9} M_\odot$/yr.
However, in a more  realistic  disk model that includes
suitable radiation and energy transport mechanisms and allows for the possibility
of time-dependent outbursts, this
 would not be the case.

In the simulations  presented here the Roche lobe of the primary is as empty as possible
initially, i.e. the
density is constant and equals the floor value.

\begin{figure}[ht]
\begin{center}
\rotatebox{0}{
\resizebox{0.48\linewidth}{!}{%
\includegraphics{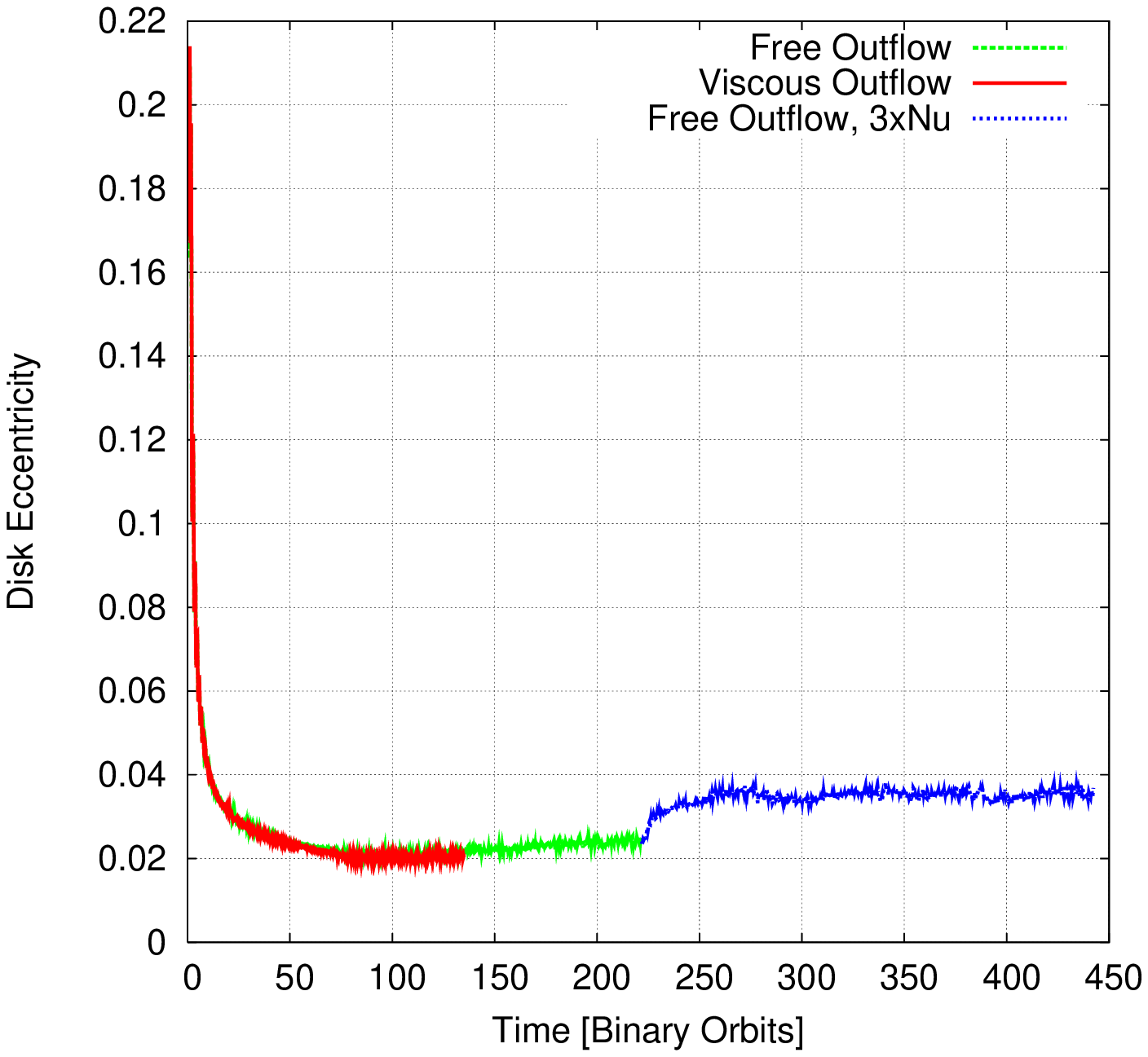}}}
\rotatebox{0}{
\resizebox{0.48\linewidth}{!}{%
\includegraphics{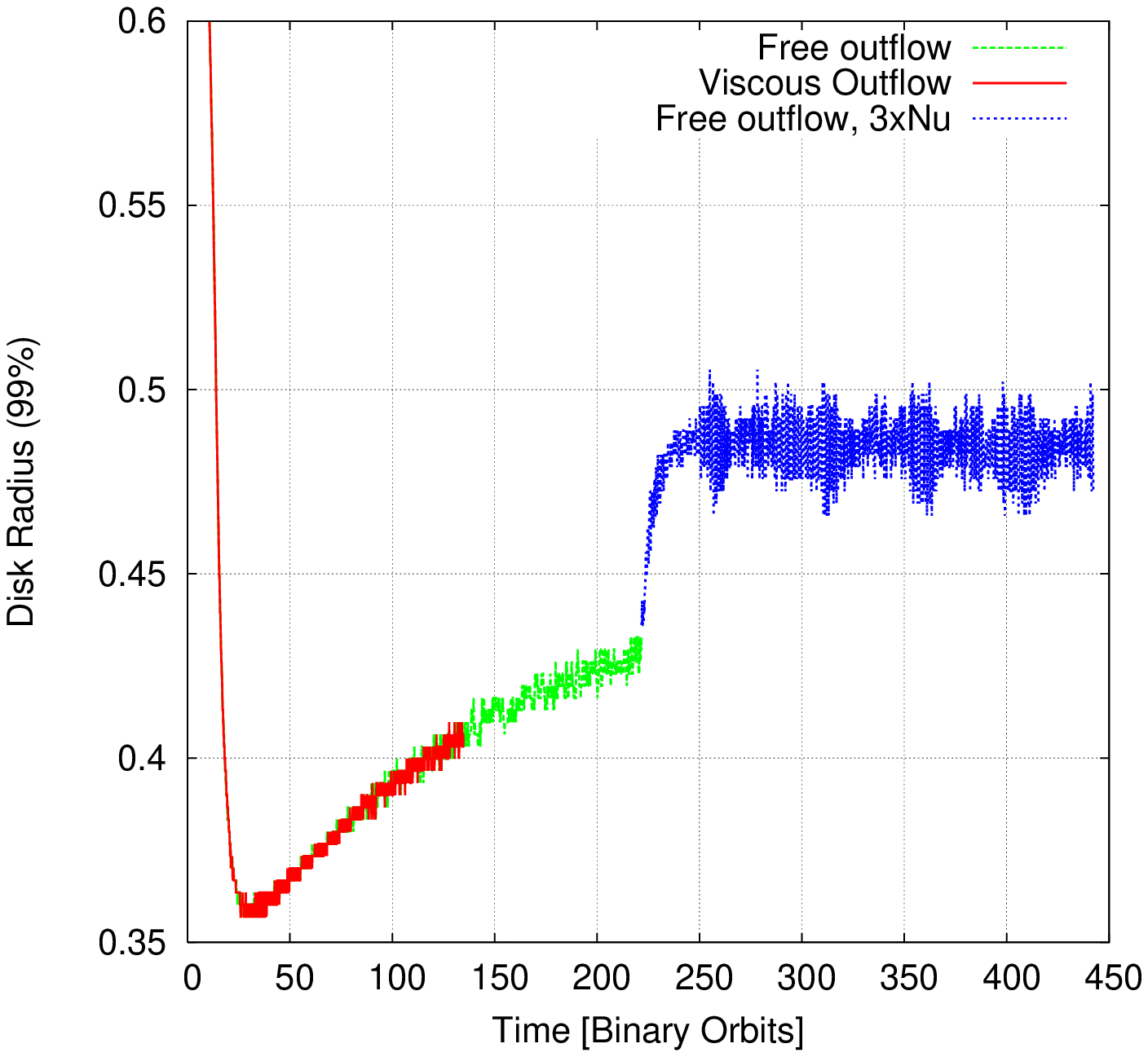}}}
\end{center}
  \caption{Mean disk eccentricity $e\subscr{d}$ and disk radius $r\subscr{d}$ as functions of time
for models that  include mass inflow through the $L_1$ point and different inner boundary
conditions and viscosities. For the first two  plots (green and red curves) the standard viscosity
$10^{-5}$ has been used,  while  the third plot (blue curve) is for a model that has been
continued from the first at time $t\approx 230$ but with $\nu = 3\times 10^{-5}.$
}
   \label{fig:oyc01-mult}
\end{figure}

The first set of simulations were performed with the standard viscosity $\nu = 10^{-5}$ in
dimensionless units which  would scale to a value of $2.2 \times 10^{13}$~cm$^2$/s for the OY~Car
system. The inner boundary condition is open to allow for accretion
of mass onto the  primary star and enable a steady state to be reached.
 We have performed simulations with both the `viscous outflow' and the `free outflow'
conditions as described in section~\ref{subsec:setupbc} above.
From Fig.~\ref{fig:oyc01-mult} it is clear that the  standard viscosity value
used does not result in an eccentric disk.
 Even an increase of the viscosity  by a factor of 3 
 does not  produce an eccentric disk in the `free outflow' case
as the final (blue) curve in Fig.~\ref{fig:oyc01-mult} indicates.
In this latter case the disk's radius is increased to  $r\subscr{d} \approx 0.48$
due to the spreading action of viscosity but the disk nevertheless does not switch
to an eccentric mode. The  apparently large initial disk radii in these models are explained
by the fact that the disk is initially empty and the 99\% radius is   contributed to by the material
that is infalling from the $L_1$ point.

\begin{figure}[ht]
\begin{center}
\rotatebox{0}{
\resizebox{0.48\linewidth}{!}{%
\includegraphics{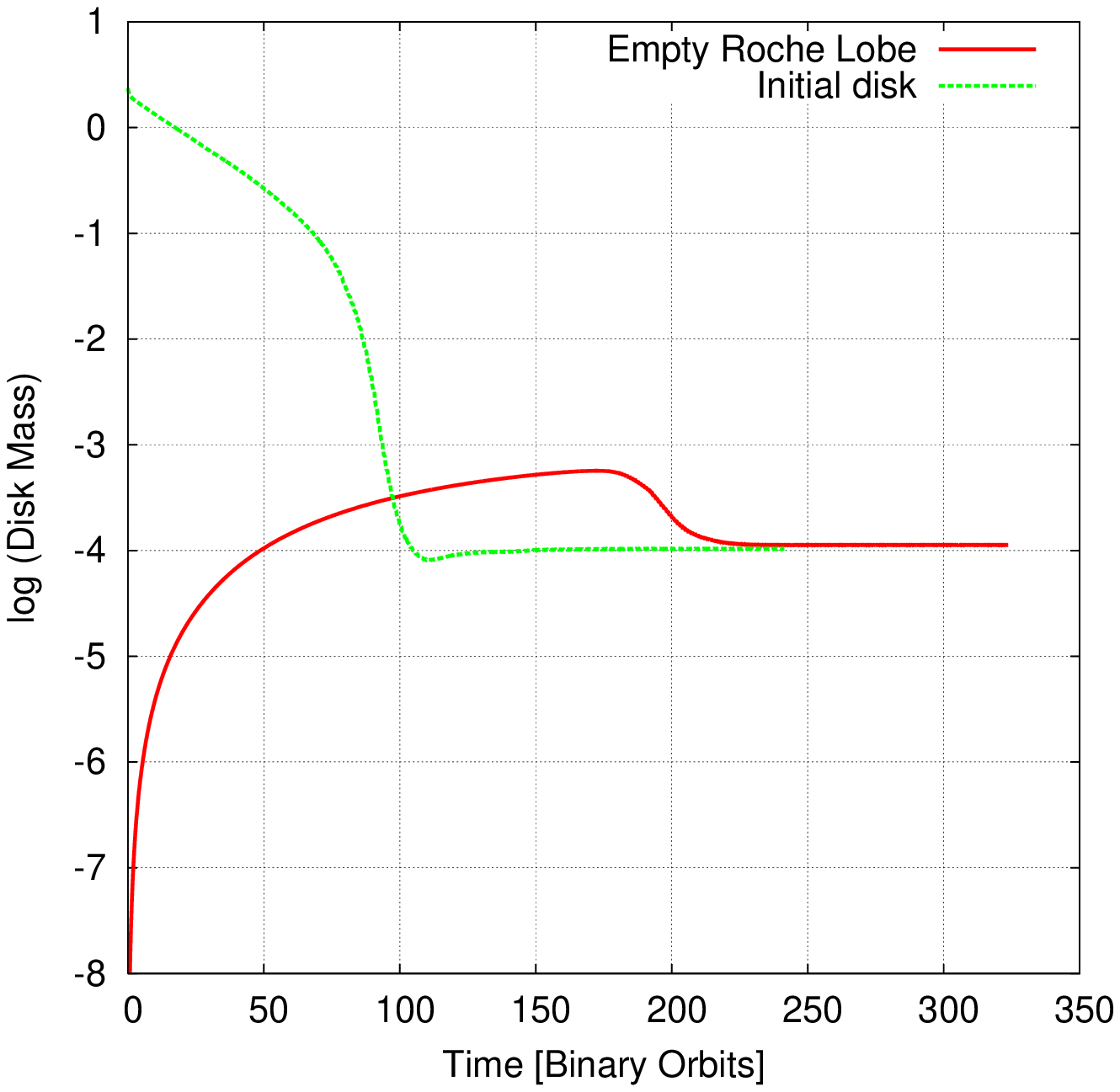}}}
\rotatebox{0}{
\resizebox{0.48\linewidth}{!}{%
\includegraphics{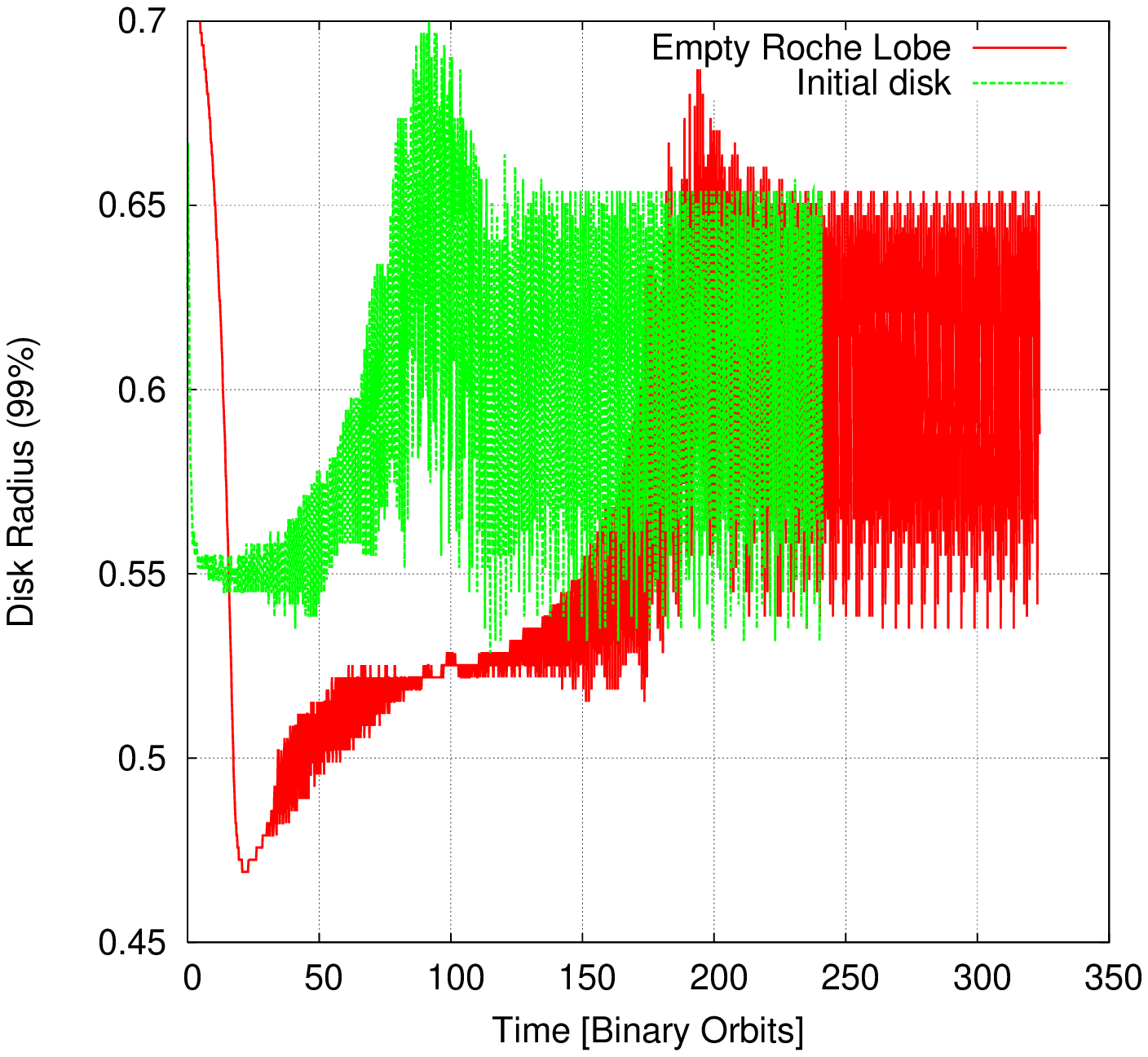}}}\\
\rotatebox{0}{
\resizebox{0.48\linewidth}{!}{%
\includegraphics{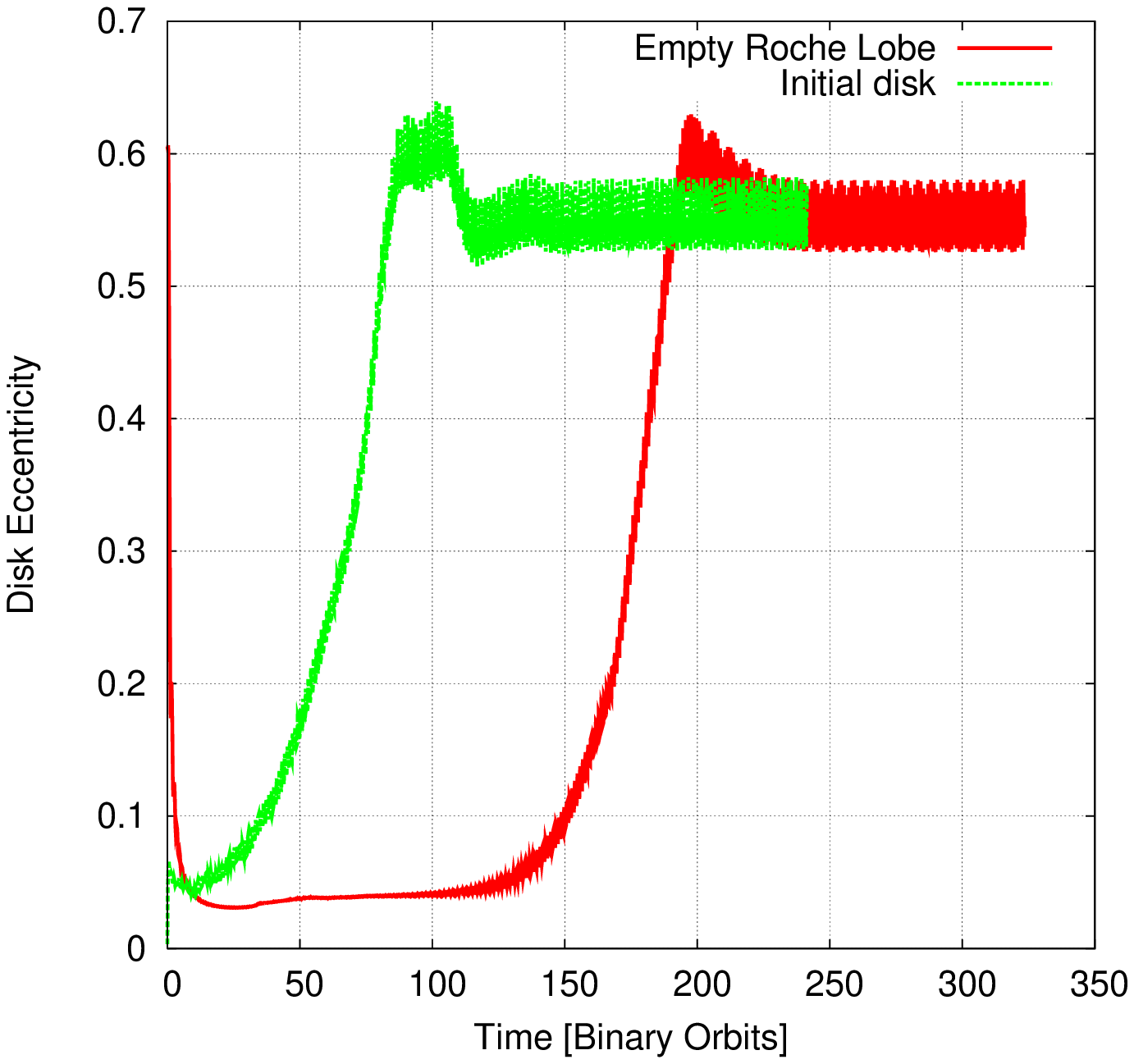}}}
\end{center}
  \caption{Disk mass $m\subscr{d}$, radius $r\subscr{d}$ and eccentricity $e\subscr{d}$
as functions of time
for models including mass inflow through the $L_1$ point with different initial
conditions.
The first (red) curve corresponds to an initially empty Roche lobe of the primary,
while in the second (green) a standard initialized disk
is used. The `viscous outflow' condition and a viscosity
$\nu=10^{-4}$ (ten times standard) are used.
}
   \label{fig:oyc01e-mult}
\end{figure}

\begin{figure}[ht]
\begin{center}
\rotatebox{0}{
\resizebox{0.98\linewidth}{!}{%
\includegraphics{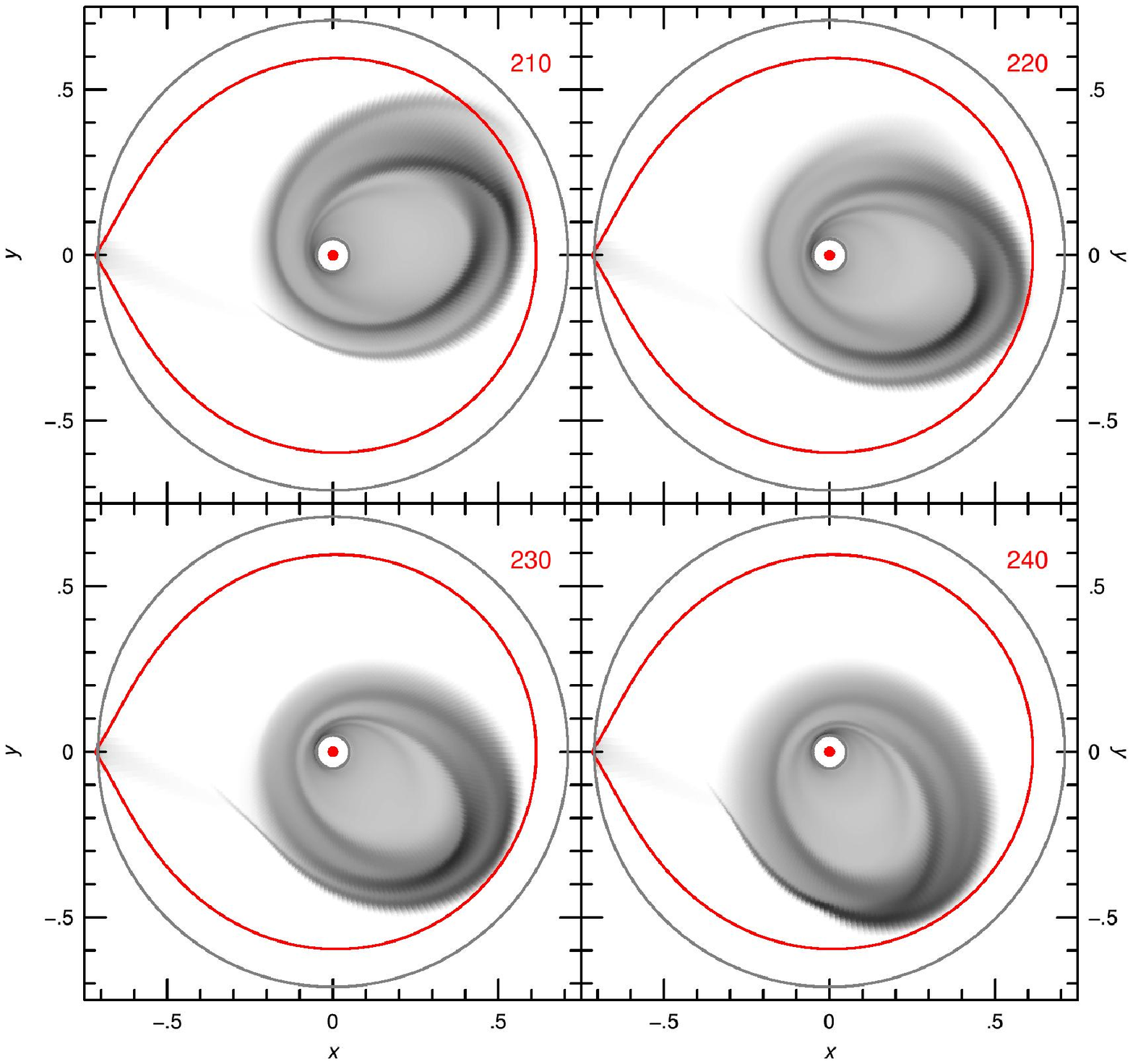}}}
\end{center}
\caption{Greyscale plot of the   surface density of the disk for the model
with stream inflow   that started with an initial disk at
$210, 220, 230$ and $240$ orbital periods. The solid (red) curve
indicates the Roche lobe of the primary star (central red dot). The `viscous outflow'
condition is used at the inner boundary.
}
\label{fig:oyc01e-xy-200ff}
\end{figure}

We then ran models with a larger viscosity $\nu = 10^{-4}$ 
 using the `viscous outflow' condition at the inner boundary. 
 The models were started  from both an empty Roche lobe and a
standard initial disk as in the previous simulations.
In these cases the disk attained an eccentric mode.
In both cases  an identical final state is reached, however much  more quickly
 for the model with
an initial disk.
The evolution of the total mass in the disk (within the primary's Roche lobe)
is displayed in the top left panel of Fig.~\ref{fig:oyc01e-mult}.
The two cases show a complementary behaviour:
in the case with an initial disk the total mass drops towards the final
equilibrium (at around $\log{m\subscr{disk}} = -4$) while for the empty case it rises initially,
overshoots the equilibrium point and settles to the equilibrium value roughly
100 orbits later.
Additional disk properties such as the disk radius, eccentricity and precession
rate are all independent of the initial condition as indicated by the other panels
in Fig.~\ref{fig:oyc01e-mult}. 

\subsection{Disk dynamics}
The dynamical influence and importance of the stream inflow can now be studied (irrespective
of initial conditions) through a direct
comparison with the corresponding  model without a stream; the results are summarized
in Table~\ref{tab:stream}. The mean disk eccentricity $e\subscr{d}$ and the disk radius $r\subscr{d}$
in the final equilibrium phase are very similar in both cases, with  an expected 
 smaller disk radius and slightly smaller eccentricity due to the incoming relatively  low angular
momentum disk material. The surprising difference lies in the
magnitude of the rate of  disk precession.  For the model without the stream the disk precesses
in a retrograde sense  quite rapidly,  performing  a full revolution in about 17 binary orbits. 
The impact of
the stream is such as to  slow down the precession such that it takes 
about 140 orbits to make a full revolution. 
\begin{table}[ht]
\centerline{
\begin{tabular}{llll}
\hline
Boundary condition   &  $e\subscr{d}$   &  $r\subscr{d}$   &  $\dot\varpi\subscr{d} [1/P\subscr{orb}]$    \\
\hline
No stream inflow     &   0.56   &  0.651   &  -0.060  \\
With stream inflow   &   0.55   &  0.610   &  -0.0072 \\
\hline
\end{tabular}
}
\caption{Comparison of two models with and without stream inflow, all other parameters being
identical. The viscosity is $\nu = 10^{-4}$ and at the inner boundary `viscous outflow'
conditions are applied.
The  model that has no stream  corresponds to the second model in Table~\ref{tab:res-bc}.
}
\label{tab:stream}
\end{table}

The   surface density distribution of the model with stream inflow and an
initial disk is shown   during the quasi-stationary 
 phase at 4  times separated by 10 binary orbits.
From Fig.~\ref{fig:oyc01e-xy-200ff} we can directly infer the  slow retrograde precession
of the disk which is also confirmed independently
by a phase analysis of the $m=1$ Fourier mode at the specific radius $r=0.37.$

\begin{figure}[ht]
\begin{center}
\rotatebox{0}{
\resizebox{0.98\linewidth}{!}{%
\includegraphics{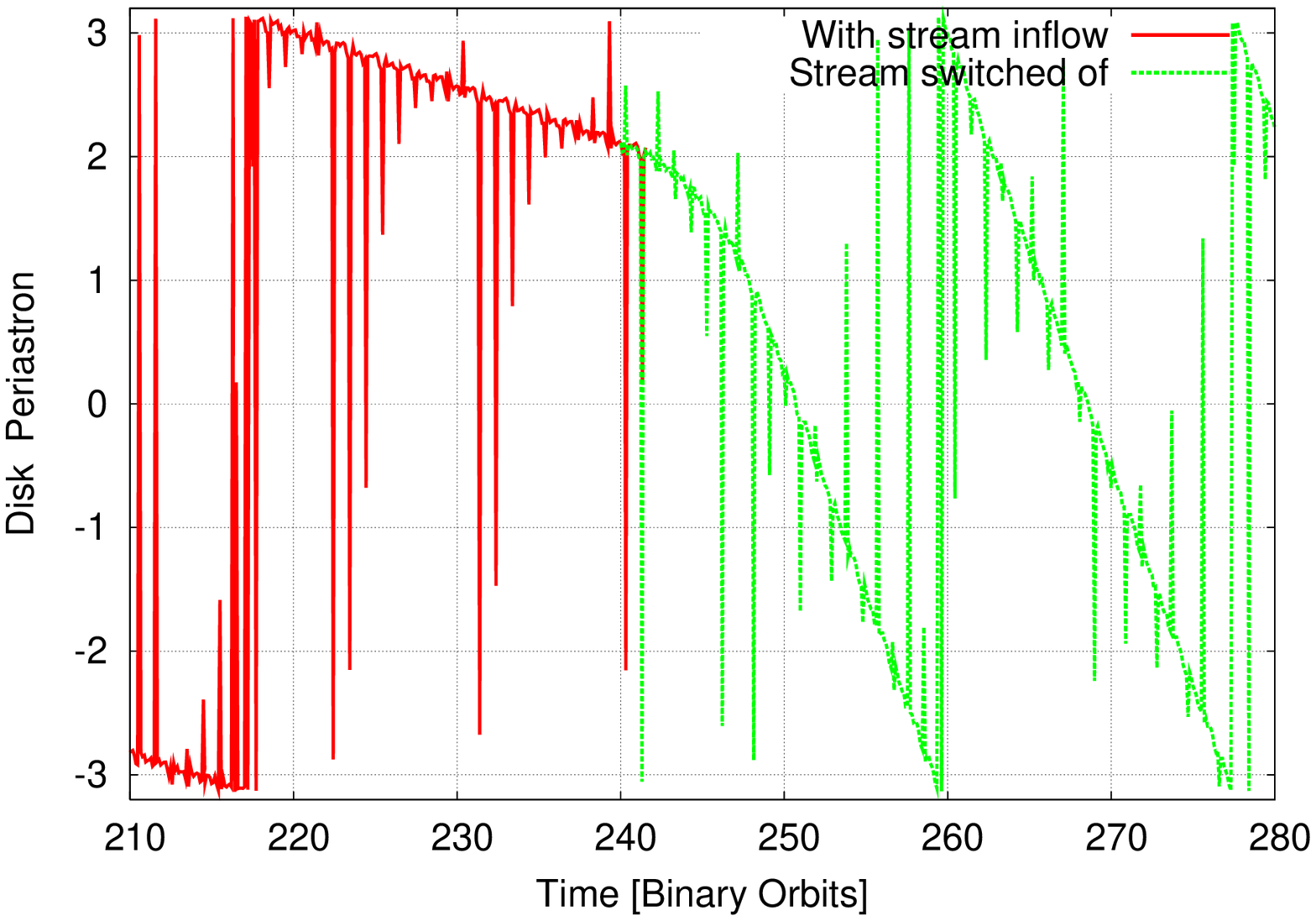}}}
\end{center}
  \caption{Longitude of the disk  pericentre 
(in the inertial frame) as a function of time. The solid (red) curve corresponds to
a disk in  a quasi-steady state 
 with stream inflow (as displayed in the previous Fig.~\ref{fig:oyc01e-xy-200ff}).
The light dashed (green) line corresponds to a model where the stream is switched off at $t=240$
and the simulation is continued.
}
   \label{fig:oyc01ex-om2}
\end{figure}

The change of the precession rate by about $\Delta \dot{\varpi}= 0.52 P\subscr{orb}^{-1}$
is also apparent from  Fig.~\ref{fig:oyc01ex-om2} where a  quasi-stationary model with  stream inflow
(from Fig.~\ref{fig:oyc01e-xy-200ff}) is continued after the stream is switched off. 
Clearly visible is the immediate transition to the state with no stream. The new
 average precession rate corresponds closely to the  case with no stream inflow.
Quite obviously, the disk precession rate is sensitive
to modifying conditions at the outer boundary
as well as at the inner boundary as we have seen above. 
Additional runs with cooler disks ($h=0.03$ and $0.04$) demonstrate that the transition from
prograde to retrograde disk precession occurs in the case with stream inflow for much larger
values around between $h=0.04$ and $0.05$, see upward triangles in Fig.~\ref{fig:hr-om}.
On the  other hand the disk radius and  eccentricity  are more robust.

\section{Conclusions}\label{Concl}

In this paper we have performed grid-based simulations of eccentric disks
in close binary systems. Adopting a  grid-based rather than a particle-based
method enables consideration of a wider and more appropriate
range of physical parameters. We have been able to consider
disk aspect ratios in the range $0.01 \le h \le 0.05$ and dimensionless
viscosities in the range $3.3 \times 10^{-6} \le \nu  \le 10^{-4},$
whereas particle-based methods are limited to the upper  limits of these ranges.
We note the relationship between $\nu$ and the  $\alpha$ parameter
of \citet{1973A&A....24..337S} $\alpha = \nu/h^2.$ Thus
$\nu=10^{-4}$  corresponds to $\alpha \sim  0.1$ for $h=0.03.$
We have also considered the effects of different inner boundary conditions, the effects of a mass-transfer
stream and considered and demonstrated numerical convergence.
However, due to heavy computational demands the grid-based simulations carried out
in this first survey  have been limited to two dimensions
and simply adopted a locally isothermal equation of state.
We summarize the main results of our simulations below.

\subsection{Instability to the formation of an eccentric disk}

We find instability to the formation of an eccentric disk in the mass range
$0.05 \le q \le 0.3$. Explorative calculations have shown instability even for 
larger mass ratios up to $q=1$ in these two-dimensional simulations, see
Fig.~\ref{fig:q-mult1}.
The most unstable cases are those with a reflecting
inner boundary, large viscosity, small aspect ratio and no mass-transfer stream.
An initially non-eccentric disk  quickly enters
 a linear phase of exponential growth of an unstable mode.

 For 
 $q=0.1$ we found growth times ranging between $162$ and $15$
binary orbits  for $\nu$  in the range  $3.3 \times 10^{-6}-10^{-4}$
 with $h$ in the range  $0.02-0.05.$
A quasi-stationary precessing eccentric disk is attained in all  cases
in about four growth times. In the most unstable cases (large $\nu$ and small $h$)
 this corresponds
to about $60$ binary orbits  or about $6$ days for superoutbursting
cataclysmic binaries. This is consistent with reported superhump development times
of a few days.

The mean eccentricities of the quasi-stationary disks
are typically in the range $0.3-0.6$ when there is no mass-transfer
stream and an inner reflecting boundary. This is in the upper part of the range
 $0.38\pm 0.1$  found for OY Car \citet{1992A&A...263..147H}
 and for IY Ma by
 \citet{2001MNRAS.324..529R}.

 However, it is important to note that results are
sensitive to the inclusion of a mass-transfer stream and the treatment
of the inner boundary condition. For the largest
viscosities, use of an inner outflow condition rather than a reflecting
boundary condition reduces the final mean  eccentricity from $0.5$ to $0.3.$ 
Incorporation of a mass-transfer stream does not affect the growth
of eccentricity for $\nu = 10^{-4}$ but we are able to see a transition to stability
to formation of an eccentric disk
at about $\nu = 3\times 10^{-5}$ in this case independently of the inner boundary condition.
This is also consistent with the idea that the effective viscosity and consequently
the mass-transfer rate through the disk
is larger during the outburst phases of cataclysmic binaries
during which superhumps are reported \citep{2005PASP..117.1204P}.

\subsection{Quasi-steady state of the eccentric disk}

Once the simulated disks go through an instability that
results in them becoming eccentric, they typically attain a 
characteristic quasi-steady state.
This was illustrated in Fig. \ref{fig:m02f-inert}. 
Viewed from an inertial frame 
the secondary has a close approach to the
 slowly precessing disk apocentre approximately once every orbit,
 pulling out a tidal tail. This later impacts back on the
disk which is not strongly affected by the secondary until
the next close approach. Thus  almost the entire  tidal dissipation may occur
as a result of these events. In this context we note that tidal
dissipation has been proposed as the source of the luminosity variations during the
main superhump phase while it has been proposed that effects due to modulation by the
mass-transfer stream become significant during the late superhump
phase
\citep{{1992A&A...263..147H},{2001MNRAS.324..529R},
{2005PASP..117.1204P}}.

\subsection{Instability mechanism}

We have investigated the mechanism responsible for the excitation
of the disk eccentricity and found this to be basically
consistent with
the mode-coupling mechanism of \citet{1991ApJ...381..259L}.
Neglecting the slow precession of the disk, viewed from the inertial frame
the  eccentricity of the disk may be considered to arise  from a stationary surface density perturbation with azimuthal mode number
$m=1.$ In addition the disk responds directly  to the tidal potential
of the companion  
producing a surface density response that corotates with it 
and consists of a sum of Fourier components  with  $m= 1,2,3,\dots$
with amplitudes that scale  
with the mass ratio, $q.$
Nonlinear
coupling between the response to the tidal potential and the $m=1$ mode
causes the excitation of secondary density waves with  pattern speeds
$m\omega/(m-1),$ $\ m= 2, 3, 4,\dots$ which exceed that of the binary
and are thus associated 
with a tidal  energy and angular momentum transfer  to it.  The 
back reaction on the $m=1$ mode causes the eccentricity to grow.
To work effectively the secondary waves should be associated with
an inner Lindblad resonance. The most favourable case corresponds to $m=3$
and the disk gas orbiting with $\Omega = 3\omega,$ at a 3:1
commensurability. In this case  the Fourier  component of the tidal potential
with $m=3$ plays a key role.

The results of simulations with $q=0.1$ supported this view with
growth being almost absent when the $m=3$ component was removed.
However, for $q=0.2$ growth still occurred in this situation  albeit at a reduced rate.
This is apparently because higher-order couplings generated through
the other Fourier components result in an effective forcing with $m=3.$
Being of higher order, this is more effective at larger $q.$ But it should be noted
that as $q$ increases, the $3:1$ commensurability is driven towards the Roche lobe
thus becoming increasingly nonlinear  and
affected by factors such as the outer boundary condition and any stream inflow.
As we indicate below, it is possible
that dissipative effects not included in our two-dimensional simulations play a
more important role. For these reasons we concentrated on the case $q=0.1$
in this  article.

\subsection{Disk precession rates and the dependence on disk aspect ratio}
The main determinants of the disk precession rate were found to be the 
disk aspect ratio and the impact of a mass-transfer stream if present.
Results for $0.01 \le h \le 0.05$  for $q=0.1$ are summarized in Fig. \ref{fig:hr-om}.

The quasi-steady  disk precession rate $\dot\varpi\subscr{d}$ was found to decrease
from positive values at small $h$ (prograde precession) to attain negative values
at larger $h.$ This is because, as expected, the pressure contribution
tends to produce retrograde precession. When no mass-transfer stream was present the
transition from prograde to retrograde occurred for $h \sim 0.025.$
We also determined the precession rates by independently solving the linear mode problem
using the average surface density profile and in view of the strong nonlinearity
present obtained surprising agreement with $h$ at the transition being $\sim 0.03.$

An  approximate parabolic
fit to the data points for the case with 
no mass-transfer stream was found to be
$\dot\varpi\subscr{d} = 0.022 -39h^2.$ This indicates that $\dot\varpi\subscr{d} \sim 0.02$
as indicated by observations \citep{2005PASP..117.1204P} only for $h \lsim 0.01.$
In view of the fact that values for $h$ estimated from
accretion disk modelling fall in the range
$0.01 \le h \le 0.03$ \citep{{2007MNRAS.378..785S}, {2006MNRAS.368.1123G}}, this is rather small.
Note that \citet{2006MNRAS.368.1123G}  also obtained similarly small estimates  from 
their one-dimensional linear formalism.

However, the simulations for which the disk was impacted by a mass-transfer stream
 show the transition for $h \approx 0.04$ and indicate the possibility of obtaining
 the observed precession rates for values of $h$ estimated from accretion disk modelling.
 Thus our results demonstrate the importance of the mass-transfer stream
 in this context as well as in the context of obtaining a transition from
 instability to stability at a reasonable value for the  disk viscosity.  On the other hand, it might be argued that the significance of the mass-transfer stream would 
be reduced during a superoutburst, when the accretion rate through the disk exceeds the mass-transfer rate from the companion.
 
 In addition we comment that there is sensitivity to the inner boundary
 condition such that changing from a reflecting condition to a dissipative
 or free outflow condition also acts to make the precession frequency
 move in the  prograde direction.
  
\subsection{Comparison with particle-based simulations}
The most recent relevant SPH simulations have been  
performed by \citet{2007MNRAS.378..785S}. Both two- and 
three-dimensional simulations have been considered.
Several important issues should be noted.
Such simulations are inevitably highly diffusive 
($\alpha \ge 0.1$) and thick ($h \ge 0.05$) so that
parameter studies at smaller $h$ and smaller $\alpha$
of the type considered here cannot be carried out.
Comparison has to be limited to cases with large $\nu$ and $h.$
The SPH simulations are performed with $\sim 10^5$ particles
and also with a mass-transfer stream. For a disk with $h =0.05$
 one can estimate that there is only about one smoothing length per
scaleheight in a three-dimensional simulation so that the
vertical thickness is poorly resolved. The three-dimensional SPH  simulations
tend to  produce a weaker phenomenon than that reported here with almost no  activity
for $q \gsim  0.2$ and
  mean disk eccentricities
 at the small end of the range implied by observation.
 In addition very long growth times approaching thousands
of orbits are generally reported. 

On the other hand   \citet{2005PASP..117.1204P}
indicate that the superhump phenomenon occurs for $q \le 0.35$
and that it is associated with high mass-transfer  events during 
superoutbursts and a short development time requiring growth times $\sim 20$ binary
orbits (see above). It may also occur in a quasi-steady manner  in systems
such as nova-like variables which are also believed to have
a high mass-transfer rate which in this case is steady. 
This is also consistent with our finding that
stability to eccentricity generation occurs at low viscosities and accordingly
low  mass-transfer rates through the disk when a mass-transfer stream is present. 

The reported two-dimensional SPH simulations are more in line with those
reported here. This suggests that the three-dimensional case
is more dissipative  though  the adequacy of the resolution in the vertical
direction should also be an issue.
It is likely that
inclusion of effects arising from the vertical structure of the disk
could result  in  processes leading to  
additional dissipation  \citep{2005A&AS..432..757P}
but these would need to be simulated with adequate resolution.
The two-dimensional grid-based simulations reported here
are possibly too
unstable to forming eccentric disks, particularly at larger $q$.
However, in those cases, the 3:1 resonance driving the instability is pushed towards the 
Roche lobe  where higher-order mode  couplings
and increasing nonlinearity occur  and numerical studies should include 
a treatment of the vertical direction if this region is to be correctly dealt with.

 Therefore an important  direction for future work will be to  develop
three-dimensional grid-based simulations which resolve the disk
vertical structure and address the issue of the importance
of possible  additional dissipation modes. 

\begin{acknowledgements}

Very fruitful discussions with Aurelien Crida and Roland Speith are  gratefully acknowledged.
W.K. acknowledges the very kind hospitality and support of  DAMTP during his
visit to Cambridge  in the spring of 2007.

\end{acknowledgements}

\bibliographystyle{aa}
\bibliography{kley,kley8}
\end{document}